\documentclass{aa}  
\usepackage{graphicx}
\usepackage{txfonts}
\usepackage{ulem}
\AtBeginDocument{
  \abovedisplayskip     =0.5\abovedisplayskip
  \abovedisplayshortskip=0.5\abovedisplayshortskip
  \belowdisplayskip     =0.5\belowdisplayskip
  \belowdisplayshortskip=0.5\belowdisplayshortskip}

\usepackage{color}

\newcommand{\Alfven}{Alfv\'{e}n }

\begin{document} 
   \title{Modeling of Magneto-Rotational Stellar Evolution}
   \subtitle{I. Method and first applications}

   \author{K. Takahashi \inst{1} \and
               N. Langer \inst{2,3}
	}
   \institute{Max-Planck-Institut f\"{u}r Gvravitationsphysik,
   		Am M\"{u}hlenberg 1, 14476 Potsdam-Golm, Germany\\
		\email{koh.takahashi@aei.mpg.de}
         \and
		Argelander-Institut f\"{u}r Astronomie, Universit\"{a}t Bonn,
   		Auf dem H\"{u}gel 71, 53121 Bonn, Germany
   		\and
   		Max-Planck-Institut f\"ur Radioastronomie, Auf dem H\"ugel 69,
   		53121 Bonn, Germany
		}

   \date{Received \today; accepted -}

  \abstract{While magnetic fields have long been considered to be important for the evolution of magnetic non-degenerate stars and compact stars, it has become clear in recent years that actually all of the stars are deeply affected.
  This is particularly true regarding their internal angular momentum distribution, but magnetic fields may also influence internal mixing processes and even the fate of the star.
  We propose a new framework for stellar evolution simulations, in which the interplay between magnetic field, rotation, mass loss, and changes in the stellar density and temperature distributions are treated self-consistently.
  For average large-scale stellar magnetic fields which are symmetric to the axis of rotation of the star, we derive 1D evolution equations for the toroidal and poloidal components from the mean-field MHD equation by applying Alfv\'{e}n's theorem, and a conservative form of the angular momentum transfer due to the Lorentz force is formulated.
  We implement our formalism into a numerical stellar evolution code and simulate the magneto-rotational evolution of 1.5\,M$_\odot$ stars.
  The Lorentz force aided by the $\Omega$ effect imposes torsional \Alfven waves propagating through the magnetized medium, leading to near-rigid rotation within the \Alfven timescale.
  Our models with different initial spins and B-fields can reproduce the main observed properties of Ap/Bp stars.
  Calculations continued to the red-giant regime show a pronounced core-envelope coupling, which reproduces the core and surface rotation periods determined by asteroseismic observations.}

   \maketitle
%
\newpage
\section{Introduction} \label{sec:intro}

A magnetic field is visible in many different types of stars. As the most evident example, the Sun shows magnetic activity such as spots, prominences, flares, and mass ejections \citep[e.g.,][]{Solanki06}. The Sun is considered to be prototypical of FGK type main-sequence stars, which are known to have convective envelopes, and most of these cool stars are known to host magnetic fields \citep{Landstreet92, Donati&Landstreet09}. These magnetic fields are thought to have a ``dynamo origin'', i.e., the field is continuously amplified through hydrodynamic induction, and would otherwise decay within the \Alfven timescale. This understanding is supported by observed correlations of the field strengths detected in these main-sequence stars with their fundamental parameters, such as the mass, age, and rotation periods \citep{Vidotto+14, See+15, See+16, Folsom+16, Folsom+18}.

On the upper main sequence, where stars have radiative envelopes, about 10\% of the stars are magnetic \citep{Landstreet92, Wade+14}. Their strong (typically $\sim$1 kG) fields are characterized by a large scale ($\sim$dipole) structure, and since neither convection nor rotation could currently produce these fields, they are thought to be stable over a significant part of the stellar lifetime \citep{Wade+00, Silvester+14}. In contrast to stars with convective envelopes, clear correlations between the field strength and major stellar parameters have not been found so far. Such properties indicate the ``fossil'' origin of the field: the strong magnetic field was somehow amplified in a past and reached a stable configuration through magneto-hydrodynamical relaxation \citep{Braithwaite&Spruit17}.

Magnetic fields are also found from evolved stars; GK-giants \citep{Auriere+15}, asymptotic-giant-branch (AGB) and post-AGB stars \citep[][and references therein]{Vlemmings14, Vlemmings19}, AFGK-yellow supergiants \citep{Grunhut+10}, and M-supergiants including $\alpha$ Ori \citep{Auriere+10, Tessore+17}, despite the field strengths in evolved stars is often small because of their large radii and slow rotation rates \citep{Auriere+08}. Furthermore, $\sim$10\% of the white dwarfs show magnetic fields, and while all neutron stars appear to have magnetic fields, some 10\% of them (known as magnetars) possess extremely strong surface magnetic fields \citep{Chanmugam92, Ferrario+15}.

Magnetic fields can have significant effects on stellar evolution. It is known for a long time that the spin evolution of solar-like stars is essentially coupled with the magnetic field evolution. The slow rotation rate of the Sun has been understood as a result of the magnetic braking \citep{Weber&Davis67}, and the solar wind itself is largely driven by the surface magnetic activities \citep[e.g.,][]{Parker58, Ofman10}. The magnetic stress of the amplified field inside the convective envelope contributes to the angular momentum transfer together with the turbulent viscosity and the mean meridional flow to determine the rotation profile in the envelope \citep{Brandenburg18}. Moreover, the convective dynamo is the result of the symmetry breaking due to stellar rotation \citep{Brun&Browning17}.

The evolution of main-sequence stars with radiative envelopes can also be significantly affected by magnetic fields. Because of the intrinsically strong stellar wind, magnetic braking for massive OB-type stars can be strong enough to become directly observable \citep[e.g., $\sigma$ Ori E;][]{Townsend+10}. A strong surface field can trap the wind material into a corotating magnetosphere, which provides detailed explanations for X-ray emission of IQ Aur \citep{Babel&Montmerle97} and phase variations of Balmer-line emissions of $\sigma$ Ori E \citep{Townsend&Owocki05}.

Theoretical works have also revealed the importance of the magnetic field in stellar evolution. For instance, magnetic stress furnished by internal fields may account for the efficient angular momentum transfer in radiatively stratified regions \citep{Spruit99}. Many stellar evolution simulations take magnetic angular momentum transfer into account, following a model originally proposed by \citet{Spruit02} and further reevaluated by other authors \citep{Maeder&Meynet03, Heger+05, Suijs+08, Denissenkov&Pinsonneault07, Fuller+19, Ma&Fuller19}. Magnetic braking, as well as magnetic wind confinement, have been considered in the context of massive star evolution \citep{Meynet+11, Petit+17, Georgy+17, Keszthelyi+19}. Sufficiently strong internal fields can modify the stellar structure via magnetic pressure and tension, and more importantly, by affecting the adiabatic indices, which modifies the efficiency of convective energy transport. Such effects have been considered in \citet{Feiden&Chaboyer12,Feiden&Chaboyer13,Feiden&Chaboyer14} for low-mass star evolution, who followed prescriptions developed by \citet{Lydon&Sofia95}.

The majority of theoretical works have modeled magnetic stars by individually considering the specific magnetic effects. However, all effects, in reality, must relate to each other, because fundamentally they are governed by the identical field. To consider the integrated effect, the global structure of the magnetic field has to be modeled. Besides, since the magnetic field should evolve in time as well as other physical quantities, the time-dependent treatment is desired. Such a treatment has recently been derived by \citet{Potter+12c}. In their framework, the global field structure is significantly simplified to have axial symmetry and dipole-like structure. Furthermore, they have formulated the evolution equation of the global magnetic field based on a mean-field MHD equation. As a consequence, important processes such as the $\alpha$- and $\eta$-effects, which express the interaction between turbulence and magnetic field, as well as the $\Omega$ effect, in which differential rotation winds up poloidal magnetic component to enhance the toroidal component, are incorporated.

Meanwhile, the prescription in \citet{Potter+12c} still has some incompleteness. For example, it is likely that their evolution equations for magnetic fields do not reproduce the magnetic flux conservation, one of the most fundamental outcomes of the ideal MHD assumptions. Similarly, their expression of the angular momentum transport by the Lorentz force does not reproduce the angular momentum conservation. These problems arise because of the ambiguity that exists in the averaging process to formulate 1D evolution equations starting from the more general 3D MHD equation.

Here, we present a new framework, in which mutual interaction of the magnetic field and the rotation during the evolution of the star is treated in a physically consistent manner. In the next section, equations that describe the evolution of the stellar magnetic field and stellar rotation, assumptions and approximations made for the formulation, and a brief description of the numerical construction are provided. In Sect.\,3, we show that our formulation including the $\Omega$ effect and the Lorentz force naturally leads to a torsional \Alfven wave, in which differential rotation and toroidal magnetic field propagates together inside the star. We further discuss that the \Alfven wave accounts for a highly efficient mechanism of angular momentum redistribution when realistic dissipation is taken into account. While we plan to apply this new formulation to general-purpose stellar evolution simulations in the future, we provide here results of magneto-rotational evolution calculations for stars of 1.5\,M$_{\odot}$ to demonstrate the capabilities and limitations of our formulation. The corresponding main-sequence evolution is analyzed in Sect.\,4.1, and the red-giant phase is presented in Sect.\,4.2. In Sect.\,5, we discuss comparisons between our simulation results and other theoretical models (Sect.\,5.1) and relevant observations (Sect.\,5.2). Conclusions are given in Sect.\,6.

\section{Methods} \label{sec:method}
To compute the time evolution of stellar models we use the 1D stellar evolution code {\it HOSHI} \citep{Takahashi+16, Takahashi+18}. The code iteratively solves the four structure equations of mass conservation, the momentum balance equation in hydrostatic or hydrodynamic form, the equation of energy conservation in the form of an entropy equation, and the energy transport equation, by the so-called Henyey method. The equation of state in the code consists of a mixture of ideal gases of photon, averaged nuclei, electron, and positron. An analytical treatment of \citet{blinnikov+96} is applied for the electron-positron gas. The free energy of the Coulomb interaction for degenerate states is included \citep{salpeter&van-horn69, slattery+82}, and ionization of hydrogen, helium, carbon, nitrogen, and oxygen is also treated by solving the Saha equation. For the opacity, the Rosseland mean opacity of the OPAL project \citep{Iglesias&Rogers96} is used together with the conductive opacity by \citet{Potekhin+06} and the molecular opacity by \citet{Ferguson+05}.

In addition to the structure equations, the evolution of the abundances of the chemical species is solved through a reaction--diffusion equation as
\begin{eqnarray}
	\frac{\partial Y_i}{\partial t} = \dot{Y_i}_{\rm ,reac}
						 + \frac{\partial}{\partial M}\left(
							(4 \pi \rho r^2)^2 D_{\rm eff} \frac{\partial Y_i}{\partial M}
						\right), \label{eq-ycons}
\end{eqnarray}
where $Y_i, \dot{Y_i}_{\rm ,reac}$, and $D_{\rm eff}$ are the mole fraction of $i$ th isotope, the rate of change of $Y_i$ due to nuclear reactions, and the effective chemical diffusivity, respectively. 49 isotopes\footnote{The list of the 49 isotopes can be found in \citet{Takahashi+19}.} are considered in this work, and the reaction rates are taken from the current version of JINA REACLIB \citep{Cyburt+10}, except for the $^{12}$C($\alpha$,$\gamma$)$^{16}$O-rate, for which we use the rate from \citet{Caughlan&Fowler88} multiplied by a factor of 1.2. The Ledoux criterion is used to evaluate convective instability. The standard mixing-length-theory \citep{Boehm-Vitense58} is applied to compute the amount of energy transported by convection.

The mixing-length theory also provides the diffusion coefficient for chemical mixing in convection zones as $D_{\rm cv} = \frac{1}{3} v_{\rm cv} l_{\rm cv}$, where $v_{\rm cv}$ is the velocity of convective eddies, $l_{\rm cv} = \alpha_{\rm MLT} {\rm min}(H_P, r)$ is the length scale of the convective flow, and $\alpha_{\rm MLT}$ and $H_P$ are the mixing-length parameter and the pressure scale height. To consider the effect of convective overshooting, the eddy velocity for regions surrounding the convective region is calculated as
\begin{eqnarray}
	v_{\rm cv} = v_{\rm cv,0} \exp \left( 
				-2\frac{\Delta r}{f_{\rm ov} h_{P, 0}}
	 \right),
\end{eqnarray}
where $f_{\rm ov}$ is an adjustable parameter determining the e-folding length scale, $v_{\rm cv,0}$ and $h_{P, 0}$ are the eddy velocity and the pressure scale height at the edge of the convective region, and $\Delta r$ is the distance from the edge. This treatment yields a similar diffusion coefficient distribution to the exponential diffusive overshoot described in \citet{Herwig00}.

In semiconvective layers, we apply a diffusion coefficient of the form
\[
	D_{\rm cv} = f_{\rm sc} D_{\rm therm} \frac{ \nabla_{\rm rad}-\nabla_{\rm ad} }{ (\phi/\delta) \nabla_{\mu}},
\]
and thermohaline convection is treated with 
\[
	D_{\rm cv} = f_{\rm thh} D_{\rm therm} \frac{ -(\phi/\delta) \nabla_{\mu}}{ \nabla_{\rm ad}-\nabla_{\rm rad} }
\]
\citep{Kippenhahn+80, Wellstein+01, Siess09}. Here 
$D_{\rm therm} \equiv (1/C_P \rho) (4acT^3/3\kappa \rho)$,
$\nabla_{\rm rad} \equiv (3\kappa/16\pi acG)( P L/T^4 M )$,
$\nabla_{\rm ad} \equiv (\partial \ln T/ \partial \ln P)|_{s=\rm{const.}}$,
$\nabla_{\rm \mu} \equiv d \log \mu/ d \log P$,
$\delta \equiv -(\partial \ln \rho / \partial \ln T)_{P, \mu}$,
$\phi   \equiv  (\partial \ln \rho / \partial \ln \mu)_{P, T}$,
are the thermal diffusivity,
the radiative temperature gradient,
the adiabatic temperature gradient,
the $\mu$-gradient,
and relevant thermodynamic derivatives, respectively, with
$\kappa$ being the Rosseland mean opacity.
For the two control parameters, $f_{\rm sc}=0.3$ and $f_{\rm thh}=1.0$ are used.

The {\it HOSHI}-code includes stellar wind induced mass and angular momentum loss. For models presented in this work with low effective temperatures of $\log T_{\rm eff} {\rm [K] } < 3.9$, an empirical mass-loss formula by \citet{deJager+88} is applied. Consequently, 1.5\,M$_\odot$ nonrotating model experiences wind mass loss of 
$\sim 3 \times 10^{-12}$ M$_\odot$ yr$^{-1}$
and 
$\sim 10^{-11}$--$10^{-8}$ M$_\odot$\,yr$^{-1}$ 
during the main-sequence and red-giant phases, respectively.

\subsection{Stellar rotation}
The effects of stellar rotation are similarly taken into account as described in \citet{Takahashi+14}. To describe a rotating star in a 1D formulation we assume shellular rotation, where all material on an isobaric surface shares the same angular velocity \citep{Zahn92}. We define the volume and mass enclosed by an isobar as either $V_P$ and $M_P$. Accordingly, the mean radius is defined as $4 \pi r_P^3 /3 = V_P$. Thus the angular velocity $\Omega$ is defined as a function of the mass coordinate $M_P$ in our simulation. The evolution equation of $\Omega$ is derived in Section \ref{sec:angmom} after we describe our treatment of the stellar magnetic field.

To consider the effect of deformation due to the centrifugal force, the isobar is assumed to have a shape described by
\begin{eqnarray}
	r (\theta) = a [ 1-\epsilon P_2 (\cos \theta) ],
\end{eqnarray}
where $P_2$ is the second-degree Legendre polynomial. Here, $\epsilon \equiv (\Omega^2 r_P^3 / 2GM_P) (a/r_P)^3 $ indicates the degree of rotation compared with the local gravity, and the length scale $a$ satisfies the relation \citep{Denissenkov&VandenBerg03},
\begin{eqnarray}
	r_P = a \left(
		1 + \frac{3}{5}\epsilon^2 - \frac{2}{35} \epsilon^3
	\right)^{1/3}.
\end{eqnarray}

The centrifugal force not only deforms the isobars but also affects the pressure balance and temperature gradient in the star. Following \citet{Endal&Sofia76}, these effects are taken into account in the structure equations by introducing parameters of $f_P$ and $f_T$ as
\begin{eqnarray}
	\frac{\partial P}{\partial M_P} &=& - \frac{GM_P}{4 \pi r_P^4} f_P
			- \frac{1}{4 \pi r_P^2} \left(  \frac{\partial^2 r_P}{\partial t^2}  \right)\\
	\frac{\partial \log T}{\partial \log P} &=& \left \{
		\begin{array}{ll}
			\nabla_{\rm rad}\frac{f_T}{f_p} \left[
						1 + \frac{r_P^2}{G M_P f_P} \left(  \frac{\partial^2 r_P}{\partial t^2} \right )
					\right]^{-1} &\mathrm{ \ (in \ radiative \ layers)} \\
			\nabla_{\rm MLT} &\mathrm{ \ (in \ convective \ layers)} 
		\end{array}
	\right.
\end{eqnarray}
where $\nabla_{\rm MLT}$ is the convective temperature gradient determined by the mixing-length theory. 
The parameters $f_P$ and $f_T$ are calculated as
\begin{eqnarray}
	f_P &=& \frac{4 \pi r_P^4}{G M_P S_P} \frac{1}{\langle g^{-1} \rangle} \\
	f_T &=& \left( \frac{4 \pi r_P^2}{S_P} \right)^2 \frac{1}{\langle g \rangle \langle g^{-1} \rangle},
\end{eqnarray}
where $g$ is the local effective gravity, in which the centrifugal force is taken into account.
The pointy brackets imply an average over the isobaric surface $S_P$ as $\langle q \rangle \equiv \frac{1}{S_P} \int q d \sigma$.

The wind mass-loss rate is enhanced when the stellar rotation is considered and the surface condition approaches the $\Omega \Gamma$ limit \citep{Langer98, Maeder&Meynet00}. We apply an enhancement according to \citet{Yoon+10, Yoon+12} as
\begin{eqnarray}
	\dot{M} = - \min \left[
		| \dot{M}(v_{\rm rot} = 0) | \times \left( 1-\frac{v_{\rm rot}}{v_{\rm crit}} \right),
		0.3 \frac{M}{\tau_{\rm KH}}
	\right], \label{eq:omegagamma}
\end{eqnarray}
where $| \dot{M}(v_{\rm rot} = 0) |$ is the mass loss rate of a nonrotating counterpart having the same luminosity and effective temperature, $v_{\rm rot}$ and $v_{\rm crit} \equiv v_{\rm K} \sqrt{1-(L/L_{\rm Edd})} \equiv \sqrt{ GM/R }\sqrt{1-(L/L_{\rm Edd})}$ are the rotation velocity and the critical rotation velocity at the stellar equator, $\tau_{\rm KH}$ is the Kelvin--Helmholtz timescale, and $L_{\rm Edd} \equiv 4 \pi c G M/\kappa$ is the Eddington luminosity.

Even without the help of a magnetic field, the angular momentum loss by stellar winds may be one order of magnitude more efficient than the accompanying mass loss when efficient angular momentum transport takes place at the subsurface region of the star \citep{Langer98}. The rate of angular momentum loss for a non-magnetic star is calculated as
\begin{eqnarray}
	\dot{J} = j_{\rm surf} \dot{M},
\end{eqnarray}
where $j_{\rm surf}$ is the specific angular momentum at the surface of the star. In our current models, $j_{\rm surf}$ is treated as a constant during the mass change, which implies the angular momentum is quickly redistributed throughout the subsurface region.

\subsection{Stellar magnetism}

As the most fundamental assumption, we assume that a large-scale stable magnetic field is embedded inside a star. Here, `stable' means that the magnetic field is in a magneto-hydrostatic quasi-equilibrium and keeps its structure, especially the geometry, for a considerably longer timescale than the \Alfven time. From a theoretical point of view, the structure, or even the existence, of such a stable magnetic field is far from trivial, which has been shown both, analytically and numerically (e.g., \citealt{Tayler73, Wright73, Markey&Tayler73, Markey&Tayler74, Braithwaite&Spruit04, Duez&Mathis10, Akguen+13}; \citealt{Braithwaite&Spruit17} and references therein). Nonetheless, our assumption may be reasonable at least for the radiative stellar envelopes close to the surface, because observations have shown that the surface magnetic field of radiative main-sequence stars displays a large scale structure and long-time stability (more than decades, i.e., longer than the \Alfven timescale of $\sim 10$ yr) \citep[e.g.][]{Landstreet92}. Indeed, 3D MHD simulations have shown that initially random fields can relax into a stable hydrostatic equilibrium in radiative stellar envelopes, after a magneto-hydrodynamical adjustment with an \Alfven time. The so obtained equilibrium states often have axial symmetry, and a comparable or stronger toroidal than poloidal component \citep{Braithwaite&Spruit04, Braithwaite&Nordlund06, Braithwaite09}. Justifications are more scarce for the convective layers since the fields are expected to vary on the convective timescale and to show a small scale structure. However, even in this case, our method may still allow following the time-averaged evolution of the mean magnetic field strength in the convective layers.

Although the large-scale stellar fields are assumed to be in a magneto-hydrostatic equilibrium, they can evolve with time responding to environmental changes. The most evident example is the density evolution due to the background thermal and/or elemental change, which leads to the field strength evolution as a consequence of the magnetic flux conservation. The next such mechanism is the $\Omega$ effect, in which differential rotation inside the star winds up the poloidal magnetic component to induce the additional toroidal component. We note that differential rotation can exist in a hydrostatic object since rotation does not require any driving force. Finally, we also treat the effects of turbulence on the magnetic fields as such a background effect. More specifically, any turbulence in our model is regarded as a perturbation that does not affect the hydrostatic pressure balance, and its effects on thermal, elemental, rotational, and magnetic transports are expressed through simple effective theories. For the turbulent effects on the magnetic field, we take into account the $\alpha$ effect, which expresses the turbulent induction of the magnetic field, and the $\eta$ effect, which shows the turbulent magnetic diffusivity, according to the mean-field MHD-dynamo equation \citep{Brandenburg18}.

Although deformation effects due to rotation are included in the stellar structure equations as described above, those are small unless the star rotates close to critical. Therefore, spherical symmetry is assumed when deriving the governing equations of the magnetic field. This treatment makes the derivation much easier. Generalization of the derivations to the case of a star with a deformed structure is beyond the scope of the current work.

\subsubsection{Simplification of the magnetic field}

The stellar magnetic field is assumed to be axially symmetric with respect to the rotation axis of the star. The field is divided into a poloidal and a toroidal component:
\begin{eqnarray}
	\vec{B}(r, \theta) &\equiv& \vec{B}_{\rm pol}(r, \theta) + \vec{B}_{\rm tor}(r, \theta)\\
	\vec{B}_{\rm pol} &=& B_r(r, \theta) \vec{e}_r + B_\theta(r, \theta) \vec{e}_\theta \\
	\vec{B}_{\rm tor} &=& B_\phi(r, \theta) \vec{e}_\phi,
\end{eqnarray}
where $B_r, B_\theta$, and $B_\phi$ are the $r, \theta$, and $\phi$ components of the magnetic fields, respectively, and they are functions of the radius, $r$, and the colatitude, $\theta$.

Because the magnetic field satisfies the solenoidal (the divergence-free) condition, one can find a vector potential $\vec{A}$ that satisfies
\begin{eqnarray}
	\vec{B} = \nabla \times \vec{A}.
\end{eqnarray}
To express the poloidal magnetic field, we utilize the toroidal component of the vector potential, $\vec{A}_{\rm tor} = A_\phi \vec{e}_\phi$, as
\begin{eqnarray}
	\vec{B}_{\rm pol} = \nabla \times \vec{A}_{\rm tor},
\end{eqnarray}
so that not only $\vec{B}_{\rm tor}$ but also $\vec{B}_{\rm pol}$ naturally satisfies the solenoidal condition. Because of the axial symmetry, the poloidal components of the magnetic field can be related to $A_\phi$ as
\begin{eqnarray}
	B_r(r, \theta) 		&=& \frac{1}{r \sin \theta} \frac{\partial}{\partial \theta}(A_\phi \sin \theta) \\
	B_\theta(r, \theta)	&=&-\frac{1}{r} \frac{\partial}{\partial r}(A_\phi r).
\end{eqnarray}

To express the magnetic field evolution by a 1D method, the latitudinal dependence of the magnetic field has to be somehow determined. As for the simplest case, we approximate that the poloidal field has the same latitudinal dependence as a dipolar field, thus
\begin{eqnarray}
	A_\phi(r,\theta) \equiv A(r) \sin \theta,
\end{eqnarray}
which yields
\begin{eqnarray}
	B_r(r, \theta) 		&=& \frac{2A}{r} \cos \theta \\
	B_\theta(r, \theta)	&=&-\frac{\sin \theta}{r} \frac{\partial (Ar)}{\partial r}.
\end{eqnarray}

To achieve a magneto-hydrostatic state, this poloidal field will be accompanied by a toroidal component, with a comparable or stronger field strength than the poloidal field \citep{Braithwaite&Nordlund06, Braithwaite09}, since otherwise, the poloidal field is unstable \citep{Wright73, Markey&Tayler73, Markey&Tayler74}. We may assume that this toroidal component has the polarity of $\sin \theta$. Besides, we take a toroidal field into account that is additionally induced by the $\Omega$ effect. Such a toroidal component should have the same polarity as the original poloidal field, and we take one of the simplest latitudinal dependencies as $\sin 2 \theta$. Thus we write
\begin{eqnarray}
	B_\phi(r, \theta) 		&=& B_{\rm stb}(r) \sin \theta + B(r) \sin 2\theta.
\end{eqnarray}
However, we will see below in Section\,\ref{sec:method-evolution} that only the toroidal component induced by the $\Omega$ effect is capable of transferring angular momentum. Therefore, the evolution of $B(r)$ only will be discussed in the subsequent sections (Sect.\,\ref{sec:wave1}, \ref{sec:result-MS}, \ref{sec:result-RG})

\subsubsection{Evolution equations for the magnetic field} \label{sec:method-evolution}

Consider a surface $S$ that is embedded in a star and moves with time with velocity field $\vec{V}(\vec{r}, t)$. Let $\vec{P}$ be the time-dependent vector field, which will be later regarded as the magnetic field or the electric current field. Then the flux of $\vec{P}$ on $S$ is defined as $\Phi \equiv \int_{S} \vec{P} \cdot \vec{n} d S = \int_{S} \vec{P} \cdot d \vec{S}$, where $\vec{n}$ is the normal vector of $S$. In this situation, Alfv\'{e}n's theorem tells that the total time derivative of $\Phi$ is written as
\begin{eqnarray}
	\frac{ D \Phi }{D t} = \int_S \left(
		\frac{\partial \vec{P}}{\partial t}
		- \nabla \times (\vec{V} \times \vec{P})
	\right) \cdot d \vec{S}, \label{eq-flux}
\end{eqnarray}
if and only if $\nabla \cdot \vec{P} = 0$ is satisfied. For interested readers, an elementary proof is given in Appendix\,\ref{sec:app-theorem}.

The macroscopic evolution of the magnetic field may be described by the mean-field MHD-dynamo equation:
\begin{eqnarray}
	\frac{\partial \vec{B}}{\partial t} =
		\nabla \times ( \vec{v} \times \vec{B} + \vec{ \alpha } \cdot \vec{B} )
		- \nabla \times \left( (\eta+\eta_t) \nabla \times \vec{B} \right), \label{eq-dynamo}
\end{eqnarray}
where $\vec{v}$ is the fluid velocity,
$\eta$ is the magnetic diffusivity, and
$\vec{ \alpha }$ and $\eta_t$ are the pseudo-tensors indicating the $\alpha$-effect
and the turbulent magnetic diffusivity \citep[][and references therein]{Brandenburg18}.
Because of the large magnetic Reynolds number in a stellar environment, $\eta$ is almost dominated by the turbulent magnetic diffusivity. Hence, hereafter we rename the total magnetic diffusivity as $\eta + \eta_t \rightarrow \eta$.

Since $\nabla \cdot \vec{B} = 0$, eq. (\ref{eq-flux}) can be applied to the magnetic field. Substituting eq. (\ref{eq-dynamo}) into eq. (\ref{eq-flux}), the evolution equation of the magnetic flux is obtained as
\begin{eqnarray}
	\frac{ D \Phi_B }{D t}
	&=& \oint_C \left(
		\vec{U} \times \vec{B} + \vec{ \alpha } \cdot \vec{B}
		- \eta \nabla \times \vec{B}
	\right) \cdot d \vec{l}, \label{eq-Bflux1}
\end{eqnarray}
in which Stokes' theorem is applied, and $\Phi_B \equiv \int_{S} \vec{B} \cdot d \vec{S}$ and $\vec{U} \equiv \vec{v} - \vec{V}$.

\begin{figure}[t]
	\begin{center}
  	\includegraphics[width=0.5\textwidth]{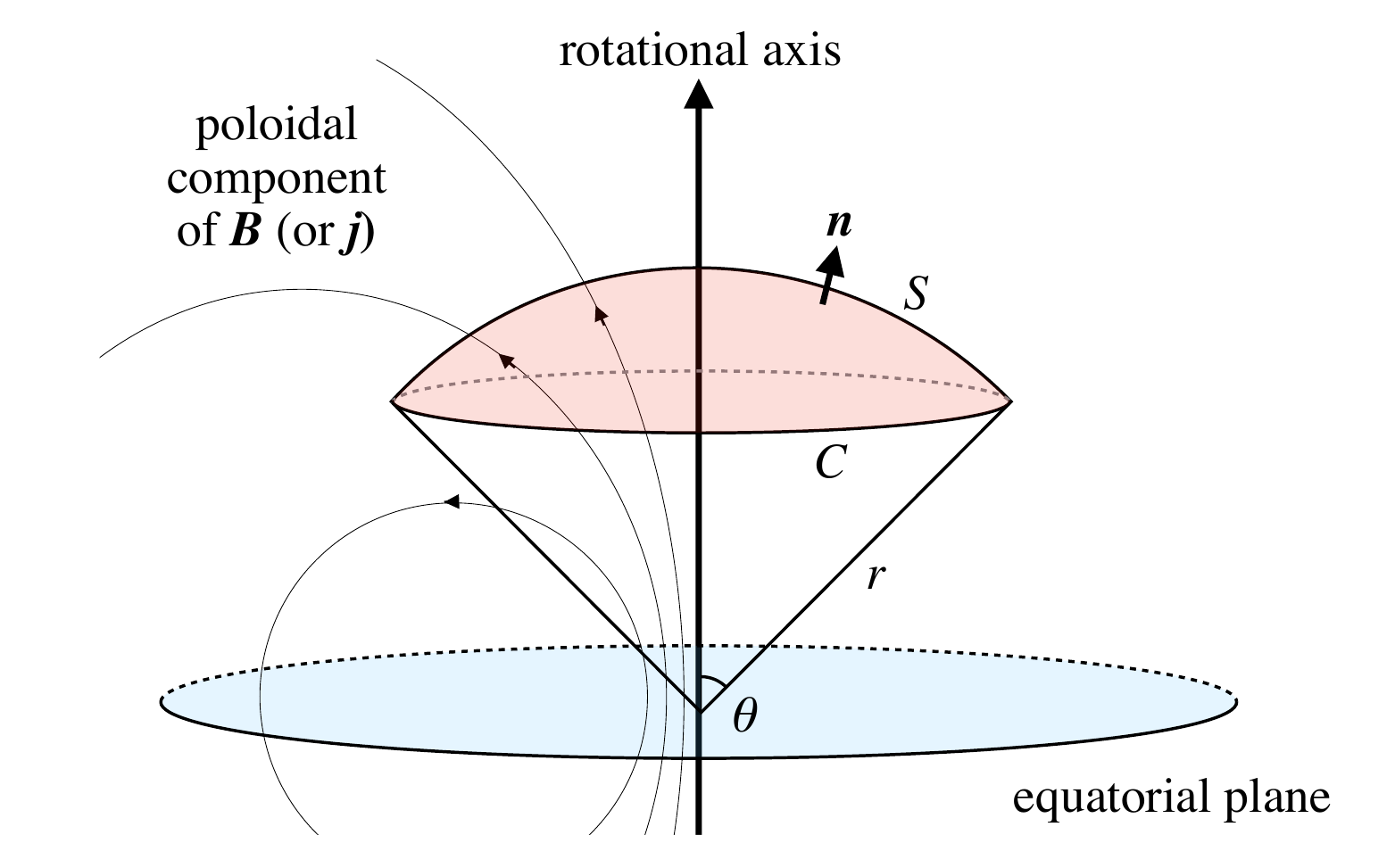}
	\caption{A schematic illustration of the geometry used to define fluxes $\Phi_B$ and $\Phi_j$. $S$, which is the surface of a polar cap with a radius $r$ and the opening angle $\theta$, is shown by the red shaded area. Fluxes are defined as $\Phi_B \equiv \int_S \vec{B} \cdot \vec{n} dS$ and $\Phi_j \equiv \int_S \vec{j} \cdot \vec{n} dS$.
	}
	\label{fig:geometry1}
	\end{center}
\end{figure}
For practical use, the surface $S$ together with the boundary $C$ has to be defined. Here we take the surface $S$ as a polar cap with radius $r$ and opening angle $\theta$, and accordingly, the boundary $C$ is taken to be a conic section that is located at the angle $\theta$ (Fig. \ref{fig:geometry1}). Considering the axial symmetry of the magnetic field, magnetic flux penetrating the surface is written as
\begin{eqnarray}
	\Phi_B	&=& 2 \pi r \sin \theta A_\phi.
\end{eqnarray}
Similarly, the right hand side of eq. (\ref{eq-Bflux1}) becomes
\begin{equation}
\begin{split}
	&\oint_C \left(
		\vec{U} \times \vec{B} + \vec{\alpha} \cdot \vec{B}
		- \eta \nabla \times \vec{B}
	\right) \cdot d \vec{l} \\
	& \quad = 2 \pi r \sin \theta \left(
		\vec{U} \times \vec{B} + \vec{\alpha} \cdot \vec{B}
		- \eta \nabla \times \vec{B}
	\right)_\phi. 
\end{split}
\end{equation}

We take the Lagrangian expansion velocity $v_m$ as the velocity of the polar cap, thus $\vec{V} = v_m \vec{e}_r$. The vector field $\vec{U} \equiv \vec{v} - \vec{V}$ now means a flow other than that due to stellar expansion or contraction. For the sake of simplicity, we consider that only the rotational flow contributes to $\vec{U}$. In other words, we tentatively neglect the advection of the magnetic field by the meridional flow. In the end, the evolution equation of $A_\phi$ is obtained as
\begin{equation}
\begin{split}
	\frac{ d (A_\phi r) }{ dt }
		&= r( \vec{\alpha} \cdot \vec{B})_\phi \\
		&\quad +	\eta \left[
				\frac{\partial^2}{\partial r^2} (A_\phi r)
			+	\frac{1}{r}	\frac{\partial}{\partial \theta}
				\left(
					\frac{1}{\sin \theta} \frac{\partial}{\partial \theta} (A_\phi \sin \theta)
				\right)
			\right].
\end{split}
\end{equation}
Here and hereafter, a time derivative at constant radius is shown by $\partial/\partial t$, while a time derivative at constant mass coordinate is shown by $d/dt$. As we assume $A_\phi = A(r) \sin \theta$, we obtain the evolution equation of $A(r)$ as
\begin{eqnarray}
	\frac{ d (A r) }{ dt }
		&=& r \frac{ (\vec{ \alpha } \cdot \vec{B})_\phi }{ \sin \theta }
		+	\eta r \frac{\partial}{\partial r}\left(
				\frac{1}{r^2} \frac{\partial}{\partial r} (A r^2)
			\right). \label{eq-Bflux2}
\end{eqnarray}

Equation (\ref{eq-flux}) can also be applied to the electric current field $\vec{j}(\vec{r}, t)$ since $\vec{j}$ is proportional to the curl of $\vec{B}$ and thus divergence-free assuming MHD. Using the evolution equation of the electric current, the evolution equation of the electric current flux,
\begin{equation}
\begin{split}
	\frac{ D \Phi_j }{D t}
	= \frac{c}{4\pi} \oint_C 
	&\{ \nabla \times ( \vec{v} \times \vec{B} + \vec{\alpha} \cdot \vec{B} ) \\
	&\quad	- \nabla \times \left( \eta \nabla \times \vec{B} \right)
		- \vec{V} \times (\nabla \times \vec{B})
	\} \cdot d \vec{l}, \label{eq-jflux1}
\end{split}
\end{equation}
is obtained.
Owing to the axial symmetry, the electric current flux can be written as
\begin{eqnarray}
	\Phi_j	&\equiv& \int_{S} \vec{j} \cdot d\vec{S} \\
			&=& \left( \frac{c}{4\pi} \right) 2 \pi r \sin \theta B_\phi.
\end{eqnarray}
The right-hand side of eq. (\ref{eq-jflux1}) is reduced to
\begin{equation}
\begin{split}
	\lefteqn{ \frac{c}{4\pi} \oint_C \left\{
		 \nabla \times ( \vec{v} \times \vec{B} + \vec{\alpha} \cdot \vec{B} )
		- \nabla \times \left( \eta \nabla \times \vec{B} \right)
		- \vec{V} \times (\nabla \times \vec{B})
	\right\} \cdot d \vec{l} } \\
	&\quad = \left( \frac{c}{4\pi} \right)
		2 \pi \sin \theta \sin 2 \theta \Biggl\{
		   Ar \frac{ \partial \Omega }{ \partial r }
		+ Br \frac{ d \ln (\rho r^2) }{ d t } \\
	&\quad\quad	+ \eta r^2 \frac{\partial}{\partial r}\left(
				\frac{1}{r^4} \frac{\partial}{\partial r} (B r^3)
			\right)
		+ \frac{\partial \eta}{\partial r} \frac{\partial Br}{\partial r}
		+ r \frac{ (\nabla \times (\vec{\alpha} \cdot \vec{B}))_\phi }{ \sin 2 \theta }
	\Biggr\} \\
	&\quad + \left( \frac{c}{4\pi} \right)
		2 \pi \sin^2 \theta \Biggl\{
		 B_{\rm stb} r \frac{ d \ln (\rho r^2) }{ d t } \\
	&\quad\quad	+ \eta r \frac{\partial}{\partial r}\left(
				\frac{1}{r^2} \frac{\partial}{\partial r} (B_{\rm stb} r^2)
			\right)
		+ \frac{\partial \eta}{\partial r} \frac{\partial B_{\rm stb} r}{\partial r}
	\Biggr\},
\end{split}
\end{equation}
in which the latitudinal dependencies of the magnetic field described above are used, and a relation $\frac{\partial v_m}{\partial r} = -\frac{1}{\rho r^2}\frac{d (\rho r^2)}{d t}$, which is valid for a spherically symmetric flow, is assumed. By taking the terms with $\sin 2 \theta$ dependence, the evolution equation of $B(r)$ is obtained as
\begin{equation}
\begin{split}
	\rho r^2 \frac{ d }{ dt }\left( \frac{B}{\rho r} \right)
		&=	   Ar \frac{ \partial \Omega }{ \partial r }
		+ \eta r^2 \frac{\partial}{\partial r}\left(
				\frac{1}{r^4} \frac{\partial}{\partial r} (B r^3)
			\right) \\
		&\quad \quad + \frac{\partial \eta}{\partial r} \frac{\partial Br}{\partial r}
		+ r \frac{ (\nabla \times (\vec{\alpha} \cdot \vec{B}))_\phi }{ \sin 2 \theta }. \label{eq-jflux2}
\end{split}
\end{equation}
The remaining terms with $\sin \theta$ dependence may be used to describe the evolution of $B_{\rm stb}$ as
\begin{equation}
	\rho r^2 \frac{ d }{ dt }\left( \frac{B_{\rm stb}}{\rho r} \right)
		=\eta r \frac{\partial}{\partial r}\left(
				\frac{1}{r^2} \frac{\partial}{\partial r} (B_{\rm stb} r^2)
			\right) 
		 + \frac{\partial \eta}{\partial r} \frac{\partial B_{\rm stb}r}{\partial r}.
\end{equation}
However, considering the function of stabilizing the poloidal component, a simpler relation
\begin{equation}
	B_{\rm stb} = f_{\rm tp} \frac{2A}{r},
\end{equation}
with $f_{\rm tp} \gtrsim \mathcal{O}(1)$ as the ratio between poloidal and toroidal field strengths, might also provide a reasonable estimate.

We have not yet determined the functional form of the electromotive force induced by the $\alpha$ effect, $( \vec{\alpha} \cdot \vec{B} )/c$. In this work, we keep this issue open and hereafter show results omitting the $\alpha$-effect. It is because $\vec{ \alpha }$ is a pseudo-tensor, which can be a function of the magnetic field strength, the rotation frequency, and the local thermodynamic quantities as well so that it will be complex in the general case. Nevertheless, the $\alpha$ effect accounts for the induction of the poloidal field from the toroidal field, which is not provided by other means considered in this work. Such a term is therefore indispensable for obtaining magnetic amplification especially in convective regions \citep{Featherstone+09, Auguston+16, Hotta+16}, and will thus be included in our future work.

In summary, the evolution of the stellar magnetic field is described by the two equations,
\begin{eqnarray}
	\frac{ d (A r) }{ d t } 
		&=&	4 \pi \rho \eta r^3 \frac{\partial}{\partial M}\left(
				4 \pi \rho \frac{\partial}{\partial M} (A r^2)
			\right) \label{eq-Bflux3} \\
	\frac{ d }{ d t }\left( \frac{B}{\rho r} \right)
		&=&	  4 \pi  Ar \frac{ \partial \Omega }{ \partial M }
		+ 4 \pi \eta r^2 \frac{\partial}{\partial M}\left(
				\frac{4 \pi \rho}{r^2} \frac{\partial}{\partial M} (B r^3)
			\right) \nonumber \\
	&& 
		+ (4 \pi)^2 \rho r^2 \frac{\partial \eta}{\partial M} \frac{\partial Br}{\partial M}. \label{eq-jflux3}
\end{eqnarray}
It is noteworthy that both $A r$ and $B/\rho r$ scale as $|B| r^2$ if we assume $\rho \propto r^{-3}$. Therefore, the terms on the left-hand side of both eqs. (\ref{eq-Bflux3}, \ref{eq-jflux3}) show the magnetic flux conservation. For eq. (\ref{eq-Bflux3}), the rest term describes the magnetic diffusion. The first, second, and third terms on the right-hand side of eq. (\ref{eq-jflux3}) account for the $\Omega$ effect, the magnetic diffusion, and the magnetic advection caused by the gradient of the magnetic diffusivity, respectively.

\subsubsection{Boundary conditions}
We solve two diffusion--advection equations for the magnetic field evolution. Hence, in total, four boundary conditions are needed for closures.

At the center of the star, we set $A = 0$ and $B = 0$ such that the magnetic field does not diverge. At the surface of the star, we set $1 + (1/A)(\partial Ar/\partial r) = 0$ and $B=0$. The first condition is obtained by approximating that the poloidal magnetic field outside the star coincides with the dipole field. The second condition is obtained by assuming that there is no radial electric current that penetrates the stellar surface to outer space.

\subsection{Evolution equation of the angular velocity} \label{sec:angmom}

The equation of fluid momentum conservation can be written as
\begin{equation}
	\frac{\partial}{\partial t} \left( \rho \vec{v} \right)
		+	\nabla \cdot ( \rho \vec{v} \vec{v} )
		= -\nabla P
		+	\rho \vec{ g }
		+	\nabla \cdot \vec{ {\mathit \Pi} }
		+	\nabla \cdot \vec{ {\mathit M} },
\end{equation}
where $\vec{g}, \vec{ {\mathit \Pi} }$, and $\vec{M} \equiv \frac{1}{4\pi} \vec{B}\vec{B} - \frac{1}{8\pi}|\vec{B}|^2\vec{I}$ are the gravity, the Reynolds stress tensor, and the Maxwell stress tensor, respectively. Here, we use a relation $\frac{1}{c} \vec{j} \times \vec{B} = \nabla \cdot \vec{ {\mathit M} }$ describing the balance between the Lorentz force and the magnetic stress, which is satisfied for MHD. We note that the viscous stress owing to the fluid viscosity is neglected because of the extremely large Reynolds number of the stellar system.

Starting from the basic equation, one can write the evolution equation of the specific angular momentum as
\begin{equation}
\begin{split}
	\rho \frac{d}{d t} (r \sin \theta v_\phi)
		&=	r \sin \theta ( \nabla \cdot \vec{ {\mathit \Pi} } )_\phi
		+	\frac{1}{r^2} \frac{\partial}{\partial r} \left( r^3 \sin\theta \frac{ B_r B_\phi }{ 4\pi } \right) \\
	&\quad	+	\frac{1}{\sin \theta} \frac{\partial}{\partial \theta} \left( \sin^2 \theta \frac{ B_\theta B_\phi }{ 4\pi } \right),
\end{split}
\end{equation}
where axial symmetry is assumed. By averaging over a sphere, we obtain
\begin{equation}
\begin{split}
		\rho S \frac{ d }{ dt }( i \Omega ) 
		&=	\left(
				\int_0^{\pi} 2 \pi r^3 \sin^2 \theta ( \nabla \cdot \vec{ {\mathit \Pi} } )_\phi d \theta
			\right) \\
		&\quad +
			 \frac{\partial}{\partial r} \left( r^3 \int_0^{\pi} \frac{B_r B_\phi}{2} \sin^2 \theta d \theta \right),
\end{split}
\end{equation}
where $S \equiv \int 2 \pi r^2 \sin \theta d \theta $ is the surface area of the sphere and $i \equiv \int 2 \pi r^4 \sin^3 \theta d \theta / S \sim 2 r^2 / 3$ is the specific moment of inertia of the sphere.

Finally, by assuming the latitudinal dependencies of the magnetic field and by applying an empirical formula of the viscous force due to shear, the evolution equation of the angular velocity is obtained;
\begin{equation}
\begin{split}
		\frac{ d }{ d t }(i \Omega )
		&=	\frac{\partial}{\partial M} \left( (4 \pi \rho r^2)^2 \nu_{\rm cv} i r^{-{n_{\rm cv}}} \frac{ \partial ( \Omega r^{n_{\rm cv}} ) }{ \partial M } \right) \\
	&\quad	+	\frac{\partial}{\partial M} \left( (4 \pi \rho r^2)^2 \nu_{\rm eff} i \frac{ \partial \Omega }{ \partial M } \right)
		+	\frac{\partial}{\partial M} \left( \frac{ 8 r^2 A B }{ 15 } \right), \label{eq-jcons}
\end{split}
\end{equation}
where $\nu_{\rm cv}$ and $\nu_{\rm eff}$ are the effective viscosities due to convective turbulence and turbulence induced by other instabilities. Here, it is noteworthy that $B_\phi$ in the magnetic tensor should omit $B_{\rm stb}$, the toroidal component that would exist for stabilization of the poloidal field, but only account for the toroidal component that is induced by the $\Omega$ effect. This is to be consistent with the assumption on the dynamical equilibrium of the original field, as otherwise nonzero Lorentz forces disrupt the original configuration. This requirement is naturally satisfied if we assume that $B_{\rm stb}$ has $\sim \sin \theta$ polarity, such that the effect on the Lorentz force cancels after averaging over the sphere.

The first term on the right-hand side of eq.(\ref{eq-jcons}) describes the angular momentum transfer owing to the Reynolds stress due to convective turbulence. In general, angular momentum distribution in a convective region is affected by the interplay between the convective advection and turbulence and rotation. However, the theoretical treatment is uncertain \citep[e.g.,][]{Tassoul00}. Thus our model incorporates the parameter $n_{\rm cv}$ that indicates which kind of angular velocity structure forms in a convective region in equilibrium. Two extreme cases are either $n_{\rm cv}=0$ favoring rigid body rotation, or $n_{\rm cv}=2$ favoring isotropic specific angular momentum. We will apply $n_{\rm cv}=0$ for our calculations unless otherwise noted. For the second term, we assume that a region reaches rigid rotation in equilibrium because shear rotation is the source of the energy to drive instability in most of the considered cases. The third term accounts for the angular momentum transfer due to the Maxwell stress.

It is noteworthy that all of these stress terms have a conservative form, i.e., only the surface term appears after integrating over the whole star. For the magnetic stress, this property originates from the fact that the electromagnetic field, assuming MHD, only possesses a negligible amount of momentum compared to the matter. We also note that, even in the case of $n_{\rm cv}=2$, the rotation law does not necessarily reach the case of isotropic specific angular momentum, because the second term counterbalances towards rigid rotation. In our code, the most efficient rotation induced viscosity in a convective region is the dynamical shear instability. Considering the huge uncertainty involved in the efficiency estimates, we assume that $\nu_{\rm DS} = \nu_{\rm MLT}$ in case of $n_{\rm cv}=2$. This yields $\Omega(r) \propto r^{-1}$, which can be checked by applying $\nu_{\rm cv} = \nu_{\rm eff}$ to eq. (\ref{eq-jcons}).

\subsection{Input physics}

\subsubsection{Diffusion coefficients}

We assume that several hydrodynamical instabilities may develop in the rotating stellar interior and that turbulence driven by such instabilities accounts for the Reynolds stress. The convective viscosity is estimated as $\nu_{\rm cv} = D_{\rm cv}$ and $\nu_{\rm eff}$ consists of 6 terms as
\begin{equation}
	\nu_{\rm eff} =	\nu_{\rm ES}
		+	\nu_{\rm DS}
		+	\nu_{\rm SS}
		+	\nu_{\rm SH}
		+	\nu_{\rm GSF}
		+	\nu_{\rm PT},
\end{equation}
where each term stands for the viscosity corresponding to
the Eddington-Sweet circulation ($\nu_{\rm ES}$),
the dynamical shear instability ($\nu_{\rm DS}$),
the secular shear instability ($\nu_{\rm SS}$),
the Solberg--H{\o}iland instability ($\nu_{\rm SH}$),
the Goldreich--Schubert--Fricke (GSF) instability ($\nu_{\rm GSF}$),
and the Pitts--Tayler instability ($\nu_{\rm PT}$), respectively.

The viscosity coefficients $\nu_{\rm ES}$, $\nu_{\rm DS}$, $\nu_{\rm SH}$, and $\nu_{\rm GSF}$ are calculated according to \citet{Pinsonneault+89} and \citet{Heger+00}, with the modification that we use the minimum of the pressure scale height $H_P$ and the radius to estimate the typical length scale for each instability. In the original works, they use the velocity scale height of the respective flow, which is further limited by the radius or the width of the unstable region, as the typical length instead. We find that this modification has only a limited effect on the overall stellar evolution. To compute $\nu_{\rm SS}$, we follow the prescription by \citet{Maeder97}. An $m=1$ instability is assumed to grow in a region with a strong toroidal magnetic field. The effective viscosity owing to this Pitts--Tayler instability, $\nu_{\rm PT}$, is estimated according to \citet{Spruit02} and \citet{Maeder&Meynet04}. For clarity, we give the corresponding equations in Appendix.\,\ref{sec:app-PTvisc}. We note that these prescriptions involve a control parameter, $f_\mu$, which is multiplied to the $\mu$-gradient. This is an influential parameter of the stellar simulation, as it affects the stability conditions.

We assume that turbulence driven by (magneto-)hydrodynamical instabilities accounts for the chemical mixing as well. Another control parameter $f_c$ is set, which indicates the ratio between the chemical diffusivity and the viscosity. Thus
\begin{equation}
	D_{\rm eff} = D_{\rm cv}
	 + f_c \times \nu_{\rm eff}
\end{equation}
is used for rotating models. Similarly, the effective magnetic viscosity is estimated as
\begin{equation}
	\eta = D_{\rm cv} + f_m \times \nu_{\rm eff},
\end{equation}
but $f_m = 1$ is set in the current work.

We note that the Eddington--Sweet circulation accounts for the most efficient ``turbulent'' diffusivity in the radiative envelope in the present models. The Eddington--Sweet circulation, which is also referred to as the Eddington--Vogt circulation, has been firstly postulated as a laminar meridional flow driven by a thermal imbalance which results from the difference of temperature gradient in a latitudinal direction in a rotating star \citep{vonZeipel24a, vonZeipel24b, Eddington25, Vogt25, Sweet50}. As the effect on the angular momentum transport could be modeled as an advection \citep[cf.][]{Maeder&Zahn98}, magnetic advection due to the Eddington-Sweet circulation would be formulated by more stringent consideration of the $\vec{U} \times \vec{B}$ term in eq. (\ref{eq-Bflux1}). However, considering the existence of the baroclinic instability, which will operate with a dynamical timescale in a radiative zone even with a very small differential rotation \citep{Fujimoto88, Kitchatinov14}, it will also be natural to assume that the Eddington-Sweet circulation in an actual star will be typically accompanied by turbulence. Bearing the uncertainties involved in the theoretical modeling in mind, we leave this discussion open in the current work.

\subsubsection{Wind-magnetic field interaction}

With a strong surface magnetic field, a part of the stellar wind blowing from a closed field region will be trapped to form a magnetosphere surrounding the star \citep{Donati+02}. Indeed, the time variation and the Balmer-line emission profiles observed in a well-known magnetic Bp star, $\sigma$ Ori E, has been reproduced by considering such a rigidly rotating magnetosphere \citep{Townsend&Owocki05, Townsend+05}. Because the net mass loss rate can be significantly reduced due to the magnetic confinement, we take this effect into account in our simulations according to \citet{udDoula+08}. The magnetic confinement parameter $\eta_*$ is estimated at first, and is used to derive the confinement efficiency $f_{\rm conf}(\eta_*) \equiv \dot{M}/\dot{M}(B=0)$, where $\dot{M}(B=0)$ is the mass loss rate of a non-magnetic star. The detailed procedure is explained in Appendix\,\ref{sec:app-windmag}.

While the strong surface field reduces the mass-loss rate, it can enhance the rate of the angular momentum loss in contrast. This is because the stellar angular momentum is not only reduced by the material flow but also by the Maxwell stress \citep{Weber&Davis67, udDoula+09}. The braking efficiency $f_{\rm break}(\eta_*) \equiv \dot{J}/\dot{J}(B=0)$, where $\dot{J}(B=0)$ is the angular momentum loss rate of a non-magnetic star, is estimated in this case. We account for the effect of the magnetic braking according to \citet{udDoula+09}. The interaction of wind material with a dipole magnetic field is assumed in their analysis, while a simple monopole geometry was assumed in the classical analysis by \citet{Weber&Davis67}, which is often used to model the evolution of solar-like stars. We give the details of our treatment also in Appendix\,\ref{sec:app-windmag}.

\subsection{Numerical settings and code test}

Equations (\ref{eq-Bflux3}, \ref{eq-jflux3}, \ref{eq-jcons}), and the diffusion part of eq. (\ref{eq-ycons}) are numerically solved with a finite-difference method. We use 1st order backward difference for the time derivative and 2nd order central difference for the space derivative. Together with the boundary conditions, these difference equations are iteratively solved simultaneously, while they are decoupled from the equations of stellar structure. One time step for the structure equations is further divided into numerous (typically $\sim$1,000) time steps for the evolution equations of the magnetic field and the angular momentum. The latter time step is controlled such that the relative differences in $A, B$, and $\Omega$ are restricted to be smaller than $\lambda$ in each step, where $\lambda$ is an arbitrarily chosen control parameter of about 10\%. The basic features of the numerical code such as magnetic flux conservation and magnetic dissipation are confirmed, and details of the code tests are described in Appendix\,\ref{sec:app-codetest}.

\subsection{Other possible magnetic effects} \label{sec:othereffects}

Several other magnetic effects are not considered in the present work. Even though they will not significantly affect the interplay between the evolution of the magnetic field and the stellar rotation, they can have a significant effect on the stellar structure in some cases. Here we briefly review these effects, which we plan to implement on top of the present formulation in forthcoming papers.

For strong magnetic fields, the stellar structure can be modified by the magnetic pressure and the magnetic tension. \citet{Feiden&Chaboyer12, Feiden&Chaboyer13, Feiden&Chaboyer14} take these effects into account for low-mass stellar evolution calculations using a geometry parameter introduced by \citet{Lydon&Sofia95}. \citet{Duez+10b} develop a more rigorous and general treatment and apply their formulation to model the young Sun.

\citet{Lydon&Sofia95} also showed how large scale magnetic fields can affect the equation of state, especially the adiabatic index, and the equations in the mixing length theory for convection. When the local magnetic field is strong enough, the pressure change during an adiabatic motion can differ from the non-magnetic case, since part of the work goes into the form of magnetic energy. The modification of the adiabatic index further affects the criterion for convective instability. Besides, it changes the specific heat and thus the efficiency of the convective energy transport.

The internal energy equation is in principle also modified when considering the magnetic effects of Jule heating and the Poynting flux. Because our formulation includes magnetic dissipation due to turbulent magnetic diffusivity (the $\eta$ effect), a corresponding Jule heating term may be taken into account in the internal energy equation for consistency in the future.

Strong and stable magnetic fields inside the star might suppress hydrodynamical flows such as convection and meridional circulations. The stabilization effect may be included in stellar evolution models by modifying the convective criterion \citep[e.g.][]{Lydon&Sofia95, Petermann+15}.

\section{Angular momentum transport via dissipating torsional \Alfven wave} \label{sec:wave1}

Differential rotation winds up the poloidal magnetic field to enhance the toroidal component, which is the $\Omega$ effect. As the toroidal component gets strong, the magnetic stress increases as well, which counteracts to reduce the differential rotation. What will happen in a stellar model when these two effects are incorporated?

We can simplify our set of equations such that the $\Omega$ effect and the magnetic stress are resolved by two linear differential equations as
\begin{eqnarray}
	\frac{ \partial ( Br^3 ) }{ \partial t }
		&=&	 \frac{B_{\rm r} r^4}{2} \frac{ \partial \Omega }{ \partial r } \nonumber \\
	\frac{ \partial \Omega }{ \partial t }
		&=& \frac{B_{\rm r}}{10 \pi \rho r^4}
			 \frac{\partial (Br^3) }{\partial r}, \nonumber
\end{eqnarray}
where the effects of magnetic diffusion and viscous angular momentum transport are neglected, the rates of change of $Ar$, radius, density, and specific moment of inertia $i$ are assumed to be small, and the relations $i \sim 2 r^2 /3$ and $A \sim r B_r /2$ are used. This hyperbolic system may be analyzed with methods used in fluid dynamics. By diagonalizing the matrix
\[ \left( \begin{array}{ccc}
 0 & B_r r^4 /2   \\
 B_r / 10 \pi \rho r^4 &     0 
\end{array} \right), \]
one may obtain a set of eigenvalues and eigenvectors as $\pm c \equiv \frac{1}{\sqrt{5}}v_A$ and $\vec{r}^{\pm} = (1 \ 1/\sqrt{5\pi \rho}r^4 )^t $. The corresponding invariant $dw^{\pm} = d\Omega \mp d(B r^3)/\sqrt{5\pi \rho}r^4$ becomes constant along the characteristic $dr/dt = \pm c$. Here, $v_A \equiv B_r/\sqrt{4 \pi \rho}$ is the \Alfven velocity of the radial magnetic field. Therefore, we expect that a wave that propagates with the \Alfven velocity forms in this system.

In this section, we analyze how this wave propagation manifests itself in our numerical models. Furthermore, we demonstrate that with the help of viscosity and magnetic dissipation, this torsional \Alfven wave serves as a highly efficient mechanism for the redistribution of angular momentum.

\subsection{Formation and propagation of torsional \Alfven wave}

\begin{figure*}[t]
	\begin{center}
	\includegraphics[width=0.7\textwidth]{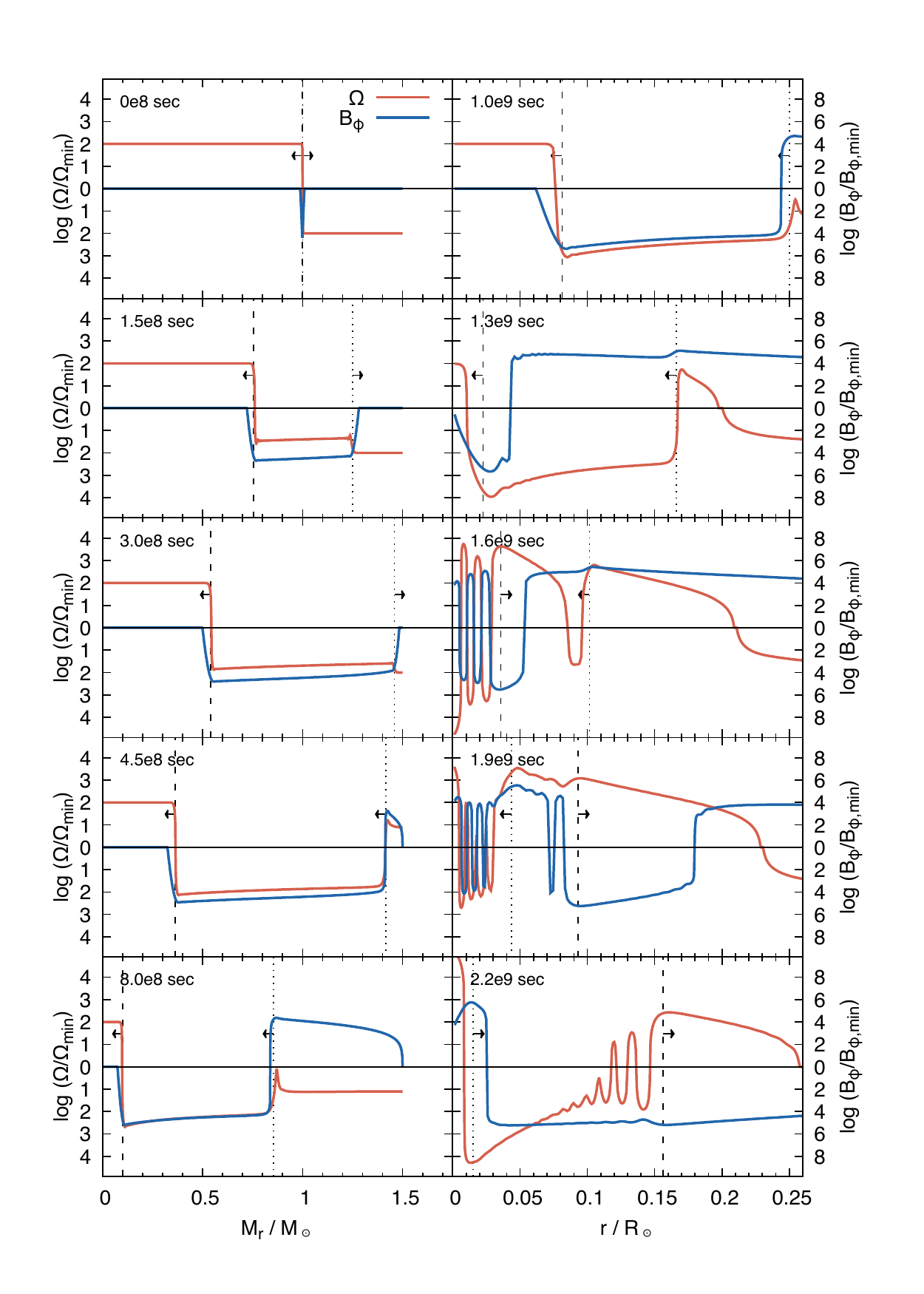}
	\caption{Angular velocity, $\Omega$ (red lines), and strength of the $\phi$-component of the magnetic field, $B_\phi$ (blue lines), for 10 different times during the evolution of a 1.5\,$M_\odot$ main-sequence stellar model, illustrating the propagation of torsional \Alfven waves in the stellar interior. Here, the $\Omega$ effect and the Maxwell stress are taken into account, but the $\eta$ effect is not. The X-axis for the left panels, showing times up to $8 \times 10^8$\,s, is the Lagrangian mass coordinate, the right panels, depicting later times, use the radius coordinate. On the Y-axis, The logarithm of $\Omega$ divided by $\Omega_{\rm min}$ (or log $B_{\phi, {\rm min}}/B_{\phi, {\rm min}}$) is plotted with the normalization values of $\Omega_{\rm min} = 1 \times 10^{-6}$\,rad\,s$^{-1}$ and $B_{\phi, {\rm min}} = 1\times10^3$\,G. Prograde rotation and positive B-field polarity are plotted in the upper half of the panels, retrograde rotation and negative polarity are plotted in the lower half. Small structure with $-\Omega_{\rm min} < \Omega < \Omega_{\rm min}$ and $-B_{\phi, {\rm min}} < B_\phi < B_{\phi, {\rm min}}$ is omitted from this plot. The black dashed and black dotted lines show the positions of the wavefronts, $r_f$, which are estimated by eq.\,(\ref{eq:wavefront}). Arrows indicate the direction of the wave propagation.
    }
	\label{fig:wave3}
	\end{center}
\end{figure*}

We explore the wave propagation using a 1.5\,M$_\odot$ main-sequence model which has a radius of $\sim 1.5$\,R$_\odot$. We have calculated the coupled evolution of the toroidal field and the stellar rotation including the $\Omega$ effect and the magnetic stress, but we set the viscosity and magnetic diffusivity to zero. A uniform radial field as $B_r (=2A(r)/r) = 1$\,kG is set in the beginning, so that the wave velocity becomes $c \sim 80$ cm s$^{-1}$, and correspondingly, the estimated wave-crossing time from the center to the surface is $\sim 1.3 \times 10^9$\,s. A step-function distribution of $\Omega = 10^{-4}$ rad s$^{-1}$ for $M \leq 1$ M$_\odot$, and $\Omega = -10^{-4}$ rad s$^{-1}$ otherwise, is imposed as the initial angular velocity distribution. The toroidal magnetic field $B_\phi (=B(r))$ is set to be zero everywhere.

Figure \ref{fig:wave3} shows how the rotation and the magnetic field evolve with time. In the beginning, two wavefronts launch from the discontinuity and start propagating towards the stellar surface and the center. A wave, which is similar to a rarefaction wave in fluid dynamics, is formed in between the two wavefronts. The up-going front reaches the surface at $\sim 3 \times 10^8$ s, and is then reflected to follow the down-going wavefront. The down-going front reaches the stellar center at $\sim 1.4 \times 10^9$ s and then is also reflected. At $\sim 1.8 \times 10^9$ s, the two wavefronts cross and penetrate each other.

Although the wavefronts have a diffuse structure, especially in the central region, Fig.\,1 shows that their propagation agrees well with the prediction, which is shown as black dashed and dotted lines. The positions of the two lines are directly calculated by the time integral of 
\begin{equation}
	r_f(t) = r_{f, {\rm ini}} \pm \int_0^{t} c(r_f) dt. \label{eq:wavefront}
\end{equation}
The wave-crossing time in the simulation is estimated to be $1.71 \times 10^9$ s, as the two wavefronts meet again at this time after traveling either through the stellar surface or the center. This agrees well with the simple estimate of $\sim 1.3 \times 10^9$ s given above. The timescale of shear rotation does not appear in the propagation timescale because the efficiency of the winding-up of the poloidal magnetic field is proportional to the required torque to affect the angular momentum of the material, and therefore they cancel out each other.

However, the timescale of the $\Omega$ effect relates to the time that the wave passes the width of the wavefront, $\lambda/c$, and therefore, the strength of the toroidal magnetic component at the wavefront can be estimated as
\begin{equation}
	B_\phi \sim B_r r \frac{\Delta \Omega}{\lambda} \frac{\lambda}{c} = \sqrt{20 \pi \rho} r \Delta \Omega. \label{eq:toroidalfield}
\end{equation}
This is proportional to the angular velocity difference at the wavefront, $\Delta \Omega$, but is independent of the strength of the poloidal magnetic component. This is because the stronger the seed poloidal magnetic field is, the shorter is the wave-crossing time of the width of the wavefront, and therefore they cancel out each other. This relation yields $B_\phi \sim 7 \times 10^7$ G at $M_r = 1$ $M_\odot$ with $\Delta \Omega = 2 \times 10^{-4}$ rad s$^{-1}$ and explains the simulation result.

In another test calculation with a 10-times stronger initial poloidal magnetic field, the wave velocity increases by a factor of 10, but the toroidal magnetic component does not change. On the other hand, in a calculation with ten times smaller initial angular velocity, the toroidal magnetic component decreases by a factor of 10, but the wave velocity stays constant.
 
The standard \Alfven wave and the $\Omega$-B wave discussed here are essentially identical, as they share the same driving force and a similar propagation velocity. Hence we refer to the wave solution in our simulation as a torsional \Alfven wave hereafter. While a small fluctuation propagates along the magnetic field in the former case, a large number of windings are required to launch the wave in the latter. This difference arises from the weak magnetic field considered in the present case. We compute $\sim10^{15}$ erg g$^{-1}$ for the gravitational and thermal energies, $\sim10^{12}$ erg g$^{-1}$ for the rotational kinetic energy, but only $\sim10^{4}$ erg g$^{-1}$ for the magnetic energy of the poloidal component. With such a weak magnetic field, a strong magnetic amplification due to the $\Omega$ effect is required for the magnetic stress to affect the dynamics of the rotating flow. The toroidal component induced in the \Alfven wave always has comparable specific energy to the rotational kinetic energy (eq.(\ref{eq:toroidalfield})). The required number of windings decreases for higher initial poloidal field strengths. The torsional \Alfven wave will converge to the standard \Alfven wave if the poloidal component is so strong that a small number of windings is sufficient to drive the wave.

\subsection{Torsional \Alfven wave with dissipation} \label{sec:wave2}

\begin{figure*}[t]
	\begin{center}
	\includegraphics[width=\textwidth]{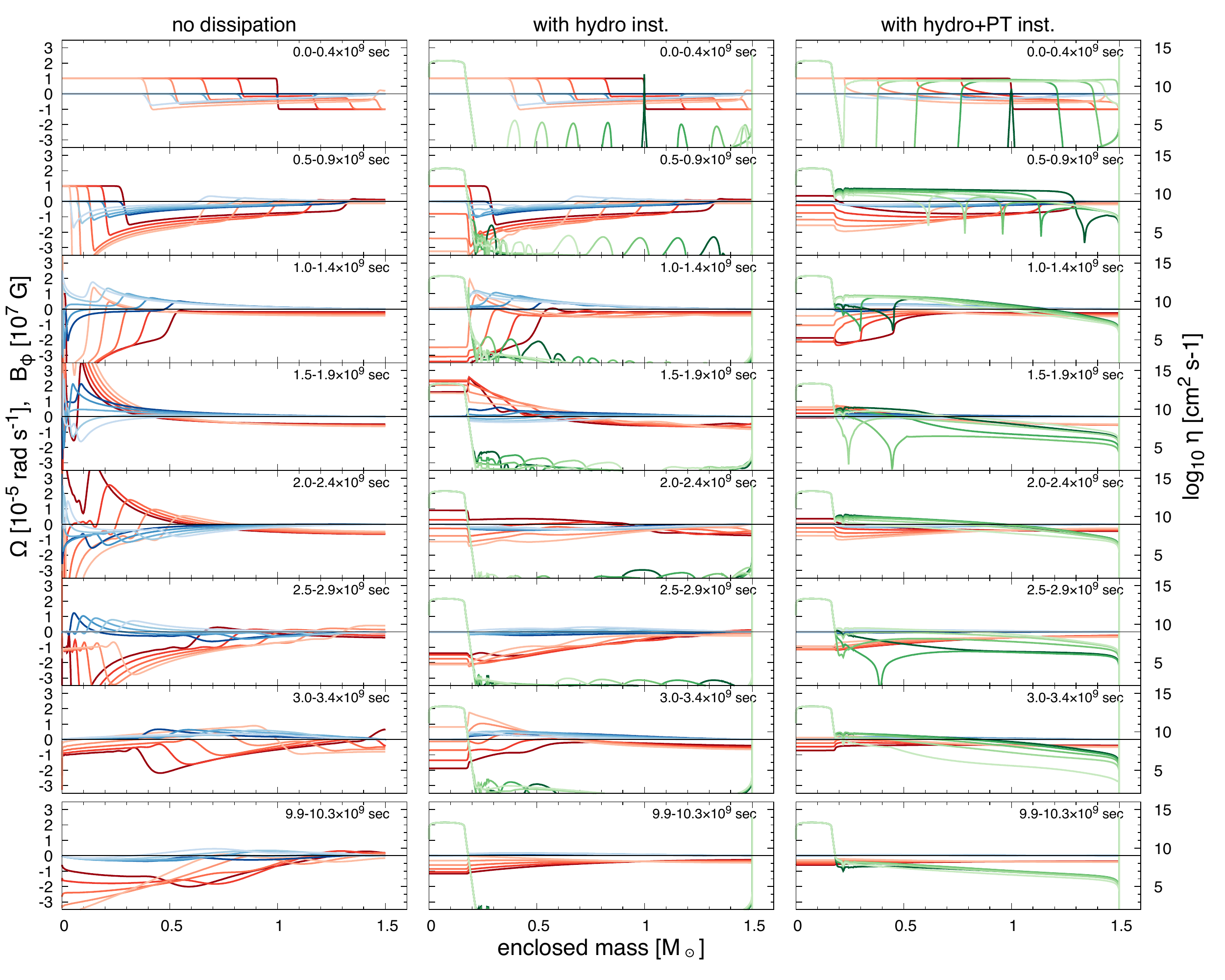}
	\caption{Angular velocity $\Omega$ (red), toroidal magnetic component $B_{\phi}$ (blue), and magnetic viscosity $\eta$ (green) as a function of the Lagrangian mass coordinate at eight different moments in time in test calculations of a 1.5 $M_\odot$ main-sequence stellar model, using different assumptions for magnetic and viscous dissipation. The left panel shows results for no dissipation, results with dissipation only due to hydrodynamic instabilities are shown in the middle, and results with hydrodynamic dissipation and Pitts--Tayler instabilities are shown in the right panel. For this stellar model the wave-crossing time is $t_{\rm wc} = 1.71 \times 10^9$ s. The top seven plots of each panel cover about two wave-crossing times ($3.4 \times 10^{9}$\,s) with 5 snapshots per plot, one every $10^{8}$ s, for the period given in the top right corner of each plot. In the bottom row of plots, results after about six wave-crossing times are shown, i.e. from $9.9 \times 10^{9}$\,s to $10.3 \times 10^{9}$\,s. The temporal evolution is available as an online movie. 
	}
	\label{fig:wave+diff}
	\end{center}
\end{figure*}

Here, we discuss the wave propagation for calculations where dissipation effects are included, using the same stellar model and initial conditions as in the previous section. Figure\,\ref{fig:wave+diff} shows the resulting evolution of angular velocity, toroidal magnetic, and magnetic viscosity. The cases of neglecting dissipation effects (left) can be compared with results including dissipation due to hydrodynamic instabilities only (middle), and considering both, hydrodynamic and magneto-hydrodynamic dissipation (right).

As shown above, without dissipation the waves travel freely through the star, and the wave-fronts meet each other once per wave-crossing time. Due to numerical diffusion, the angular velocity distribution after two wave-crossing times (seventh plot in the first column), which should coincide with the initial distribution, has become somewhat more diffuse. Nevertheless, we can follow the back-and-forth sloshing of the wave for more than 30 crossing times.

\begin{figure}[t]
	\begin{center}
	\includegraphics[width=0.5\textwidth]{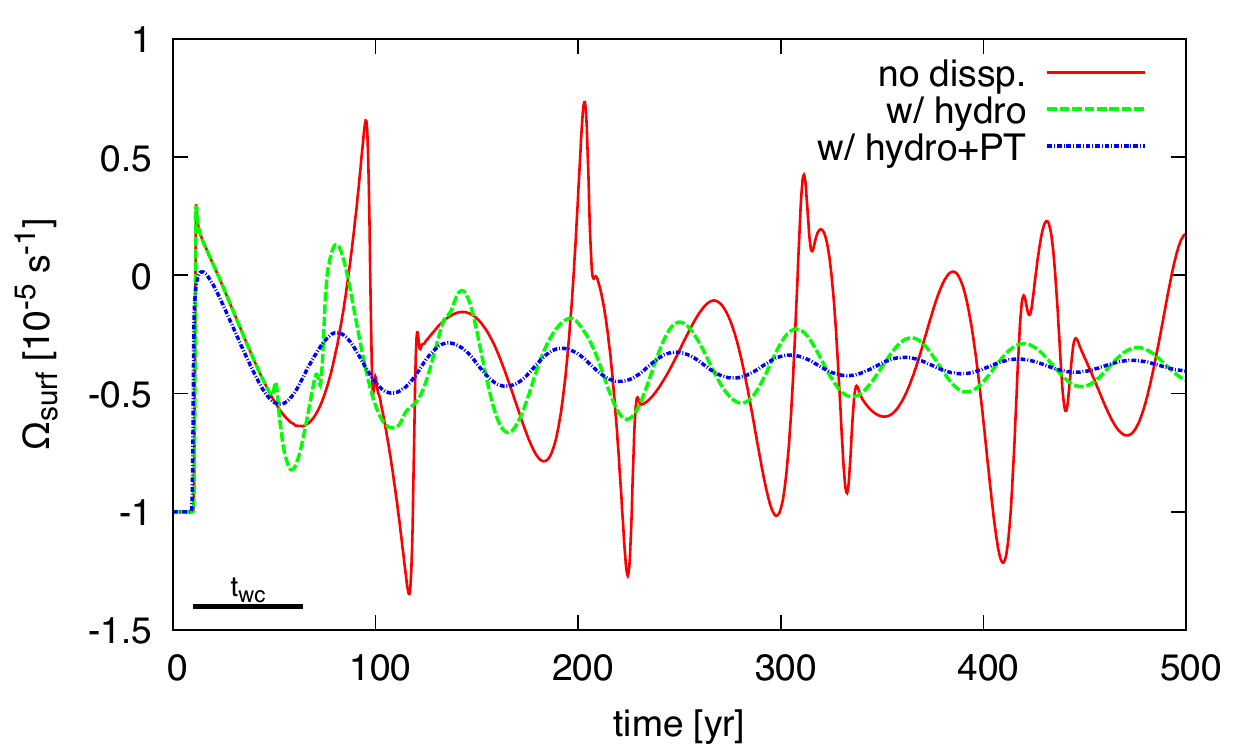}
	\caption{Surface angular velocity as a function of time in test calculations of our 1.5 $M_\odot$ magneto-rotational main-sequence model, in which constant radial magnetic field of $B_r =$ 1 kG and a step function with an arbitrary amplitude for the initial angular momentum distribution are imposed as initial conditions. The star starts to oscillate on the \Alfven time scale. Lines correspond to the case without dissipation (red, solid), including only hydrodynamic instabilities (green, dashed), and the case with hydrodynamic and magneto-hydrodynamic instabilities (blue, dash-dotted), respectively. The wave-crossing time of $t_{\rm wc} = 1.71 \times 10^9$ s is referenced by the thick black bar at the bottom left corner.
	}
	\label{fig:surface}
	\end{center}
\end{figure}
The middle column of Fig.\,\ref{fig:wave+diff} shows that turbulent dissipation due to hydrodynamic instabilities affects wave propagation. Whereas turbulence due to secular shear and GSF instabilities accounts for some viscosity in the radiative envelope, this is too small to matter here. However, our 1.5\,M$_\odot$ model has a hydrogen-burning convective core, in which rigid rotation is established by the large convective viscosity (note that we apply $n_{\rm cv}=0$ in this calculation). This has a big impact on the angular momentum redistribution. Since waves with shorter wavelengths have shorter dissipation time, the convective region effectively filters out waves with shorter wavelengths, which originally compose the step function used for the initial condition. As a result, a standing wave is quickly formed in the model, with a wavelength of twice the stellar radius. Hence, the standing wave corresponds to the $n=1$ fundamental mode oscillation, which has one node at $\sim 1.1$ M$_\odot$ (see an online movie of Fig.\ \ref{fig:wave+diff}).

In the calculation including hydrodynamic and magneto-hydrodynamic dissipation, the latter works effectively only during the first one or two wave-crossing times. First, shear rotation propagates inside the star with the velocity of $\sim v_A/\sqrt{5}$. In regions passed by this shear wave, a toroidal field is induced by the $\Omega$ effect, which soon triggers the Pitts--Tayler instability. As a result, the large diffusivity due to the Pitts--Tayler instability covers the entire star at $t \sim t_{\rm wc}$, which diffuses the angular velocity very effectively, together with the convective diffusivity. At later times, however, the diffusivity due to the Pitts--Tayler instability becomes weak since there is no more strong shear rotation amplifying the toroidal field component. Thus, the overall evolution becomes comparable to the case that only considers the hydrodynamic instabilities.

Figure \ref{fig:surface} shows the first 500 yr evolution of the surface angular velocity of our model. The damped, quasi-sinusoidal variation for the models including dissipation is due to the torsional oscillation. The period is comparable to the wave-crossing time $t_{\rm wc}$, which is given by
\[
    t_{\rm wc} = \int_0^{R} \left( \frac{1}{\sqrt{5}} \frac{B_r}{\sqrt{4 \pi \rho}} \right)^{-1} dr
\]
with the stellar radius $R$. If such waves were excited, such oscillations could exist in real stars. Because the oscillation timescale relates to the internal density and poloidal magnetic field distributions, observations of the changing surface rotation frequency may allow deriving the internal magnetic field strength by observing the change in the surface rotation frequency. We will later discuss the comparison between our model and relevant observations in Section \ref{sec:obs-rotperiod}.

The torsional \Alfven wave oscillation gradually decreases its amplitude. The model including only  hydrodynamic instabilities approaches a trivial stationary state of $\partial \Omega / \partial r = B_\phi = 0$ with a decay timescale of $t \sim 10$ $t_{\rm wc}$. This means that the angular momentum in the star is effectively redistributed to achieve rigid rotation, as torsional \Alfven wave propagates and dissipates throughout the star.

It is noteworthy that an integrated evolution of the toroidal magnetic field and the rotational flow in a radiative region of the Sun has been modeled by 2D axisymmetric simulations in \citet{Charbonneau&MacGregor92, Charbonneau&MacGregor93}. In their simulations, the phase shift across poloidal field lines can be followed. This leads to efficient wave dissipation because large gradients in the toroidal field are formed \citep{Charbonneau&MacGregor92}. Although there is a difference in the detailed mechanism of the dissipation, their results show that the radiative region in the sun approaches rigid rotation as the dissipative \Alfven wave propagates, which is consistent with our model.

The onset of Pitts--Tayler instability could affect the wave propagation if the turbulent dissipation modifies the wavefront structure. Let us consider a differentially rotating region in a star, where considerable strength of the poloidal field exists but initially zero toroidal component. The condition for the Pitts--Tayler instability to grow within a propagation time of the \Alfven wave is $\tau_{\rm PT} < \tau_{\rm wc}$, where $\tau_{\rm PT}$ is the growth time of the Pitts--Tayler instability and $\tau_{\rm wc}$ is the wave-crossing time. When the toroidal component is induced by the $\Omega$ effect with a differential rotation parameter $q \equiv \partial \ln \Omega/\partial \ln r$, the growth time of the Pitts--Tayler instability can be estimated as $\tau_{\rm PT} = \sqrt[3]{ \tau_{\rm wc}^2/(\Omega q) }$. Hence the condition above can be expressed as $\sqrt[3]{ \Omega \tau_{\rm wc} q } > 1$. Our simulation with a strong poloidal field of $B_{\rm r} \sim 1$ kG has $\tau_{\rm wc} \sim 10^9$ s. Therefore, the condition can be well satisfied with a canonical value of $\Omega \sim 10^{-5}$ s$^{-1}$, although $q$ can have a variety of value of $\lesssim 1$.

However, the condition for the wavefront to be disturbed by the growing turbulence is normally not satisfied in a radiative stellar envelope. For the turbulence driven by the Pitts--Tayler instability to affect the wavefront, $l_v > v_{\rm A} \tau_{\rm PT}$ may be required. The vertical length scale of the turbulence, $l_v$, is estimated as $r (\omega_A/N)$, using the toroidal \Alfven angular frequency, $\omega_{A} \equiv B_{\rm tor}/\sqrt{4 \pi \rho} r$, and the Brunt-V\"{a}is\"{a}l\"{a} frequency, $N$ \citep{Spruit02}. Thus the condition becomes $\Omega q/N > 1$. For our 1.5 M$_\odot$ model, we find $N \sim 10^{-2}$ s$^{-1}$ in the radiative envelope, which implies that in most of the case this condition will not be satisfied. Also, this condition does not depend on the poloidal field strength. Therefore, in our simplified 1D picture, although the Pitts--Tayler instability will grow at the wavefront, the wave propagation will not be affected by the turbulence induced by the instability because the unstable region will be too thin.

\section{Main-sequence evolution of 1.5M$_\odot$ stars} \label{sec:result-MS}

\begin{table*}[ht]
    \centering
    \caption{Characteristics of 1.5 M$_\odot$ magneto-rotational models. $\tau_{\rm MS}$ is the main-sequence lifetime, $\Delta M_{\rm MS}$ is the total mass lost during the MS phase, $\langle \dot{M} \rangle_{\rm MS} = \Delta M_{\rm MS}/\tau_{\rm MS}$ is the averaged mass loss rate for the MS phase, $\tau_{\rm break,ZAMS} = -J_{\rm ZAMS}/\dot{J}_{\rm ZAMS}$ is the braking timescale measured at ZAMS, $\eta_{\rm *,ZAMS}$ is the magnetic confinement parameter at ZAMS, and $P_{\rm rot,TAMS}$, $B_{\rm p, TAMS}$, and $J_{\rm TAMS}$ are the rotation period, the field strength at the pole, and the total angular momentum at TAMS.  
    }
\scalebox{1.0}{
    \begin{tabular}{ccccccccccc}
        \hline
        \hline
       $P_{\rm rot,ini}$
    & $B_{\rm p,ini}$ 
    & $\tau_{\rm MS}$ 
    & $\Delta M_{\rm MS}$ 
    & $\langle \dot{M} \rangle_{\rm MS}$ 
    & $\tau_{\rm break,ZAMS}$
    & $\eta_{\rm *,ZAMS}$
    & $P_{\rm rot,TAMS}$ 
    & $B_{\rm p, TAMS}$ 
    & $J_{\rm TAMS}/J_{\rm ZAMS}$
    \\
        d 
    & G 
    & Gyr 
    & M$_\odot$ 
    & M$_\odot$ yr$^{-1}$
    & Gyr
    & -
    & d 
    & G 
    & -
    \\
        \hline
$      1.00$ & $     10.00$ & $      2.73$ & $ -1.33$e-02 & $ -4.86$e-12 & $     20.12$ & $      0.24$ & $      2.24$ & $      0.05$ & $      0.82$ \\
$      1.00$ & $    100.00$ & $      2.67$ & $ -8.26$e-03 & $ -3.09$e-12 & $      3.90$ & $     23.74$ & $      2.59$ & $      0.62$ & $      0.73$ \\
$      1.00$ & $  1.00$e+03 & $      2.58$ & $ -1.78$e-03 & $ -6.91$e-13 & $      0.45$ & $  2.64$e+03 & $     55.68$ & $     23.93$ & $      0.04$ \\
$      1.00$ & $  1.00$e+04 & $      2.88$ & $ -3.43$e-04 & $ -1.19$e-13 & $      0.03$ & $  7.51$e+05 & inf.         & $  9.76$e+02 & 0.           \\
\\
$     10.00$ & $     10.00$ & $      2.70$ & $ -8.30$e-03 & $ -3.07$e-12 & $      9.44$ & $      2.89$ & $     22.91$ & $      0.56$ & $      0.80$ \\
$     10.00$ & $    100.00$ & $      2.63$ & $ -2.87$e-03 & $ -1.09$e-12 & $      1.45$ & $  2.81$e+02 & $     69.81$ & $      8.35$ & $      0.28$ \\
$     10.00$ & $  1.00$e+03 & $      2.49$ & $ -6.47$e-04 & $ -2.60$e-13 & $      0.15$ & $  3.54$e+04 & $  1.22$e+10 & $    182.46$ & $  2.18$e-09 \\
$     10.00$ & $  1.00$e+04 & $      2.94$ & $ -2.18$e-04 & $ -7.41$e-14 & $  1.02$e-02 & $  6.41$e+06 & inf.         & $  2.40$e+03 & 0.           \\
\\
$    100.00$ & $     10.00$ & $      2.71$ & $ -6.04$e-03 & $ -2.23$e-12 & $      6.72$ & $      7.74$ & $  2.63$e+02 & $      1.40$ & $      0.70$ \\
$    100.00$ & $    100.00$ & $      2.59$ & $ -1.87$e-03 & $ -7.23$e-13 & $      0.87$ & $  8.68$e+02 & $  3.30$e+03 & $     18.41$ & $      0.06$ \\
$    100.00$ & $  1.00$e+03 & $      2.44$ & $ -5.31$e-04 & $ -2.18$e-13 & $      0.09$ & $  1.11$e+05 & $  8.47$e+14 & $  2.30$e+02 & $  3.73$e-13 \\
$    100.00$ & $  1.00$e+04 & $      2.95$ & $ -2.07$e-04 & $ -7.01$e-14 & $  7.26$e-03 & $  1.31$e+07 & inf.         & $  2.50$e+03 & 0.           \\
\\
$  1.00$e+03 & $     10.00$ & $      2.71$ & $ -5.85$e-03 & $ -2.16$e-12 & $      6.47$ & $      8.97$ & $  2.68$e+03 & $      1.48$ & $      0.69$ \\
$  1.00$e+03 & $    100.00$ & $      2.58$ & $ -1.83$e-03 & $ -7.08$e-13 & $      0.81$ & $  1.02$e+03 & $  3.72$e+04 & $     19.33$ & $      0.05$ \\
$  1.00$e+03 & $  1.00$e+03 & $      2.44$ & $ -5.30$e-04 & $ -2.17$e-13 & $      0.08$ & $  1.34$e+05 & $  1.07$e+16 & $  2.30$e+02 & $  3.10$e-13 \\
$  1.00$e+03 & $  1.00$e+04 & $      2.96$ & $ -2.09$e-04 & $ -7.05$e-14 & $  7.09$e-03 & $  1.38$e+07 & inf.         & $  2.48$e+03 & 0.           \\
  	  \hline
    \end{tabular}
}
    \label{tab:MSmodels}
\end{table*}

\begin{figure*}[t]
	\centering
	\includegraphics[width=\textwidth]{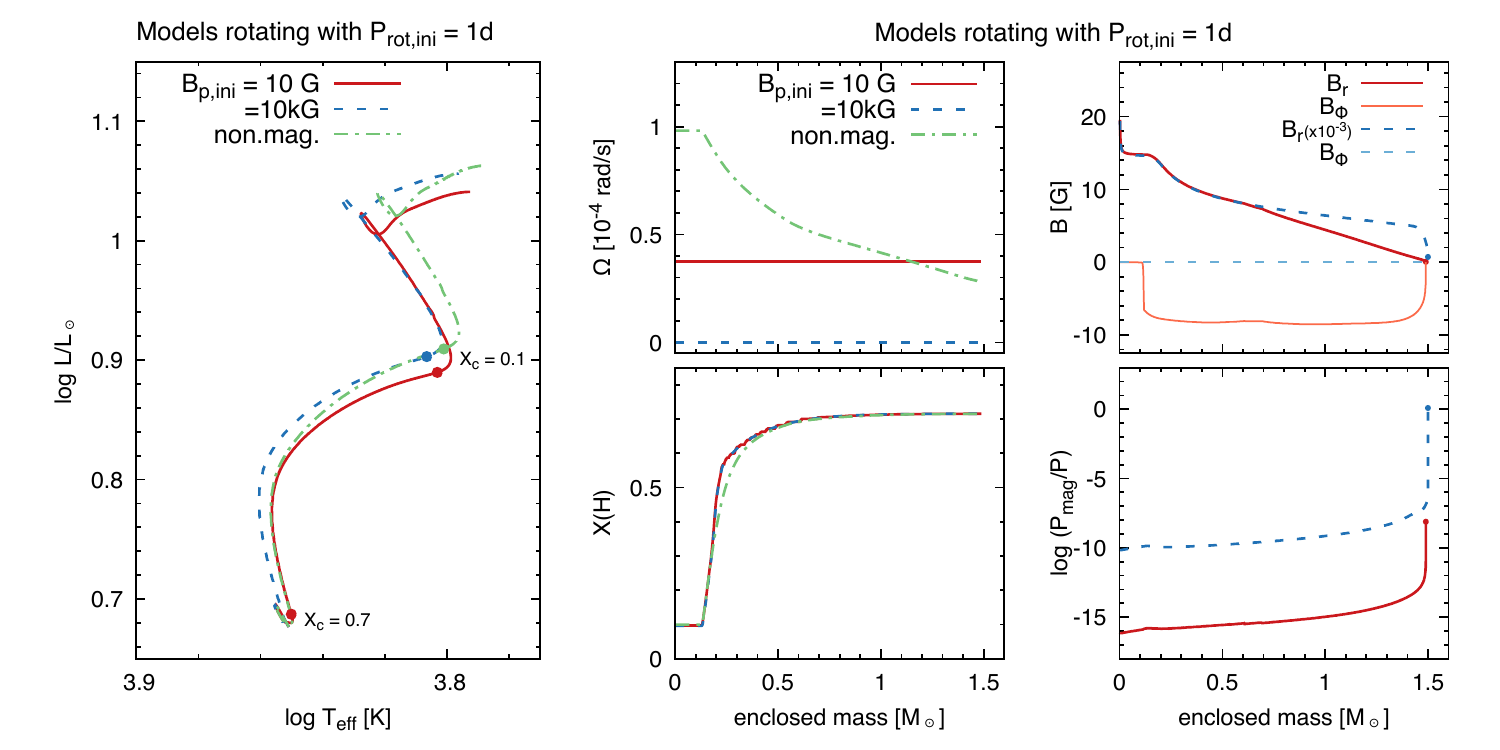}
	\caption{Left panel) Main-sequence evolution of 1.5 M$_\odot$ models in the HR diagram, computed with $n_{\rm cv} = 0$ and $P_{\rm rot, ini} =$ 1 d. The result of magnetic models with an initial magnetic field strength of $B_{\rm p,ini} =$ 10 G and 10 kG are shown by the red solid and blue dashed lines, while the result of the nonmagnetic model is shown by the green dash-dotted line. Epochs for which the central hydrogen become $X_c = 0.1$ (and 0.7 for the $B_{\rm p,ini} =$ 10 G model) are indicated by dots. Right panels) Profiles of internal angular velocity (top left), radial and toroidal magnetic field strength (top right), hydrogen mass fraction (bottom left), and the ratio between the magnetic pressure and the pressure (bottom right) at $X_c = 0.1$ as a function of Lagrangian mass coordinate. Again, profiles of the $B_{\rm p,ini} =$ 10 G and 10 kG models are shown by the red solid and the blue dashed lines, while that for the nonmagnetic model are by the green dash-dotted lines. In the top right panel, radial and toroidal field strength are shown by the thick and thin lines, respectively, and the radial component for the $B_{\rm p,ini} =$ 10 kG model is multiplied with a factor of $10^{-3}$. The surface values are highlighted by dots in the right-top and right-bottom panels.
	}
	\label{fig:MS}
\end{figure*}

\begin{figure*}[t]
	\begin{minipage}{0.5\hsize}
		\begin{center}
		\includegraphics[width=\textwidth]{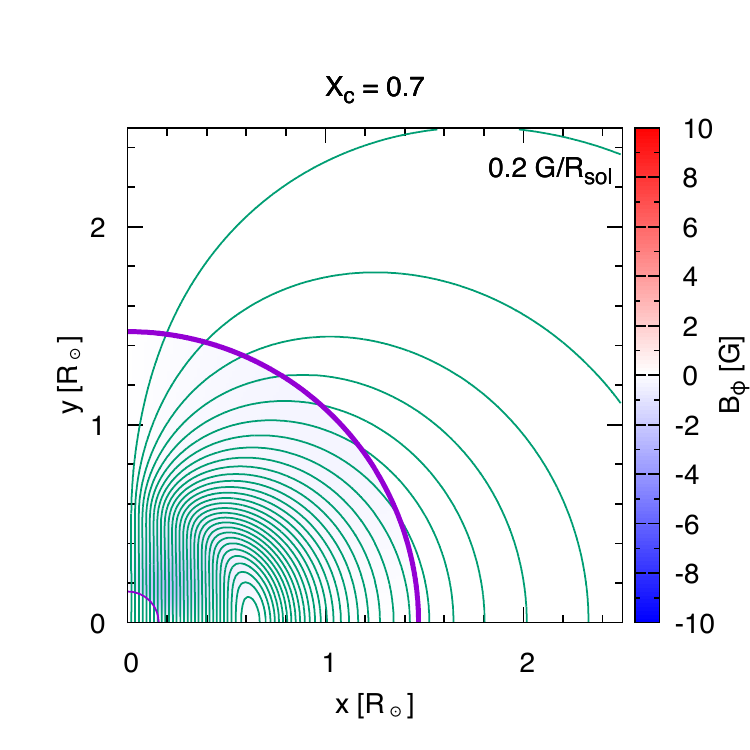}
		\end{center}
	\end{minipage}
	\begin{minipage}{0.5\hsize}
		\begin{center}
		\includegraphics[width=\textwidth]{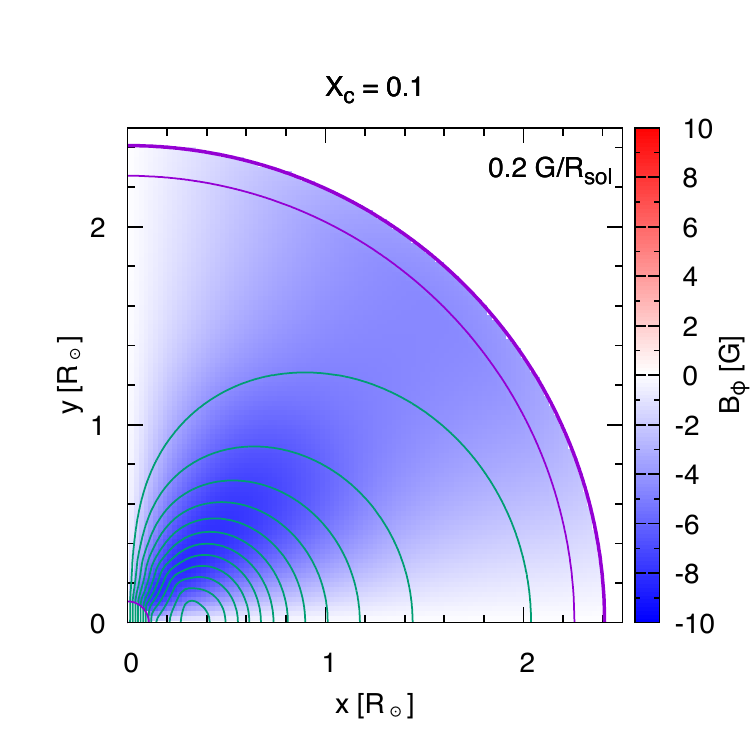}
		\end{center}
	\end{minipage}
	\caption{Reconstructed 2D magnetic field structures at core hydrogen mass fractions of X$_{\rm c}$ = 0.7 (left) and 0.1 (right), for our 1.5 M$_\odot$ model with an initial magnetic field strength of $B_{\rm p,ini} =$ 10 G and an initial rotation period of $P_{\rm rot, ini} =$ 1 d. The X-axis goes through the equatorial plane, while the Z-axis corresponds to the stellar rotation axis. Thick purple lines show the stellar surface, while thin purple lines designate the convective core boundary. Green lines indicate magnetic field lines, the interval of which is taken such that 1 line in 1 R$_\odot$ corresponds to the field strength of 0.2 G. The field lines outside of the star are constructed assuming a dipole structure. The color indicates the strength of the toroidal field. Note that deformation due to centrifugal forces and structure change of the circumstellar magnetic field due to wind interaction are not taken into account here.
	}
	\label{fig:2Dfield}
\end{figure*}
To explore the capabilities of our new modeling approach, we calculate the evolution of solar metallicity 1.5 M$_\odot$ models with the full framework as described in Sect.\,2. We set $\alpha_{\rm MLT} = 1.8$, $f_{\rm ov} = 0.01$, $f_{\mu} = 0.1$, and $f_c = 0.125$, for the mixing-length parameter, the overshoot parameter, the $\mu$-barrier parameter, and the chemical diffusion/viscosity ratio parameter, respectively. Initially, rigid rotation is applied, and the initial rotation period is chosen from $P_{\rm rot, ini} =$ 1, 10, 100, and 1000 d. For the magnetic field, we apply the simplest possible functions as
\begin{eqnarray}
	A(r) &=& \frac{ B_{\rm p,ini} r }{2} , \\
	B(r) &=& 0  .
\end{eqnarray}
This type of vector potential yields a uniform poloidal magnetic field along the rotation axis inside the star. The strength of the surface magnetic field at the pole, $B_{\rm p,ini}$, is chosen as 10, 100, 1000, or 10,000 G. Model characteristics are summarized in Tab.\,\ref{tab:MSmodels}.

In the left panel of Fig.~\ref{fig:MS}, evolution in the HR diagram are compared for rapidly rotating ($P_{\rm rot, ini} =$ 1 d) 1.5 M$_\odot$ models with $B_{\rm p,ini} =$ 10 G and 10 kG and without magnetic effects. In the right panels, internal profiles of angular velocity, hydrogen mass fraction, radial and toroidal magnetic field strength, and the ratio between the magnetic pressure and the pressure for the same models at central hydrogen mass fractions of X$_{\rm c}$= 0.1 are shown. In Figure \ref{fig:2Dfield}, the reconstructed 2D magnetic field structures for the weakly magnetic ($B_{\rm p,ini} =$ 10 G) model at X$_{\rm c}$= 0.7 and 0.1 are shown as well.

The fast rotation with $P_{\rm rot, ini} =$ 1 d, which accounts for $\sim 20$\% of the Keplerian rotation at the surface, supports the stellar surface, reducing the effective temperature. However, the surface velocity of the strongly magnetic ($B_{\rm p,ini} =$ 10 kG) model quickly decreases because of the efficient magnetic braking. This explains the offset in the HR diagram in the early main sequence, in which the strongly magnetic model shows a slightly higher effective temperature than the others.

During the main-sequence evolution, the stellar envelope expands, whereas the convective core shrinks. As a result, in the model without magnetic fields, significant differential rotation develops in the radiative envelope. The hydrogen-burning core is slightly extended by rotation induced mixing due to the secular shear instability. On the other hand, magnetic models evolve remaining close to rigidly rotating due to the highly efficient angular momentum redistribution by the dissipating torsional \Alfven wave. Even in the weakly magnetic model with $B_{\rm p,ini} =$ 10 G, which develops a toroidal field of only $\sim$-10 G in its radiative envelope, only a tiny amount of differential rotation of $\partial \Omega / \partial \ln r \sim 10^{-10}$ s$^{-1}$ develops at the core-envelope boundary. This behavior appears consistent with the results from asteroseismology which finds only a small deviation from rigid rotation in main-sequence stars \citep[e.g.][]{Aerts+17}. Consequently, no rotation induced mixing develops close to the hydrogen-burning core in the magnetic model. This explains the $\sim 5$\% fainter terminal age main-sequence (TAMS) luminosity of the weakly magnetic model. The difference in the TAMS luminosities between $B_{\rm p,ini} =$ 10 G and 10 kG models will be explained by the slight difference of the total TAMS masses.

The poloidal magnetic field strength in the magnetic model decreases with time, partly due to the envelope expansion, but more importantly, as a result of magnetic dissipation due to rotation induced turbulence. Here, turbulence triggered by the Eddington-Sweet circulation, the efficiency of which is roughly proportional to the square of the rotation frequency but not to the shear rotation, accounts for the magnetic dissipation. The ratio between the magnetic pressure, $B_{\rm mag} = (B_r^2 + B_\theta^2 + B_\phi^2)/8 \pi$, and the pressure (or the inverse of the plasma-$\beta$) is shown in the right-bottom panel of Fig.~\ref{fig:MS}. This ratio will also indicate the significance of the influence of the magnetic pressure and tension and magnetic modification of the adiabaticity to the stellar structure, which are currently not treated in our magnetic models (Sec.~\ref{sec:othereffects}). This figure shows this ratio is fairly small in particular in the inner region of the star, justifying the current treatment. Meanwhile, a strong magnetic field may influence the surface structure, which will indirectly affect the stellar evolution by changing the mass-loss and angular-momentum-loss histories. Also, it is noteworthy that material mixing due to the magnetic Pitts--Tayler instability does not take place during the main-sequence evolution in magnetic models, because it requires differential rotation. Therefore, in our current prescription, magnetic fields with different strength affect the evolution only by changing the rotation velocity and thus the rotation induced mixing.

\begin{figure*}[t]
	\begin{minipage}{0.5\hsize}
		\centering
		\includegraphics[width=0.9\textwidth]{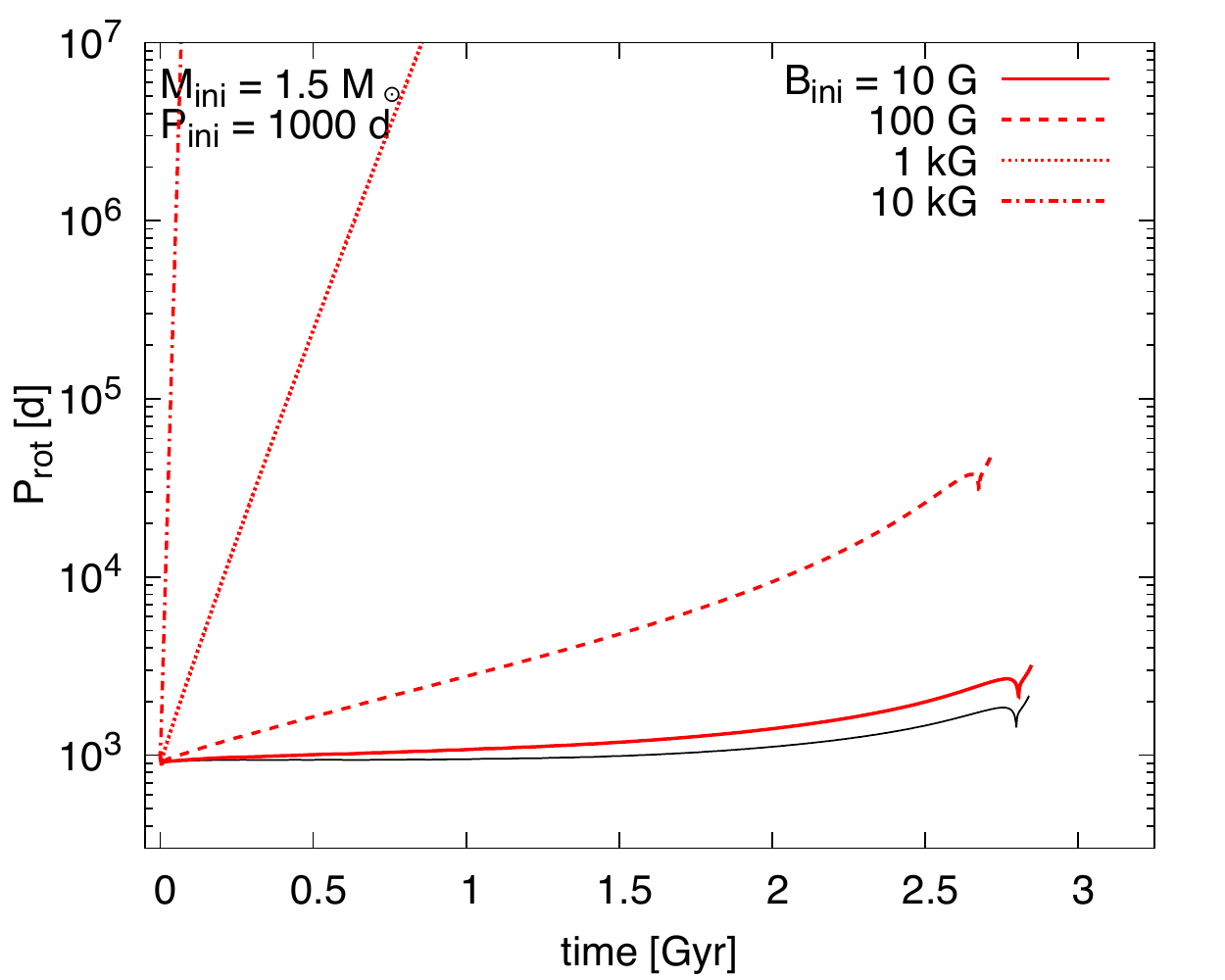}
	\end{minipage}
	\begin{minipage}{0.5\hsize}
		\centering
		\includegraphics[width=0.9\textwidth]{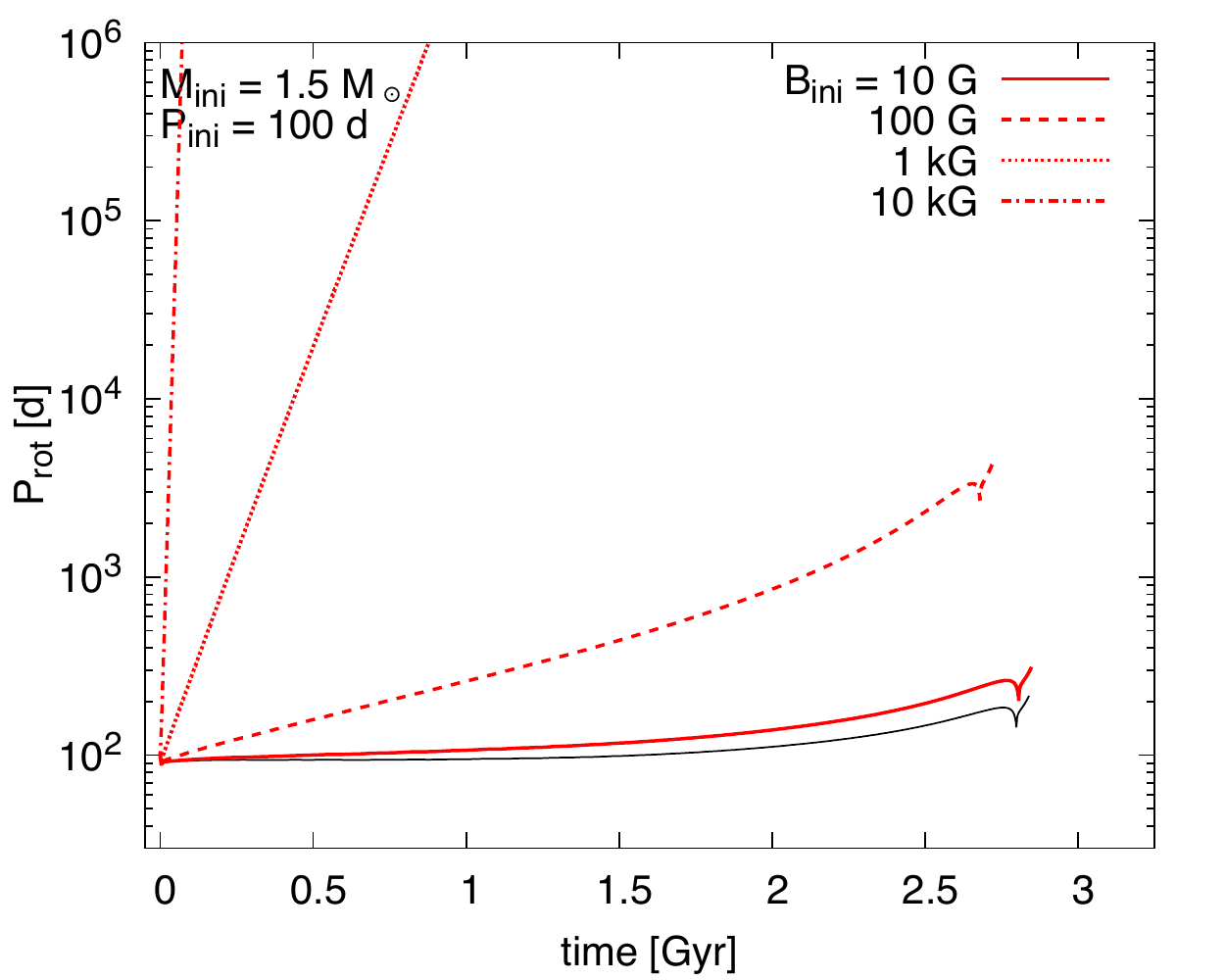}
	\end{minipage}
	\begin{minipage}{0.5\hsize}
		\centering
		\includegraphics[width=0.9\textwidth]{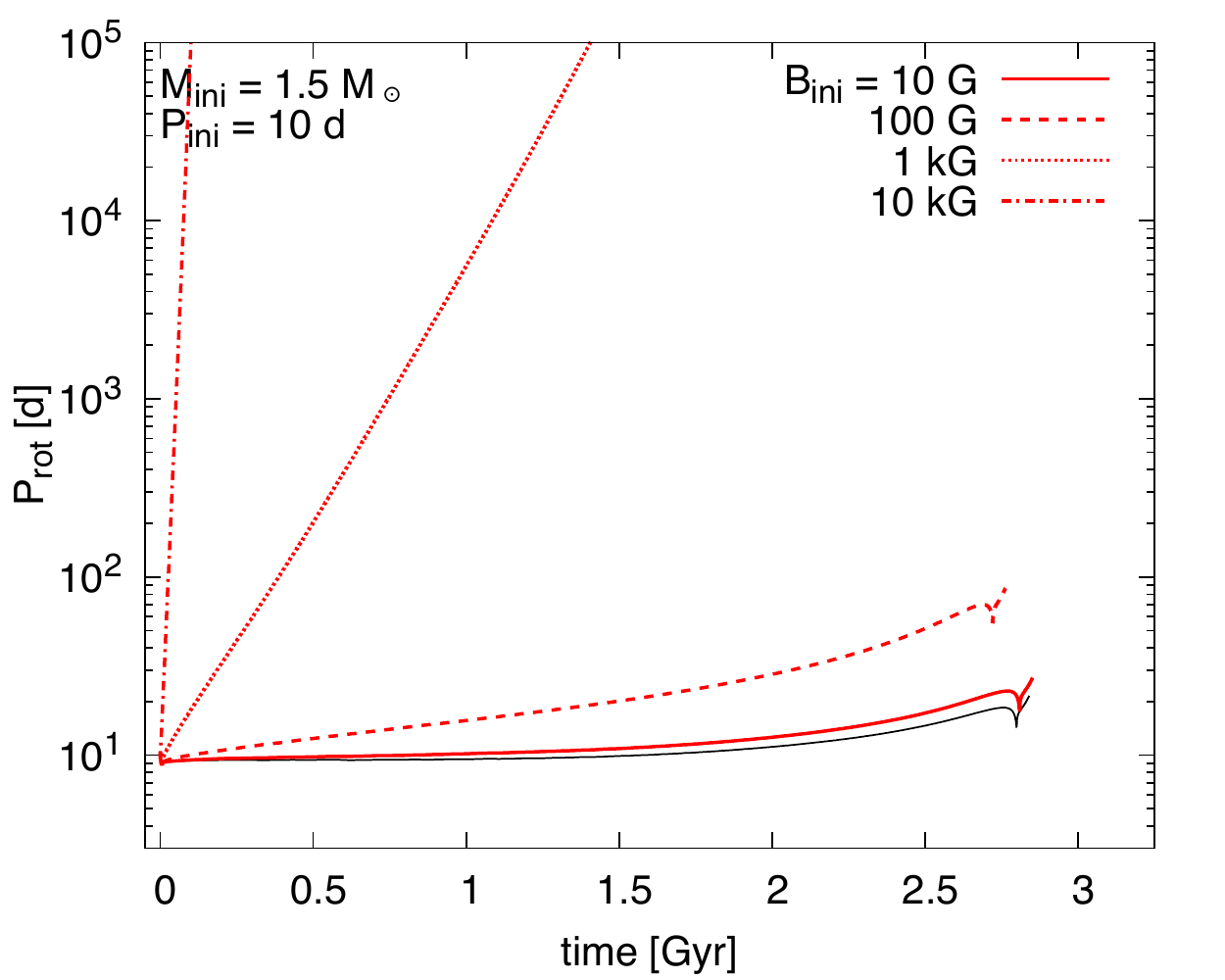}
	\end{minipage}
	\begin{minipage}{0.5\hsize}
		\centering
		\includegraphics[width=0.9\textwidth]{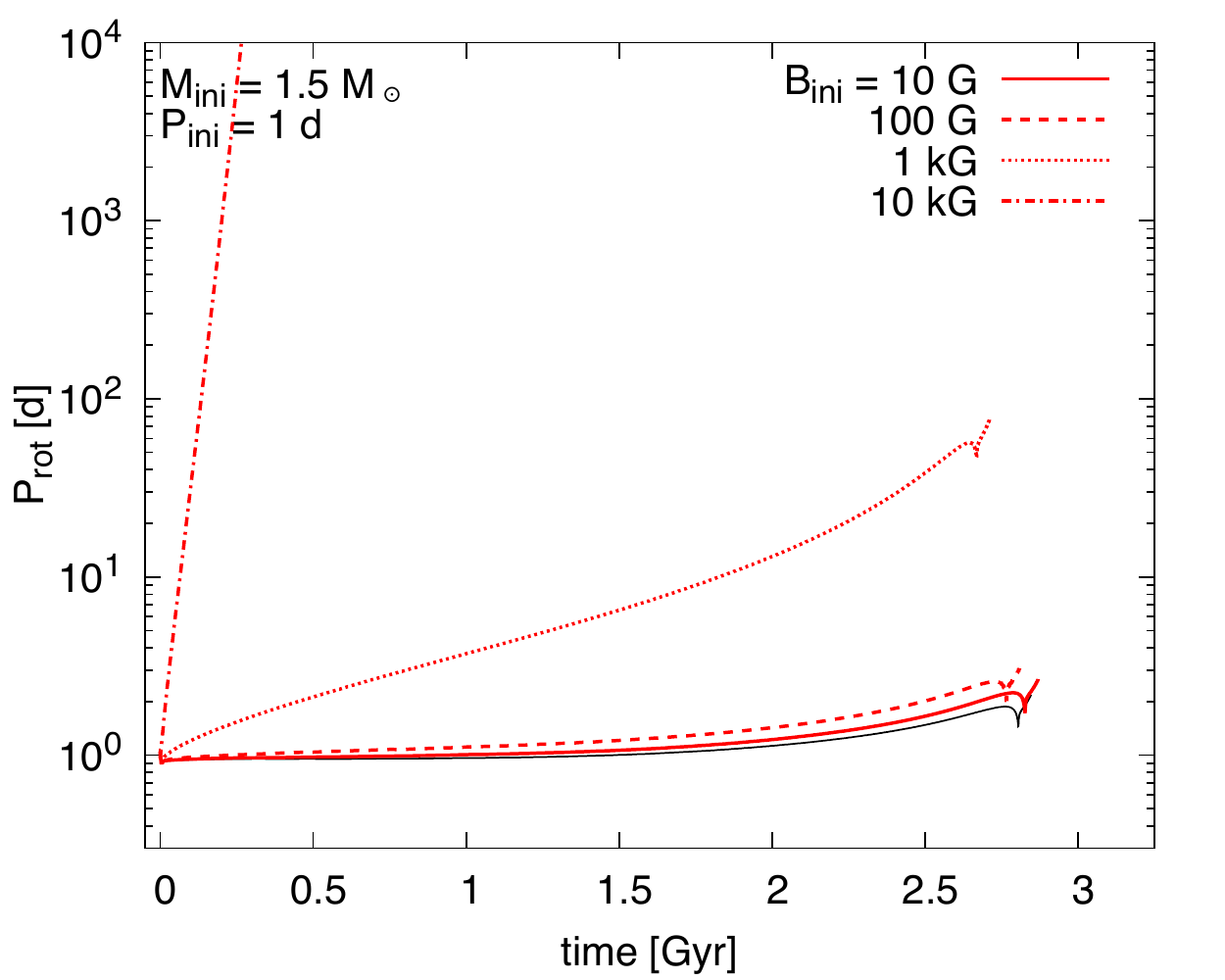}
	\end{minipage}
	\caption{Evolution of the surface rotation periods as a function of time during core hydrogen burning for our 1.5 M$_\odot$ models. Models with $B_{\rm p,ini} =$ 10, 100, 1000, and 10,000 G are shown by red solid, dashed, dotted, and dash-dotted lines, respectively. Models with $P_{\rm rot, ini} =$ 1000, 100, 10, and 1 d are shown in the top-left,  top-right, bottom-left, and bottom-right panels, respectively. The black-solid lines correspond to models without stellar wind mass loss and with $B_{\rm p,ini} =$ 10 G for each initial rotation period.
	}
	\label{fig:Protevol}
\end{figure*}
The evolution of the surface rotation periods for our models with different initial spins and magnetic field strengths is shown in Fig. \ref{fig:Protevol}. Models without wind mass loss (black solid lines) conserve their initial angular momenta. This figure shows that the stronger the surface magnetic field, the stronger the magnetic braking takes place. With the adopted mass loss rate, a B-field of $\sim$100 G is strong enough to spin down the stars by an order of magnitude within their main-sequence lifetime, except for our fastest rotating model. The efficiency of the magnetic braking scales with the strength of the surface magnetic field. Thus the spin-down of the models with $B_{\rm p,ini} =$ 1000, and 10,000 G is $\sim$10 and 100 times faster than that of the models with $B_{\rm p,ini} =$ 100 G. However, the magnetic braking gets weaker in our fastest rotating models ($P_{\rm rot, ini}$ = 1 d). This is because the surface magnetic fields are dissipated due to the efficient $\eta$ effect.

\begin{figure*}[t]
	\begin{minipage}{0.5\hsize}
		\centering
		\includegraphics[width=0.9\textwidth]{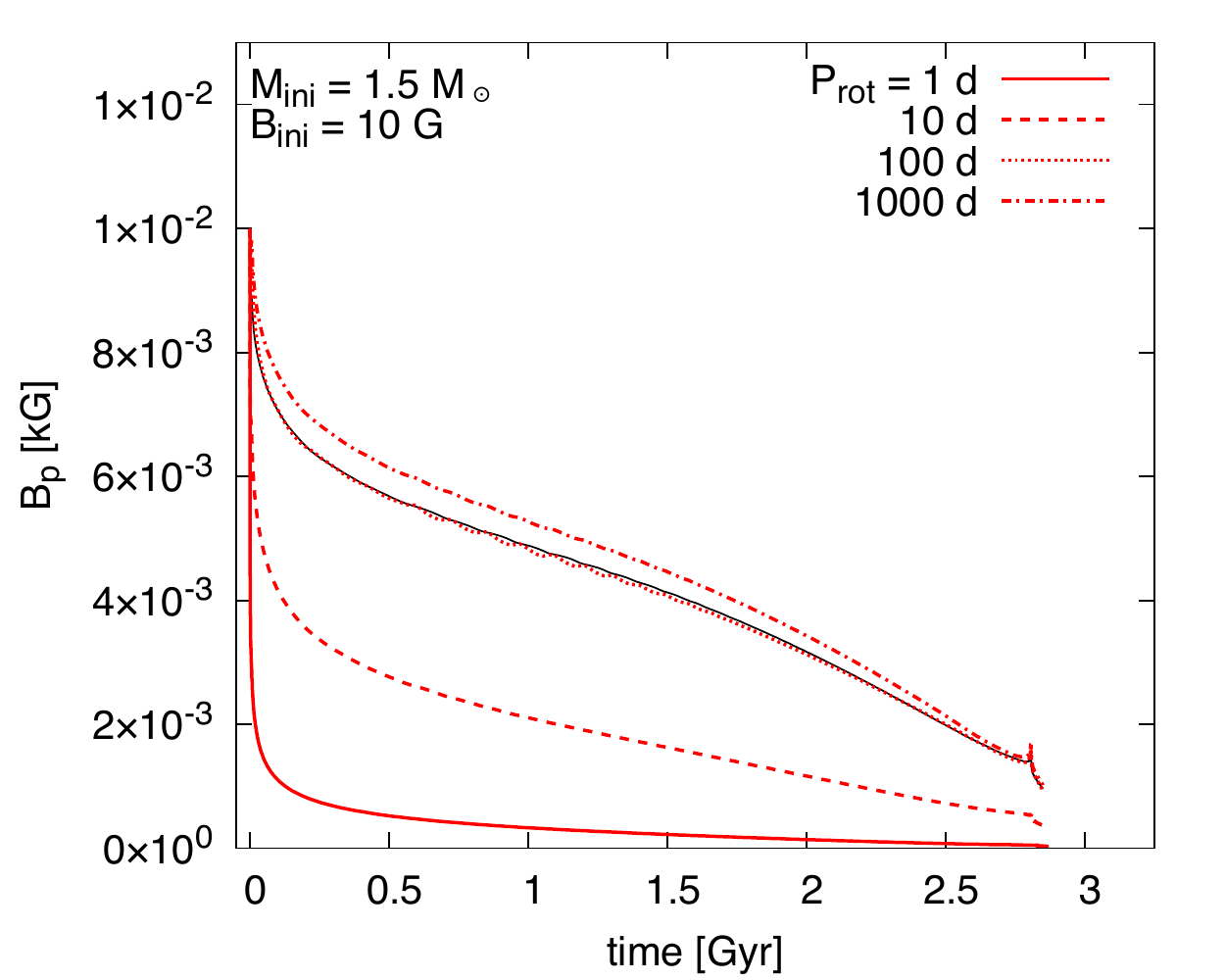}
	\end{minipage}
	\begin{minipage}{0.5\hsize}
		\centering
		\includegraphics[width=0.9\textwidth]{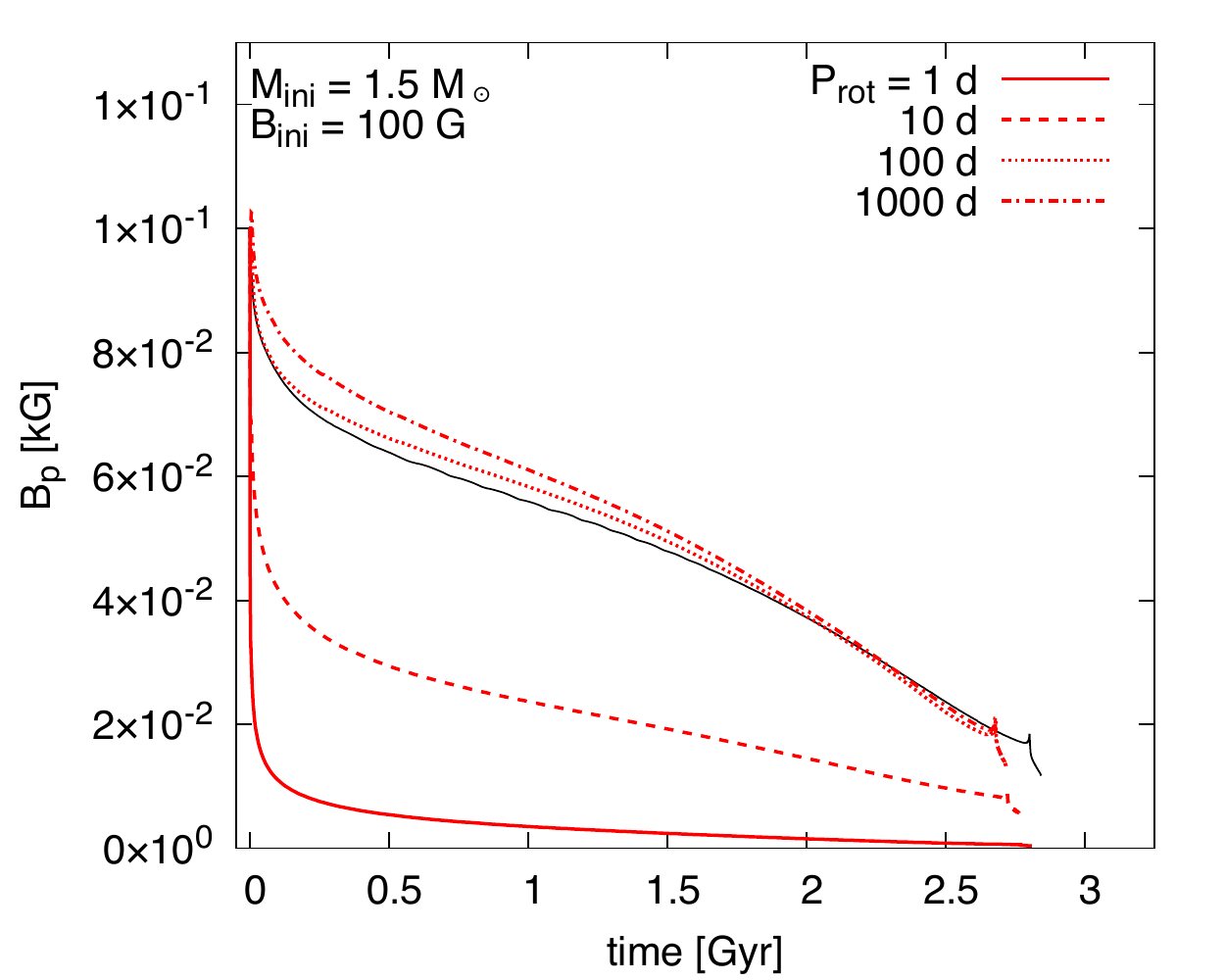}
	\end{minipage}
	\begin{minipage}{0.5\hsize}
		\centering
		\includegraphics[width=0.9\textwidth]{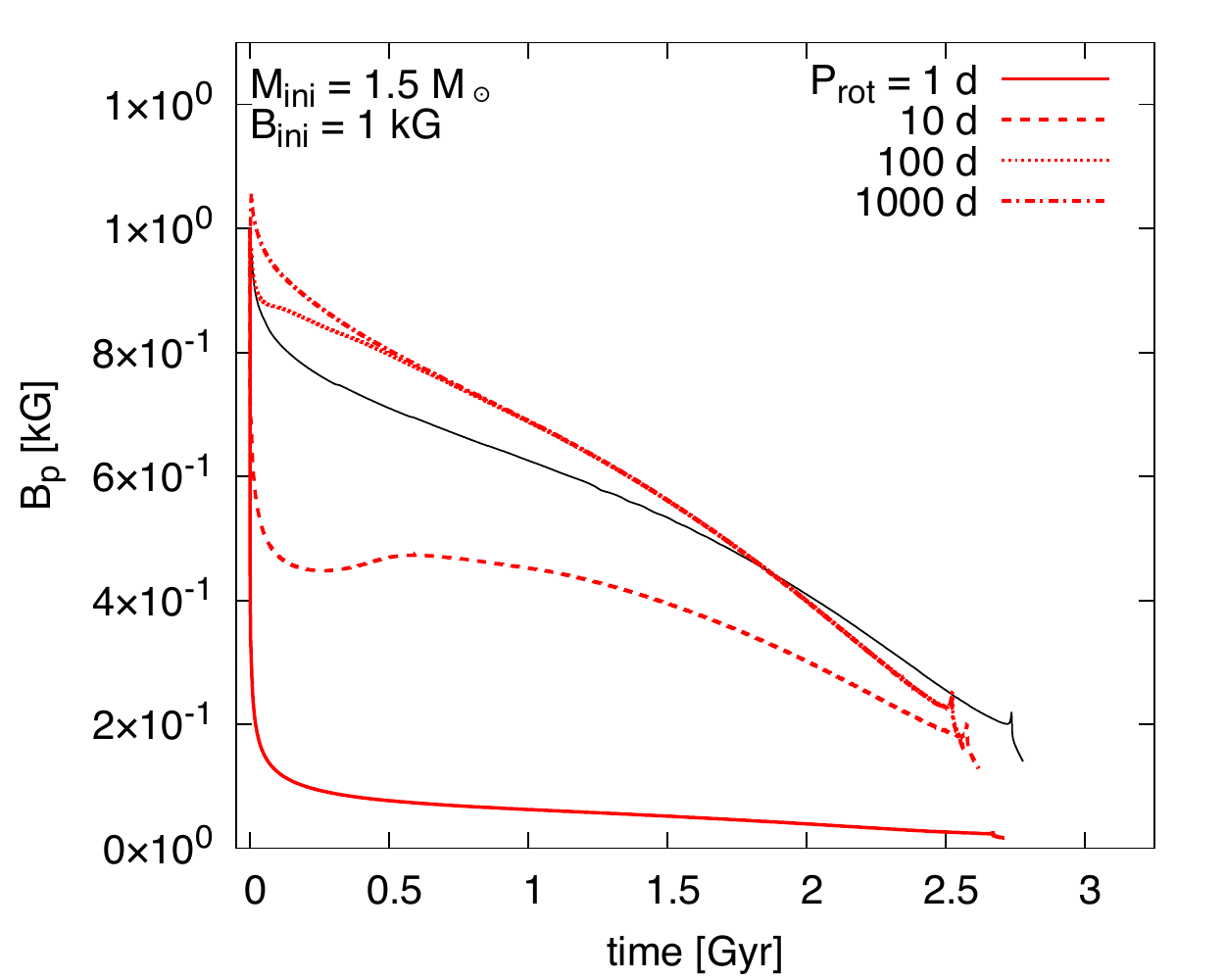}
	\end{minipage}
	\begin{minipage}{0.5\hsize}
		\centering
		\includegraphics[width=0.9\textwidth]{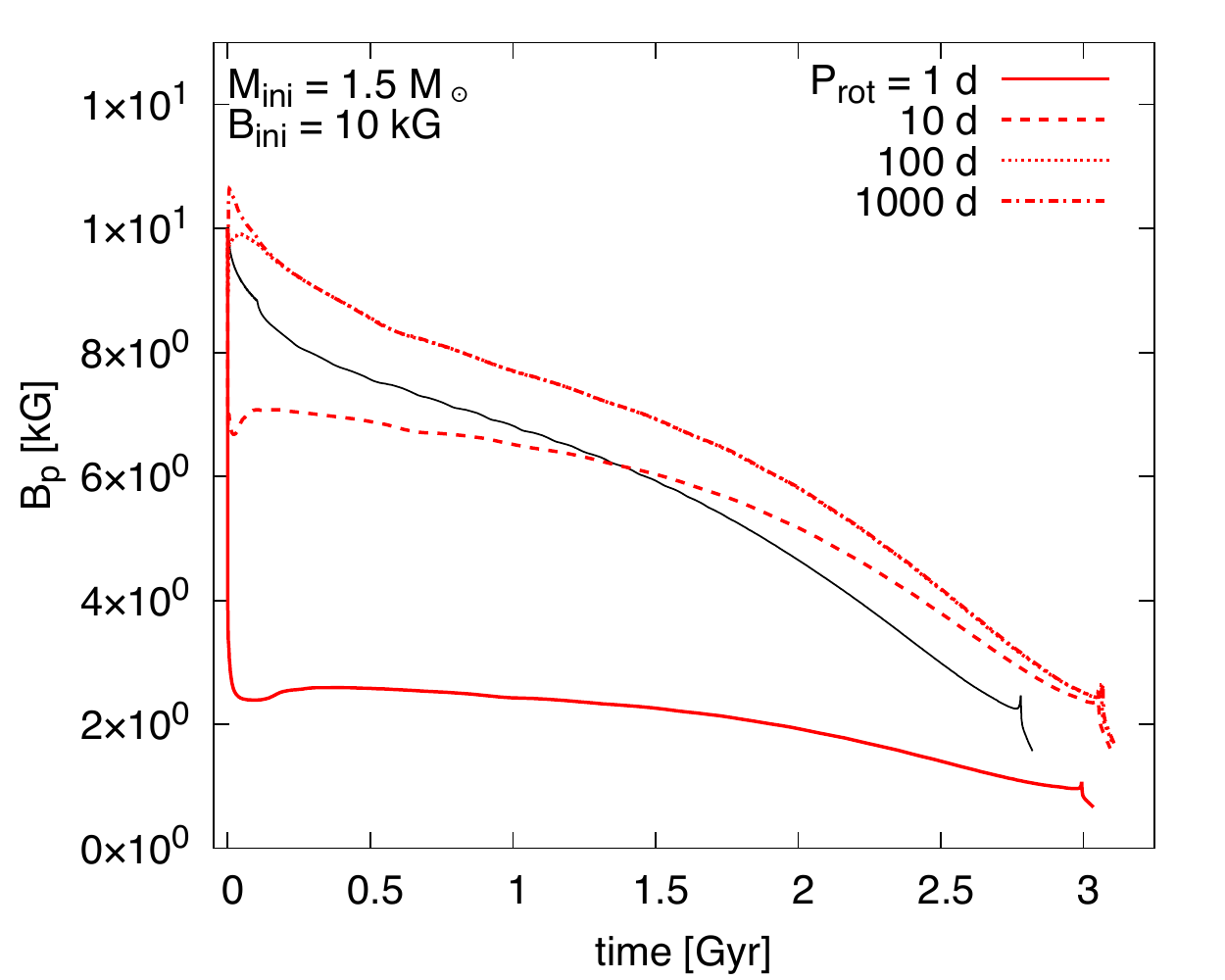}
	\end{minipage}
	\caption{Evolution of the polar surface magnetic field strength as a function of time for the same models as those shown in Fig. \ref{fig:Protevol}. Models with $P_{\rm rot, ini} =$ 1, 10, 100, and 1000 d correspond to red solid, dashed, dotted, and dash-dotted lines, respectively. Models with $B_{\rm p,ini} =$ 10 G are shown in the top left panel, those with $B_{\rm p,ini} =$ 100, 1000, and 10,000 G are respectively shown in the top-right, bottom-left, and bottom-right panels. The black solid lines depict models without the $\eta$ effect with $P_{\rm rot, ini} =$ 1000 d for each initial magnetic field strength.
	}
	\label{fig:Bradevol}
\end{figure*}
Figure \ref{fig:Bradevol} shows the time evolution of the polar surface magnetic strength $B_{\rm p}$. For comparison, the results of models without the $\eta$ effect, which have $P_{\rm rot, ini} =$ 1000 d, are also shown (black solid lines). In these, the surface magnetic field decreases due to mass loss. When a mass is lost, the layers below the surface expand, and magnetic flux conservation leads to weaker fields. A much more significant drop in magnetic field strength takes place in models with the $\eta$ effect. We see that the field strengths decrease faster for faster-rotating models, which argues for dissipation driven by rotation induced turbulence. For example, for $B_{\rm p,ini} =$ 10 G (top-left panel), the strongest dissipation takes place in the model with $P_{\rm rot, ini} =$ 1 d, but for models with $P_{\rm rot, ini} =$ 1000 and 10,000 d, the surface magnetic field evolves as in a model without $\eta$ effect. Similar behavior is obtained for the models with $B_{\rm p,ini} =$ 100 G.

Our most magnetic models ($B_{\rm p,ini} =$ 1000 and 10,000 G) follow a more complicated evolution. The two slowly rotating models maintain stronger surface fields than the model without the $\eta$ effect. This is so because the strong surface magnetic fields suppress stellar wind mass loss by the magnetic confinement. This also takes place in the model with $P_{\rm rot, ini} =$ 10 d, but not in the fastest rotating one ($P_{\rm rot, ini} =$ 1 d). Here, the surface magnetic field is quickly dissipated by rotation induced turbulence. We conclude that the surface magnetic field strength is affected by wind mass loss and by rotation induced dissipation. However, at the same time, these two effects are influenced by a strong surface magnetic field, i.e., the wind mass loss is suppressed by magnetic confinement, and the rotation is also slowed down by magnetic braking.

\begin{figure*}[t]
    \begin{center}
	\includegraphics[width=0.9\textwidth]{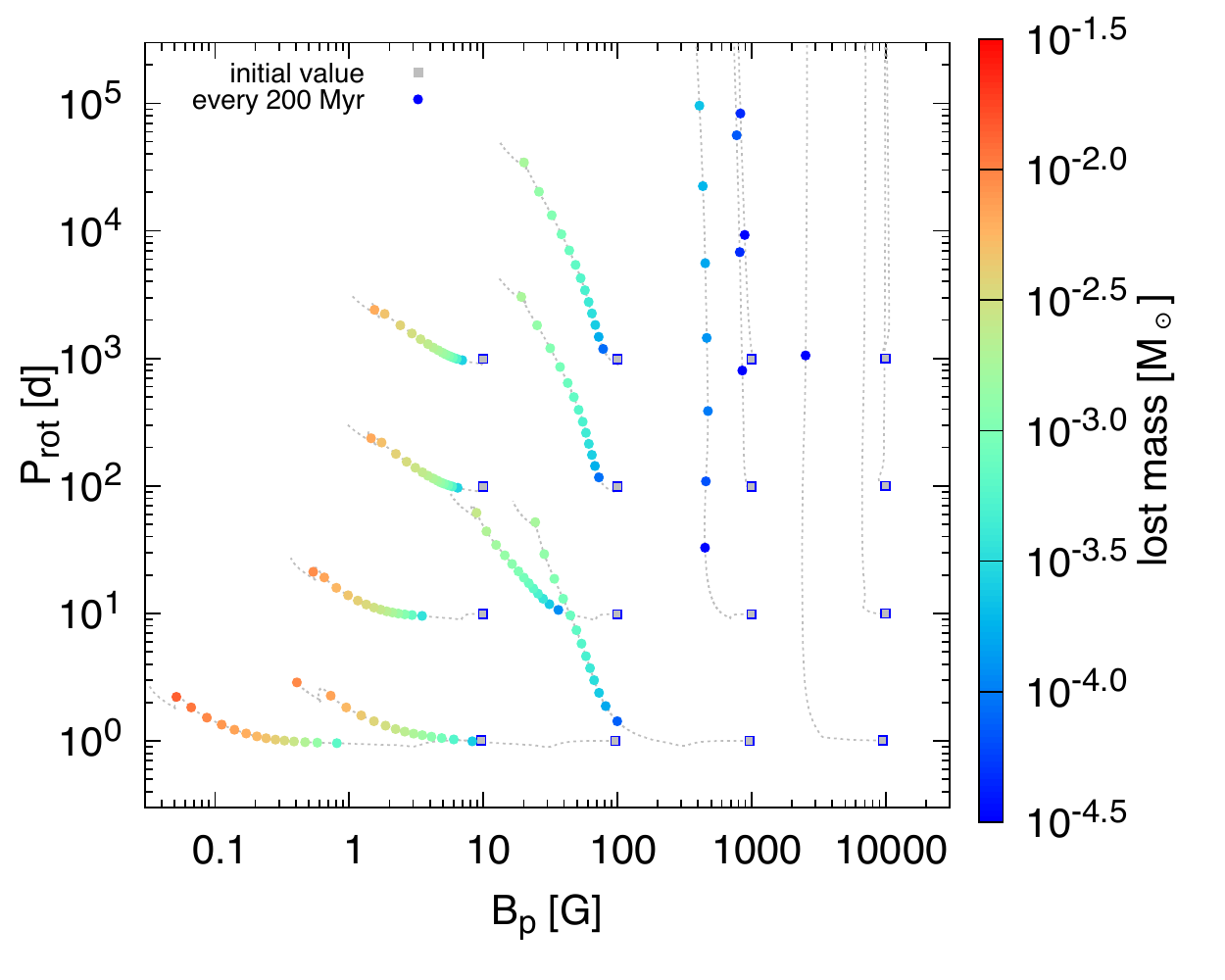}
	\caption{Surface rotation period ($P_{\rm rot}$) as function of the polar surface field strength ($B_{\rm p}$) during the main-sequence evolution of our 1.5 $M_\odot$ models. The corresponding ZAMS values are shown by gray squares with $B_{\rm p}=$10, 100, 1000, and 10,000 G and $P_{\rm rot}=$ 1, 10, 100, and 1000 d. As evolution proceeds, the models increase their rotation periods and decrease their magnetic field strengths as indicated by the gray dashed lines. Colored dots are placed on these evolutionary tracks every 200\,Myr, with the color indicating the total amount of mass lost during the previous evolution (see color bar to the left). Kinks close to the last points in models with $B_{\rm p,ini} \leq 100$ G indicate the TAMS turnoff.
	}
	\label{fig:Bp-Prot}
	\end{center}
\end{figure*}
The evolution of $B_{\rm p}$ and $P_{\rm rot}$ is shown together in Fig. \ref{fig:Bp-Prot}. As dots are placed on the tracks after a constant elapsed time, the density of dots in this figure is somewhat representative of the likelihood of observing a star at a given location. The very low density in the lower right corner of Fig. \ref{fig:Bp-Prot} implies that the likelihood to find a strongly magnetized star ($B_{\rm p} > 1000$ G) with a rotation period below $\sim 1000$\,d is very small. This is so since for the considered field strength, the magnetic spin-down timescale becomes shorter than $\sim$ 1\% of the main-sequence lifetime. At the same time, the likelihood to find a rapidly rotating $P_{\rm rot} < 1$\,d strongly magnetized star ($B_{\rm p} > 100$\,G) is also low, because fast rotation results in efficient magnetic dissipation.

Figure \ref{fig:Bp-Prot} also indicates the evolution of the total amount of mass lost by the models due to stellar winds. The model with $B_{\rm p,ini} = 10$ G and $P_{\rm rot, ini} = 1000$ d is understood the easiest, as it has the most modest magnetic confinement and negligible rotational mass loss enhancement. This model loses $5.9 \times 10^{-3}$ M$_\odot$ during its main-sequence phase. Comparably, the models with  $B_{\rm p,ini} = $10,000 G lose only $2.1 \times 10^{-4}$ to $3.4 \times 10^{-4}$ M$_\odot$, because of efficient magnetic confinement. On the other hand, the model with $B_{\rm p,ini} = 10$ G and $P_{\rm rot, ini} = 1$ d experiences magnetic dissipation, and therefore the magnetic confinement on the model is minimal. Its wind mass loss is enhanced by the fast rotation, and it loses $1.33 \times 10^{-2}$ M$_\odot$. The rotational enhancement also takes place for other models with $P_{\rm rot, ini} = 1$ d, but, with the stronger initial surface fields of  $B_{\rm p,ini} = $ 100 and 1000 G, the lost amounts are reduced because either the magnetic confinement or the reduction of the rotational enhancement due to the magnetic braking also happen. The lost masses are $8.3 \times 10^{-3}$ M$_\odot$ for the model with $B_{\rm p,ini} = 100$ G and $P_{\rm rot, ini} = 1$ d and $1.8 \times 10^{-3}$ M$_\odot$ for the model with $B_{\rm p,ini} = 1000$ G and $P_{\rm rot, ini} = 1$ d. We conclude that by considering a surface magnetic field, the total wind mass loss displays a complex behavior because of the interplay between wind mass-loss rate, rotation, magnetic confinement, and magnetic spin-down, even for a fixed stellar mass and metallicity.

\section{Red-giant branch evolution of 1.5 M$_\odot$ stars} \label{sec:result-RG}

Here we describe the results of our model calculation of a 1.5 M$_\odot$ star of solar metallicity from the zero-age main-sequence (ZAMS) up to core helium ignition at the tip of the red-giant branch, where the model experiences a violent helium flash and the calculation is ended. We include the same physics as that used in the models of the previous section. Here, we use an initial rotation period of $P_{\rm rot} = 1.4$\,d, a zero toroidal field, and a uniform poloidal field with $B_{\rm p,ini}=$ 10 G. These values may be representative of normal A-type stars. Our fiducial model is computed using $n_{\rm cv} = 0$, and later we will discuss results obtained with $n_{\rm cv} = 2$ for an otherwise identical model.

\begin{figure*}[t]
	\begin{center}
	\includegraphics[width=\textwidth]{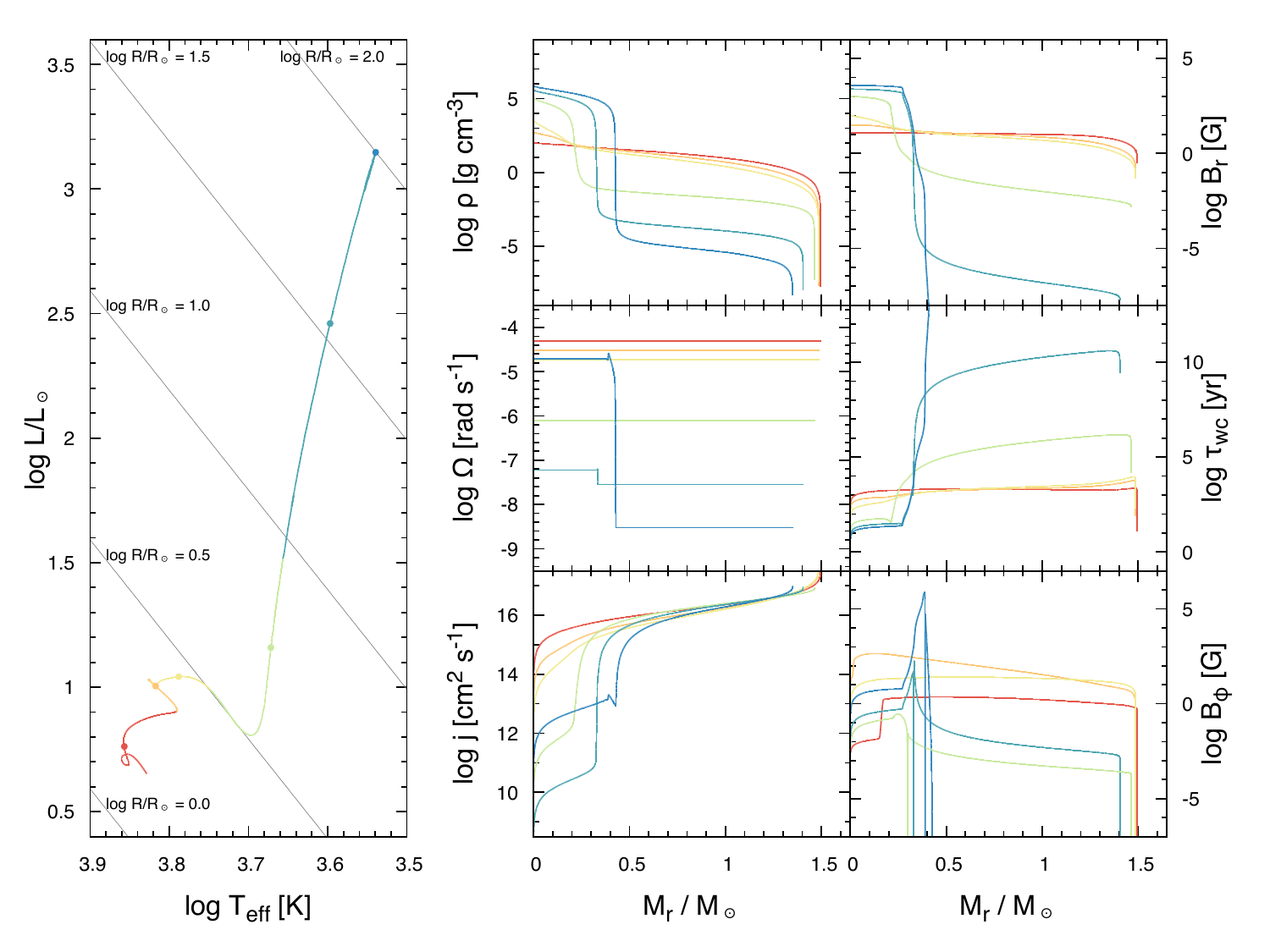}
	\caption{Left panel) Evolution of a 1.5 M$_\odot$ magneto-rotating model in the HR diagram, computed with $n_{\rm cv} = 0$. The line colors indicate the different evolutionary phases. Grey lines are iso-radius lines for $\log R/R_{\odot} = 0.0, 0.5, 1.0, 1.5$, and $2.0$. Right panels) Profiles of internal density (top left), radial field strength (top right), rotation frequency (middle left), local wave-crossing time (middle right) specific angular momentum (bottom left), and toroidal field strength (bottom right), as a function of the Langrangian mass coordinate. In the left panel, the six epochs for which profiles are shown in the right panels are indicated by colors and dots that have the same color as the profiles.
	}
	\label{fig:RGevol}
	\end{center}
\end{figure*}

The evolution of the fiducial model in the HR diagram is shown in the left panel of Fig.\ref{fig:RGevol}. We divide the evolution into 5 phases, which are indicated by lines with different colors (red, orange, yellow, green, and blue, from the beginning) in the figure. Representatives including the model at helium ignition are selected from each phase, which are shown by dots on the HR diagram. Corresponding internal profiles of physical key properties are shown in the right panels using the same line color (dark-blue for the model at helium ignition).

The top-left figure shows the increasing density contrast between the core and the envelope. The angular velocity profiles demonstrate that rigid rotation is maintained even after the core contraction (yellow and green), until the core growth phase (blue), due to the efficient magnetic angular momentum transfer. This efficiency can be estimated through the crossing time of the torsional \Alfven wave, which is defined as $\tau_{\rm wc} =  r \sqrt{20 \pi \rho} / B_r$ (middle-right panel). Profiles of the radial magnetic field strength are shown in the top-right panel. The radial component keeps $\gtrsim 0.1$ G until envelope convection develops, and consequently, the local wave-crossing time remains to be shorter than $\lesssim 10^4$ yr.

The magnetic diffusivity is significantly enhanced in the convective envelope. Hence it decreases the radial component after the development of envelope convection, increasing the local wave-crossing time (green and blue). Finally, it becomes longer than the evolutionary timescale of $\sim 10$ Myr. Consequently, the model starts to develop differential rotation during the core growth phase (blue).

The rotation of the red-giant core significantly accelerates during this phase. The more than two orders of magnitude acceleration can be evaluated as follows. At first, the matter accreted by the helium core has a large specific angular momentum of
\[
	j_{\rm acc} \sim 3.3 \times 10^{13} 
	\left( \frac{R_{\rm base}}{1 \ R_\odot} \right)^2 
	\left( \frac{\Omega_{\rm base}}{10^{-8} \ \mathrm{rad} \ \mathrm{s}^{-1} } \right) 
	\ \mathrm{cm}^2 \ \mathrm{s}^{-1},
\]
where $R_{\rm base}$ and $\Omega_{\rm base}$ are the radius and the angular velocity of the base of the convective envelope, respectively. Then, the total accreted angular momentum can be estimated as
\[
	J_{\rm acc} = 9.2 \times 10^{45} 
	\left( \frac{j_{\rm acc}}{ 3.3 \times 10^{13} \ \mathrm{cm}^2 \ \mathrm{s}^{-1} } \right)
	\left( \frac{M_{\rm acc}}{ 0.14 \ M_\odot } \right)
	\ \mathrm{g} \ \mathrm{cm}^2 \ \mathrm{s}^{-1},
\]
where $M_{\rm acc} \sim$ 0.14 M$_\odot$ is the accreted amount of mass. This accreted amount of angular momentum largely exceeds the original core angular momentum of $\sim 10^{43}$ g cm$^2$ s$^{-1}$. Since the magnetic field maintains rigid rotation in the helium core, the accreted angular momentum is quickly redistributed throughout the core. Assuming that the accreted amount of angular momentum is redistributed within a core of $M_{\rm core} = 0.43$ M$_\odot$ and $R_{\rm core} = 1.6 \times 10^{-2}$ R$_\odot$, a core angular velocity of 
\begin{eqnarray}
	\Omega_{\rm core} \sim &2.2 \times 10^{-5}&
	\left( \frac{J_{\rm acc}}{ 9.2 \times 10^{45} \ \mathrm{g} \ \mathrm{cm}^2 \ \mathrm{s}^{-1} } \right) \nonumber \\
&&\times
	\left( \frac{ 4.2 \times 10^{50} \ \mathrm{g} \ \mathrm{cm}^2 }{(2/5)M_{\rm core}R_{\rm core}^2} \right)
	\ \mathrm{rad} \ \mathrm{s}^{-1} \nonumber
\end{eqnarray}
is expected, which corresponds well to the simulation result. As a consequence of the accelerated core rotation, strong differential rotation occurs at the core-envelope boundary. Consequently, a strong toroidal field of $\sim10^6$ G is induced.

\section{Comparison with previous results} \label{sec:comp-theory}

\subsection{Models including the Tayler--Spruit dynamo} \label{sec:TSdynamo}

As the large impact of magnetic fields on the internal angular momentum distribution of stars has become more and more evident during the last two decades (cf., Sect.\,\ref{sec:intro} and Sect.\,\ref{sec:comp-obs} below), many stellar evolution models attempted to account for this by incorporating magnetic angular momentum diffusion as proposed by \citet{Spruit99, Spruit02} in various versions \citep{Maeder&Meynet03, Maeder&Meynet04, Heger+05, Yoon&Langer05, Denissenkov&Pinsonneault07, Suijs+08, Brott+11a, Yoon+12, Fuller+19}. This model relies on a dynamo picture, the so-called Tayler--Spruit dynamo, which is assumed to operate in differentially rotating, radiative layers in stars.

The original picture of the Tayler--Spruit dynamo consists of three steps. First, due to the $\Omega$ effect, a strong toroidal magnetic field develops in a radiative layer due to differential rotation. Second, the Pitts--Tayler instability induces turbulence in this region. Finally, a radial magnetic field with considerable strength is induced by the turbulent stretching of the toroidal field. The induced radial component plays a role as the next seed field of the $\Omega$ effect such that it closes the dynamo loop. Following this picture, one may estimate the strength of local magnetic components of $B_r$ and $B_\phi$ as well as the magnetic stress $S_{\rm TS} = B_r B_\phi / 4 \pi$. The magnetic stress is further rewritten in terms of the viscosity, $\nu_{\rm TS}$, by equating $\rho \nu_{\rm TS} r \partial_r \Omega = S_{\rm TS}$. This magnetic viscosity is then used in evolutionary calculations in the form of a diffusion coefficient for angular momentum transport.

The Tayler--Spruit picture is incompatible with our modeling in two aspects. First, the time evolution of the toroidal component $B_\phi$ is significantly simplified such that $B_\phi$ is directly proportional to $\partial_r \Omega$. This simplification leads to a diffusion approximation of the magnetic stress, so that angular momentum redistribution takes place locally as a form of diffusion, while the most natural consequence of the Lorentz force and the $\Omega$ effect is the formation of the torsional \Alfven wave (Sect.\,\ref{sec:wave1}). Second, the Tayler--Spruit dynamo relies on the $\alpha$ effect of the Pitts--Tayler instability, such that the ``radial'' component that contributes to the Lorentz force is dominated by the secondary generated field. In contrast, the poloidal component that contributes to the wave propagation in our case is the original field, and a hydrodynamic induction to reproduce the poloidal component is not assumed.

It is evident that, when a star has a structured poloidal magnetic field, the dominant phenomenon occurring after the $\Omega$ effect will be wave propagation, at least for the simplified 1D geometry adopted here. For a structured but weak poloidal field, the field strength only affects the timescale, and the torsional \Alfven wave will still form. This is because the vertical length scale of the Pitts--Tayler instability is always smaller than the wave-crossing length scale (see Sect.\,\ref{sec:wave2}).

Then the important question will be that what will happen in a star that has a very weak and unstructured initial poloidal field (as a strong and unstructured poloidal field would be unstable and does not exist). Also, in this case, a strong toroidal component may develop due to the $\Omega$ effect in a region with a differential rotation. Because the initial poloidal is weak and unstructured, the torsional \Alfven wave launched by the $\Omega$ effect will be locally trapped. Then, the region will be predominantly affected by the Pitts--Tayler instability. Turbulence may affect the magnetic field through both, the $\alpha$ and the $\eta$ effect.

If $\eta$ effect wins, then the initially weak poloidal component will soon dissipate, so that the region will have a pure toroidal magnetic field. Because purely toroidal fields are unstable \citep[such that instabilities like the Pitts--Tayler instability develop; e.g.][]{Tayler73}, eventually, all the magnetic energy will dissipate into heat. Thus the magnetic field does not affect the differential rotation in the region. If the $\alpha$ effect wins, the poloidal field is amplified to have a complicated 3D structure that is embedded within the instability region. Having a short crossing time, torsional \Alfven waves will now propagate along the amplified poloidal field, and non-linear interaction (e.g. phase mixing) may efficiently dissipate the waves. Since there will be no preferred direction for the induced poloidal field, the dissipation will take place nearly chaotically. This might result in a similar outcome to what is discussed in the Tayler--Spruit dynamo.

Therefore, the balance between the $\eta$ and $\alpha$ effects is of importance. Whether the $\alpha$ effect works in this situation as assumed in the Spruit-Taylor picture is debated, as contradictory results based on multi-dimensional MHD simulations have been reported \citep{Braithwaite06, Zahn+07}. Unfortunately, multi-dimensional simulations with realistic thermal and magnetic diffusivities, which are desirable to test the dynamo picture, are challenging with the current computational resource \citep[see discussion in][]{Braithwaite&Spruit17}. Whether the Tayler--Spruit dynamo works in a star remains unclear at this time. We compare both, results obtained with the Taylor--Spruit dynamo, and our results, with observations in Sect.\,\ref{sec:comp-obs}.

\subsection{Other approaches}

Several works have incorporated other magnetic effects than the magnetic viscosity based on the Tayler--Spruit dynamo into stellar evolution simulations.

The interaction of a surface magnetic field and the stellar wind has been considered in the context of modeling the evolution of massive stars \citep{Meynet+11, Petit+17, Georgy+17, Keszthelyi+19}, in particular, to study the expected spin-down due to their intrinsically large mass loss rate. While these works have revealed the significance of the effects of magnetic braking and/or magnetic confinement, they do not account for the evolution of the surface magnetic field but assumed either a constant magnetic field strength or constant magnetic flux during the evolution. Our simulation implies that neither assumption may be realistic, because the surface magnetic field may change due to other mechanisms in addition to the flux conservation. In particular, the mass loss will affect the surface magnetic field for massive stars, as it replaces the surface material with a matter that originally stayed below the surface, which has a different magnetic flux compared to the original surface. The Ohmic decay of the field can also be significant. Therefore, a self-consistent global simulation for the stellar magnetism may improve the current estimate of the field interaction with the stellar wind in massive stars.

Furthermore, in some of the above-quoted calculations, the assumptions for the internal magnetic fields are unrelated to the assumptions for the surface field. E.g., \citet{Meynet+11} present massive star models with a strong surface field and spin-down but assuming that no B-field is present inside the star, with the consequence of strong internal differential rotation. Such inconsistencies are avoided with the present approach. 

The models of \citet{Feiden&Chaboyer12, Feiden&Chaboyer13, Feiden&Chaboyer14, Feiden16} take into account several magnetic field effects, such as the magnetic pressure and the magnetic tension, and especially the efficiency change of the convective energy transport. The stellar structure of low mass stars with convective envelopes is shown to be sensitive to these effects. However, also these models lack a detailed theory of the evolution of the stellar magnetic field. Instead, they assume a constant surface magnetic field with a simple radial profile for the internal field strength. The large effects on the stellar structure shown by their models argue for the importance of incorporating an appropriate evolution theory for the stellar magnetic field in stellar evolution calculations, which satisfies the essential MHD conditions such as flux conservation and divergence-free magnetic field configurations.

Such global simulations have been performed for the first time by \citet{Potter+12c}, and their formalism has also been used in their later works \citep{Quentin&Tout18}. Although the physics included in the modeling is similar to ours, their formulation has two fundamental shortcomings. Their evolution equations for the magnetic field do not reproduce the magnetic flux conservation, and similarly, the angular momentum conservation is not guaranteed with their expression of the Lorentz force. It would be because their choice of the surface-averaging of the original 3D expressions is too simple: it is likely that simple weighted surface-averaging is applied for both the mean-field MHD-dynamo equation and the 3D Lorentz force.

Interestingly, Potter's models successively reproduce the observed population of slowly rotating but nitrogen-enhanced massive stars \citep{Hunter+08}. These stars can not be obtained by standard rotating single stars models \citep{Brott+11b}, whereas binary evolution can produce such stars \citep{Langer+08, Marchant16}. Their magnetically braked models not only slows down the surface rotation but also allows strong differential rotation to develop inside the star. Then the Pitts--Tayler instability aided by the $\Omega$ effect develops, which allows efficient chemical diffusion to account for the surface nitrogen enhancement. The assumptions in Potter's work are comparable to those in the Tayler--Spruit dynamo and thus draw a common picture with other evolutionary models \citep[e.g.][]{Meynet+11}. In contrast, our model predicts nearly rigid rotation inside a magnetic star, and efficient matter mixing due to any instabilities powered by differential rotation does not take place. Considering the large impact of chemical mixing on stellar evolution, we will investigate the relationship between nitrogen enhancement, rotation, and magnetic field in the future.

Another effect that is omitted from our present modeling is the magnetic suppression of hydrodynamic flows. Its relevance to accounting for the stability of atmospheres of Ap/Bp stars, which has been required for atmospheric diffusion processes to take place, has been discussed by \citet{Michaud70}. Similarly, a small macroturbulence of only $\sim$ a few km s$^{-1}$ was measured in the O type star NGC\,1624-2, which has a strong dipolar surface field of $\sim$20 kG. While usually, the macroturbulent velocities, thought to be caused by pressure waves emitted by sub-photospheric convection zones \citep{Grassitelli+15}, are at least one order of magnitude larger in such stars \citep{Simon-Diaz+17}, \citet{Sundqvist+13} show quantitatively that the magnetic pressure in this star is strong enough to suppress the sub-surface convection. The impact of the magnetic inhibition of the core convection on blue supergiant evolution has been investigated by \citet{Petermann+15}, where convectively unstable regions in a star are artificially reduced by modifying the convective criterion. They have indeed shown that this modification significantly affects the stellar structure to reproduce the enigmatic surface temperature of the progenitor of supernova 1987A, which otherwise will require a stellar merger during the evolution \citep{Menon&Heger17, Urushibata+18}.

In a convection criterion proposed by \citet{Lydon&Sofia95}, a magnetic field can both stabilize and destabilize the region, depending on the radial gradient of the field strength. In a 3D radiation magneto-hydrodynamic simulation by \citet{Tremblay+15}, which has a cuboid-shaped computational domain embedded in an atmosphere of a white dwarf (`box-in-a-star'), it has been shown that convective transport is significantly impeded when the plasma-$\beta$ (the ratio between the thermal and the magnetic pressure) is less than the unity. However, it is unclear whether such a strong field can remain within a convectively unstable layer because theoretical works strongly indicate that there is no stable magnetic structure in a marginally convective unstable barotropic region \citep[e.g.,][]{Reisenegger09, Mitchell+15}. Further investigation of convective instability under a strong magnetic field will be demanded.

\section{Comparison with observations} \label{sec:comp-obs}

\subsection{ApBp stars}

Ap/Bp stars are main-sequence A and B type stars that show enhancements in surface chemical abundances of elements such as Si, Cr, Fe, and Eu. Observationally, there is a strong coincidence of the peculiar surface abundance pattern and the strong surface magnetic field \citep{Babcock58}. E.g., a surface magnetic field stronger than $100$ G has been detected for 41 out of 97 Ap/Bp stars, while no magnetic field was found in 138 normal AB-type stars \citep{Bagnulo+06}. The contemporary understanding of this finding is that the strong surface magnetic field stabilizes the stellar subsurface layers such that peculiar chemical abundance at the surface can result from long-term gravitational settling and radiative levitation \citep{Michaud70}.

Possibly, there are two different ways to improve the stability of stellar subsurface layers. The first possibility is that strong enough magnetic fields will regulate the fluid flow such that the flow that erases the subtle chemical imprints in Ap/Bp stars is prevented. In particular, the quantitative assessment in \citet{Sundqvist+13} indicates that a strong magnetic field can indeed suppress sub-photospheric convection. In our 1.5 M$_\odot$ models, this subsurface convective turbulence is estimated to have the energy density of $\frac{1}{2} \rho v_{\rm cv}^2 \sim 6.3 \times 10^3$--$1.6 \times 10^5$ erg cm$^{-3}$, which yields an equipartition field strength of $B_{\rm eq} \sim 300$--$1400$ G. Such strong fields are maintained during the greater part of the main-sequence evolution in our magnetic models with $B_{\rm p, ini} \geq 1$ kG and $P_{\rm rot, ini} \geq 10$ d (Fig.\,\ref{fig:Bradevol}). Similarly, meridional flows that would exist in a rotating star have been postulated to disrupt the chemical inhomogeneity. By applying an estimate of \citet{Kippenhahn74}, our nonmagnetic model with rapid rotation of $P_{\rm rot, ini} = 1$ d is estimated to have $v_{\rm ES} \sim 1$ cm s$^{-1}$ close to the surface. It is $\sim 10^{-4}$ times slower than the turbulence velocity of the subsurface convection,  and hence it could be significantly affected by very weak fields with strengths of $\lesssim 10^{-5}$ G.

The second possibility is more indirect: because magnetic stars are also known to be slowly rotating stars, their subsurface layers will be less affected by rotation induced flows than non-magnetic stars. For instance, the flow velocity of the Eddington-Sweet circulation is assumed to be proportional to the square of the rotation frequency. \citet{Michaud70} has discussed that the subsurface layers of Ap/Bp type stars need to lack flows faster than $\sim 10^{-3}$ cm s$^{-1}$. Considering that a model rotating with a period of $P_{\rm rot} \sim 1$ d forms meridional flow with $v_{\rm ES} \sim 1$ cm s$^{-1}$, long rotation periods with $P_{\rm rot} \gtrsim 30$ d would be required to achieve such slow flow velocities. With the help of magnetic braking, this condition is again satisfied in our models with  $B_{\rm p, ini} \geq 1$ kG and $P_{\rm rot, ini} \geq 10$ d (Fig.\,\ref{fig:Protevol}). Besides that, turbulence powered by instabilities due to differential rotation, such as dynamical and secular shear instabilities and the Pitts--Tayler instability, may disrupt the chemical inhomogeneity. Our magnetic model is also compatible with that since our stellar models keep nearly rigidly rotating during the whole main-sequence phase (Fig.\,\ref{fig:MS}).

Correlations among the stellar age ($\tau$), mass ($M$), rotation period ($P_{\rm rot}$), and surface magnetic field strength ($B_z$) of Ap/Bp stars have been studied by several authors \citep{Mathys+97, Hubrig+00, Bagnulo+06, Kochukhov&Bagnulo06, Landstreet+07, Landstreet+08, Mathys17, Netopil+17}. This showed that magnetic Ap/Bp stars are, at the same time, slow rotators. The peak rotation velocity of $\sim 40$ km s$^{-1}$ \citep{Netopil+17} is considerably slower than the major peak at $\sim 200$ km s$^{-1}$ of the wide distribution of rotation velocities of normal AB-type stars. Some of the Ap/Bp stars even show super-long rotational periods of $P_{\rm rot} > 1000$ d \citep{Kochukhov&Bagnulo06, Netopil+17, Mathys+19, Mathys+20}. Interestingly, a similar tendency is also observed for pre-MS stars known as Herbig Ae/Be stars: magnetic Herbig Ae/Be stars are concentrated to have slow rotation velocities of $\lesssim 50$ km s$^{-1}$, while normal Herbig Ae/Be stars obey a wide distribution with a typical velocity of $50$--$250$ km s$^{-1}$ \citep{Alecian+13}.

Our results of the evolution of the surface rotation period with magnetic braking and magnetic dissipation (Fig. \ref{fig:Bp-Prot}) is qualitatively consistent with these observations. For strong enough initial surface fields of 100 G, the magnetic braking efficiently reduces the stellar angular momentum to increase the rotation period by a factor of 10 or even more, whereas the rotation period stays nearly constant within a factor of $\sim 2$ difference for the whole main-sequence phase for models with weaker fields. Moreover, models with the stronger initial field of 1000 G can account for the super-slow rotators with $P_{\rm rot} > 1000$ d if they have initial rotation periods of $\gtrsim 10$ d. Furthermore, since magnetic dissipation becomes more efficient with faster rotation, surface fields of models with fast initial rotation velocities of $\sim 100$ km s$^{-1}$ with $P_{\rm rot, ini}$=1 d rapidly decrease below $\sim 100$ G, thus contributing to the lack of fast-rotating magnetic stars.

\begin{figure}[t]
	\begin{center}
	\includegraphics[width=0.5\textwidth]{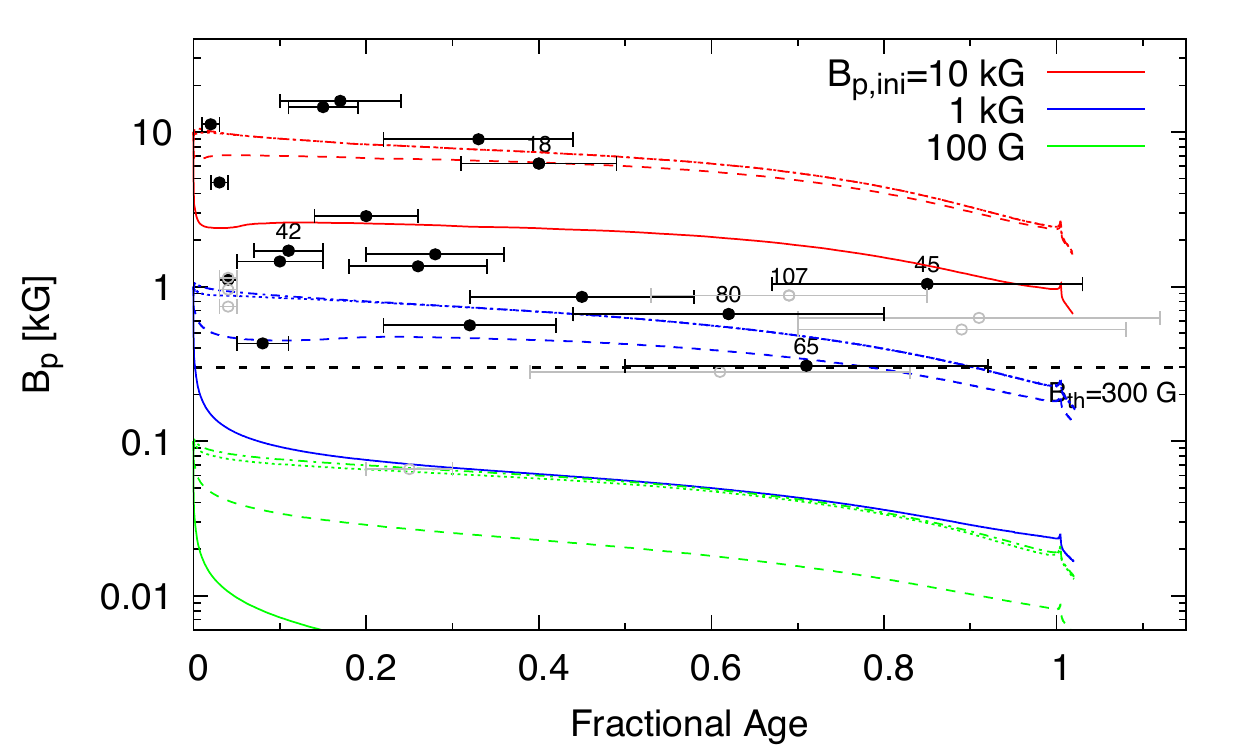}
	\caption{Comparison of the surface magnetic field evolution of our 1.5\,M$_{\odot}$ models and observational results for 2--3 $M_\odot$ Ap stars analysed by \citet{Landstreet+07, Landstreet+08}. Our models with $B_{\rm p, ini} = 10$ kG, 1 kG, and 100 G are respectively shown by red, blue, and green lines (see legend). Similar to Fig.\ref{fig:Bradevol}, solid, dashed, dotted, and dash-dotted lines indicate initial periods of $P_{\rm rot, ini}$ of 1, 10, 100, and 1000 d, respectively. Dots with error bars correspond to observed stars, where black filled symbols show stars for which a field is detected, while gray open symbols are probable magnetic stars. Numbers associated with several stars denote $V \sin i$ in units of km s$^{-1}$. The threshold magnetic field of $B_{\rm th}\sim$300 G, below which essentially no Ap/Bp stars are found \citep{Auriere+07}, is shown by the black dashed line. A numerical factor of 3.3 is multiplied to the field strengths reported in the literature to convert $B_{\rm rms}$ into $B_{\rm p}$.
	}
	\label{fig:magstars}
	\end{center}
\end{figure}

Observational tests of the evolution of the surface magnetic field are more complicated. An absence of young magnetic stars has been reported by \citet{Hubrig+00}. However, later studies have not confirmed this conclusion but identified some young magnetic stars in their sample \citep{Landstreet+07}. The authors have attributed the inconsistency to the different ways of bolometric and effective temperature corrections, which implies a significant uncertainty in the age determination by fitting the position on the HR diagram. Moreover, \citet{Landstreet+07, Landstreet+08} have collected their stellar sample from open star clusters, which allows an age determination by isochrone fitting with considerably better accuracy. \citet{Landstreet+08} suggested that a strongly magnetized star with rms (root-mean-square) fields larger than 1 kG only appears close to the ZAMS in the HR diagram. This is in line with the finding by \citet{Fossati+16} for O-type stars that the fraction of magnetic stars decreases for larger evolutionary age.

Our current simulations do not yet intend to explain any of the observations. Also, our 1.5 $M_\odot$ models are not completely consistent with the mass range of the Ap stars of 2--3 $M_\odot$ observed by \citet{Landstreet+07, Landstreet+08}. Nevertheless, as a demonstration of the capability of our magnetic stellar evolution code, our results are compared with observational results in Fig. \ref{fig:magstars}. Note that here, we multiply a factor of 3.3 to the observed field strength of $B_{\rm rms}$, which is the median rms of the so-called mean longitudinal field, $B_l$ \citep{Landstreet88, Landstreet92}, in order to compare with our simulation results of the polar field strength $B_{\rm p}$. We refer to \citet{Auriere+07} for the factor 3.3\footnote{This simple conversion will still provide a reasonable comparison with the theoretical results, although it is not $B_{\rm rms}$ but $B_{l}^{\rm max}$ that is used in the original estimate, and this conversion will actually yield a lower limit of the polar strength.}.

The majority of Ap stars appear to have a polar field strength of $\sim$1 kG during the whole main-sequence phase. This may be compatible with our models with an initial field strength of $B_{\rm p,ini} \sim$1 kG. On the other hand, although not (yet) statistically significant, the dearth of strongly magnetized ($B_{\rm p} \gtrsim$10 kG) evolved (fractional age $\gtrsim $ 0.5) Ap stars is not reproduced by our models. It implies that the magnetic flux conservation is insufficient to explain the observation. However, since efficient magnetic braking stops rotations for those strongly magnetized models, no efficient rotational turbulence is expected from the current modeling (see models with $\gtrsim 1$ kG in Fig.\ \ref{fig:Bp-Prot}). Furthermore, several Ap stars, including evolved ones (fractional age $\gtrsim $ 0.5), still show certain surface rotation velocities. This is inconsistent with our current models, since our models with $B_{\rm p,ini} \gtrsim$ 1 kG (except for the model with the fastest initial rotation) essentially stop rotation during the early main-sequence phase (see also Fig.\,\ref{fig:Bp-Prot} and Table\,\ref{tab:MSmodels}). This might imply an inaccuracy of our treatment of the wind mass loss and/or the magnetic braking.

Due to the weakness, wind mass-loss rates of AB-type stars are highly uncertain \citep[see discussion in][]{Krticka14}. Using the finite rotational periods of magnetic stars, it may be possible to constrain the unknown mass-loss rates in the less massive stars. For example, by assuming the moment of inertia and the mass loss rate are constant during the main-sequence phase, it is expected that the surface angular velocity of a magnetic star exponentially decreases with time as
\[
	\Omega(t) = \Omega_{\rm init}  e^{-t/\tau_{\rm break}},
\]
where $\Omega_{\rm init}$ is the initial angular velocity and $\tau_{\rm break} = J/\dot{J}$ is the braking timescale. The angular momentum loss rate linearly correlates with the wind mass-loss rate of a nonmagnetic star and the magnetic braking efficiency, which also linearly correlates with the surface field strength in a strong field limit (see Appendix\,\ref{sec:app-windmag}). Under this simplified case, $\sim 97$\% reduction of the currently applied value of $\sim 10^{12}$ M$_\odot$ yr$^{-1}$, which is estimated according to \citet{deJager+88}, will be required to stay within the rotation period of HD 50169 of 10,600 d, the slowest rotating Ap star with accurate magnetic field determinations so far discovered, for a magnetic stellar model with $P_{\rm rot, ini} = 100$ d and the field strength of HD 50169 of $B_{\rm p} \sim 4300$ G \citep{Mathys+19, Mathys+20}. Where a magnetic star of $B_{\rm p} \sim 100$ G is assumed to acquire 40 times slower rotation period at TAMS phase with \citet{deJager+88} mass loss rate (Fig.\,\ref{tab:MSmodels}). We note that such a reduction of the mass-loss rate would not significantly change our present prediction for weakly magnetic models ($B_{\rm p,ini} \lesssim$ 10 G; Fig.\ref{fig:Bp-Prot}).

Another important issue is to understand the threshold of $B_{\rm thr}\sim$300 G below which no Ap/Bp stars with definite field detection have been found. According to \citet{Auriere+07}, the threshold does not result from the observational bias of limiting the sample to Ap/Bp stars that satisfy the condition to stabilize the surface layers. In this case, non-Ap stars with large scale fields below 300 G should exist, which are, however, not observed \citep[e.g.][]{Bagnulo+06}. \citet{Auriere+07} discusses that differential rotation can form in the subsurface region of weakly magnetized stars, and accordingly the field can decay through the Pitts--Tayler instability. However, our model is incompatible with this hypothesis, since a magnetic model develops essentially no differential rotation even for a case with a weak initial field strength of $B_{\rm p, ini} = 10$ G (Fig.\,\ref{fig:MS}). \citet{Jermyn&Cantiello20} investigate the relation between the surface magnetic field and the subsurface convection and propose a possible explanation of a bimodal distribution of the field strength, which would be observed in O-type stars \citep{Grunhut+17}. They consider that strong fields suppress subsurface convection, otherwise, the emergence of the subsurface convection erases the magnetic field. Their critical field strength to suppress the subsurface convection might correspond to the 300 G threshold. As discussed earlier, our simple estimate also shows that magnetic fields with $\gtrsim 300$ G may be sufficiently strong to affect the subsurface convection.

On the other hand, the origin of this threshold could relate to the formation of magnetic stars. For example, magnetic stars might be predominantly formed via stellar mergers \citep{Schneider+19} that yield magnetic fields stronger than this threshold, otherwise, star formation could form only non-magnetic stars. Indeed, for O-type stars, several magnetic stars below the 300 G threshold have been fond \citep{Fossati+15a}, despite the detection limit of $\lesssim 600$ G for this type of stars \citep{Fossati+16}. The threshold may be present and may have lower values for massive stars \citep[see also][]{Jermyn&Cantiello20}. However, it is also possible that a lower threshold is the result of magnetic field decay during stellar evolution, which could have different efficiencies depending on the stellar mass, and does not relate to the stability of the stellar fields \citep{Fossati+16}. If this is the case, it is not inconsistent with our models when suitable birth probability distributions for magnetic field and rotation were adopted.

\subsection{Rotation period changes in Bp stars} \label{sec:obs-rotperiod}

Some magnetic stars show the variability of their rotation frequency on timescales of $\sim10$--100 yr, which is found spectroscopically, photometrically, or through magnetic field measurements \citep[][and references therein]{Shultz+18, Pyper&Adelman20}. The rotation period of the magnetic star can be determined by analyzing the variations \citep[e.g.,][]{Krticka+09, Krticka+15}, and in some cases, even a time evolution of the rotational period can be measured. For instance, comparing the long base-line photometric data, an increasing rotation period, thus a decreasing rotational rate has been detected for a magnetic B type star, $\sigma$ Ori E, which has been explained by the magnetic braking \citep{Townsend+10}.
 
For other B type stars of CU Vir, HD 37776, HD 142990 \citep{Pyper+98, Mikulasek+08, Mikulasek16, Shultz+19}, and more recently 13 And and V913 Sco \citep{Pyper&Adelman20}, even {\it decreasing} rotational periods are found, i.e., these stars accelerate their surface rotation. Obviously, this is inconsistent with the prediction of the magnetic braking theory. Considering the high occurrence rate of spinning-up, possibly the period derivative in $\sigma$ Ori E may also change its sign in the future \citep{Shultz+19}.

Although the mechanism for the observed spin-up has not yet been clarified, an interesting explanation has been proposed by \citet{Krticka+17}. They have considered a star in magneto-hydrostatic equilibrium and derived an incompressible wave equation describing the propagation of a torsional magnetohydrodynamic wave. By assuming axial symmetry and applying a simple poloidal field structure, they found that the periodic cycle of 67.6(5) yr estimated for CU Vir can be reproduced well by the basic resonant frequencies of 51 yr in their model. Hence, they have shown that the torsional surface oscillation, which results from the propagation of the torsional \Alfven wave, can provide a possible explanation for the observed period decrease \citep[see also][]{Stepien98}. For other possible mechanisms, see discussions in \citet{Shultz+19} and references therein.

Assuming similar physics parameters as Krticka et al. leads to quite similar magneto-hydrodynamical waves in our model. As shown in Section \ref{sec:wave2}, our model starts to oscillate torsionally when an initial perturbation is added to the background equilibrium state (Fig.\ref{fig:surface}). In both, Krticka's and our model, the propagation of the torsional \Alfven wave accounts for the oscillation. The magnetic strength distribution assumed in Krticka et al. is simpler than ours: they have considered a field configuration of ($B_R$, $B_\phi$, $B_z$) = (B, 0, -zB/R) with constant $B$ in cylindrical coordinates $R$, $\phi$, and $z$. On the other hand, the poloidal component in our model has a general radial dependency. In addition, we consider the long term evolution of the magnetic field. In fact, our model predicts that the field close to the surface can evolve to have a different field strength than the internal region. Although models presented by \citet{Krticka+17} cannot reproduce the period change of HD 37776, it might be possible to explain the rapid rotational period by considering more realistic radial distributions of the stellar magnetic field. If this hypothesis is correct, then the period of the variation of the rotation rate will roughly correlate with the strength of the surface magnetic field.

This oscillation period, which has a similar timescale to the wave-crossing time of the torsional \Alfven wave, depends on the strength of the internal poloidal field component. Therefore, it will be fundamentally possible to determine the internal field strength distribution by observing the high-order rotational period change at the surface of the star. The idea to determine the internal magnetic field by observing the surface oscillations is reminiscent of asteroseismology. However, instead of gravity or pressure waves, here it is the propagation of \Alfven waves, which produces the observable signal.  

Estimates for the presence of an internal magnetic field are provided indirectly, by considering that a strong enough magnetic field can reduce the amplitude of non-radial dipolar ($l=1$) mode oscillations \citep{Fuller+15, Stello+16, Cantiello+16}. However, so far this method only provides the information of fields close to red-giant cores, which result from long and complicated evolutionary histories. Perhaps, torsional oscillation, if observed in reality, can reveal the field distribution in the radiative envelope of a main-sequence star, which can be directly compared with corresponding stellar evolution models.

Applying asteroseismology to magnetic star still requires further developments of the theory \citep{Keifer+17, Loi&Papaloizou17, Loi&Papaloizou18, Keifer&Roth18, Loi&Papaloizou20}. Also, 2D modeling of the background stellar structure will be important to consider a realistic poloidal field configuration \citep{Rincon&Rieutord03, Reese+04, Prat+19}. Nonetheless, MHD-oscillations may play a crucial role in detecting and analyzing internal stellar magnetic fields in the future.

\subsection{Core-envelope decoupling in red-giant star}

\begin{figure*}[t]
	\begin{center}
	\includegraphics[width=\textwidth]{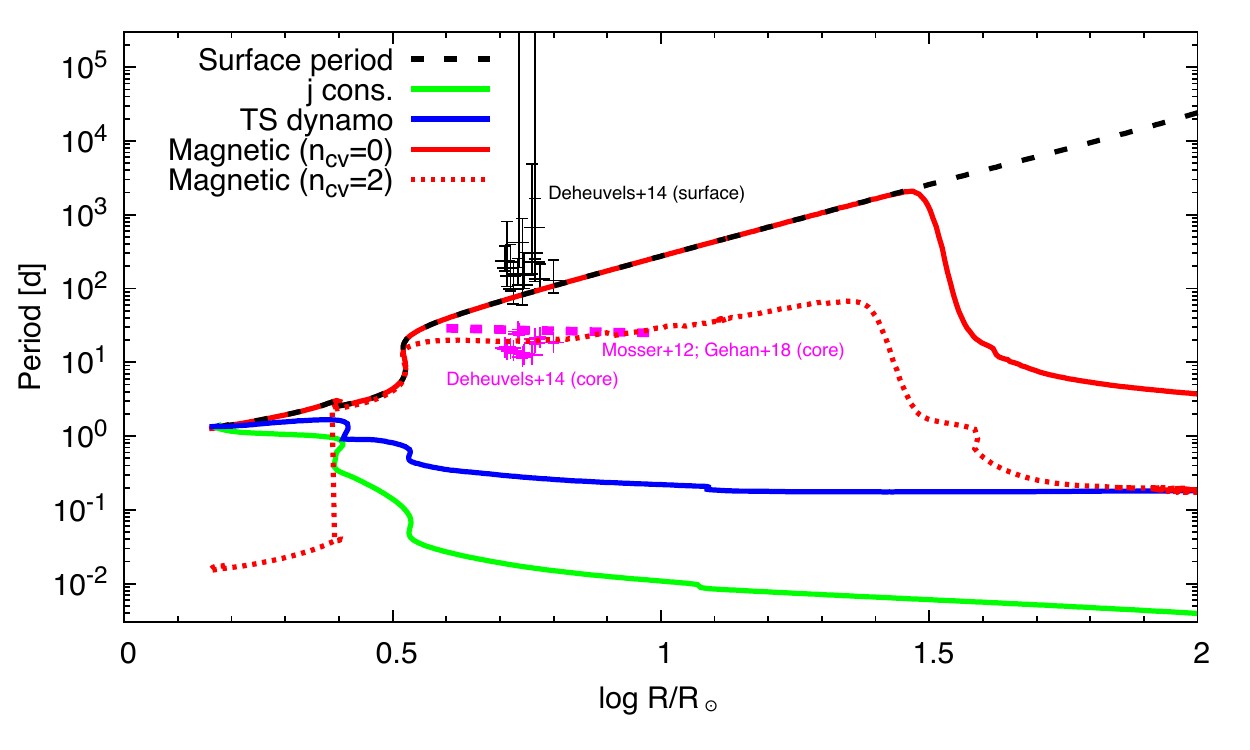}
	\caption{Rotation period at the surface (black dashed line) and near the center (red line) of our fiducial 1.5 M$_\odot$ model ($n_{\rm cv} = 0$) as function of its radius. The central rotation period of a similar model with $n_{\rm cv} = 2$ is shown by the red dotted line. For comparison, the central rotation periods of models with no angular momentum transfer (green line) and with transfer due to the Tayler--Spruit dynamo (blue line) are also shown. Theoretical models are compared with the results of asteroseismic observations. Core rotation periods obtained by asteroseismic observations \citep{Mosser+12, Gehan+18} are shown by the magenta dashed line. Surface and core periods obtained by \citet{Deheuvels+14} are shown by black and magenta pluses.
    }
	\label{fig:omega}
	\end{center}
\end{figure*}

Without angular momentum transport, the spin periods of the cores of red-giant stars would be expected to be very short, $\sim 10^{-2}$ d. Longer periods would result if an effective angular momentum transport is assumed, but the naive expectation was still that red-giant core rotation periods would be much shorter than those of the surfaces. However, during the last decade, asteroseismology has revealed rotation periods of red-giant cores of only $\sim10$ d \citep{Beck+12, Mosser+12, Deheuvels+14}. This implies that is even the most efficient angular momentum transport mechanism proposed at that time (the Tayler--Spruit dynamo) left a discrepancy of predicted and observed core rotation periods of more than one order of magnitude \citep{Cantiello+14, Spada+16, Eggenberger+17}. Recently, \citet{Fuller+19} have revised the diffusion coefficient of the TS dynamo, discussing that the saturation of the dynamo cycle takes place much later than assumed in \citet{Spruit02} such that a stronger toroidal component and magnetic torque are obtained. The red-giant core periods have been well reproduced in their simulation, however, the basic picture of the TS dynamo theory still remains uncertain (Sect.\ \ref{sec:TSdynamo}). Although other mechanisms such as angular momentum transfer by internal gravity waves \citep{Fuller+14, Pincon+17} have been proposed, these are generally not efficient enough to overcome the problem.

The rotation periods at the center and the surface of our magnetic models are compared with observations in Fig. \ref{fig:omega}. Central spin periods of a model with $n_{\rm cv}=0$ is shown by the red solid line, and that with $n_{\rm cv}=2$ is by the red dotted line. The surface period of the $n_{\rm cv}=0$ model is shown by the black dashed line, while that of the $n_{\rm cv}=2$ model is omitted since they are nearly identical. In addition to the magnetic models, central spin periods of a model with no internal angular momentum transfer (labeled as `j cons.', green) and a model with the Tayler--Spruit dynamo (`TS dynamo', blue) are shown as well.

The central period of our fiducial magnetic model with $n_{\rm cv}=0$ coincides with its surface period up to the point of $\log R/{\rm R}_{\odot} = 1.5$. In other words, the model sustains near-rigid rotation even after the formation of the red-giant envelope. While this model is incompatible with the observations, we stress the importance of this result: a stellar evolution calculation that self-consistently accounts for the interaction between differential rotation and magnetic field, obtains the enigmatically large efficiency of the angular momentum transfer needed to understand the red-giant core periods.

\begin{figure}[t]
	\begin{center}
	\includegraphics[width=0.5\textwidth]{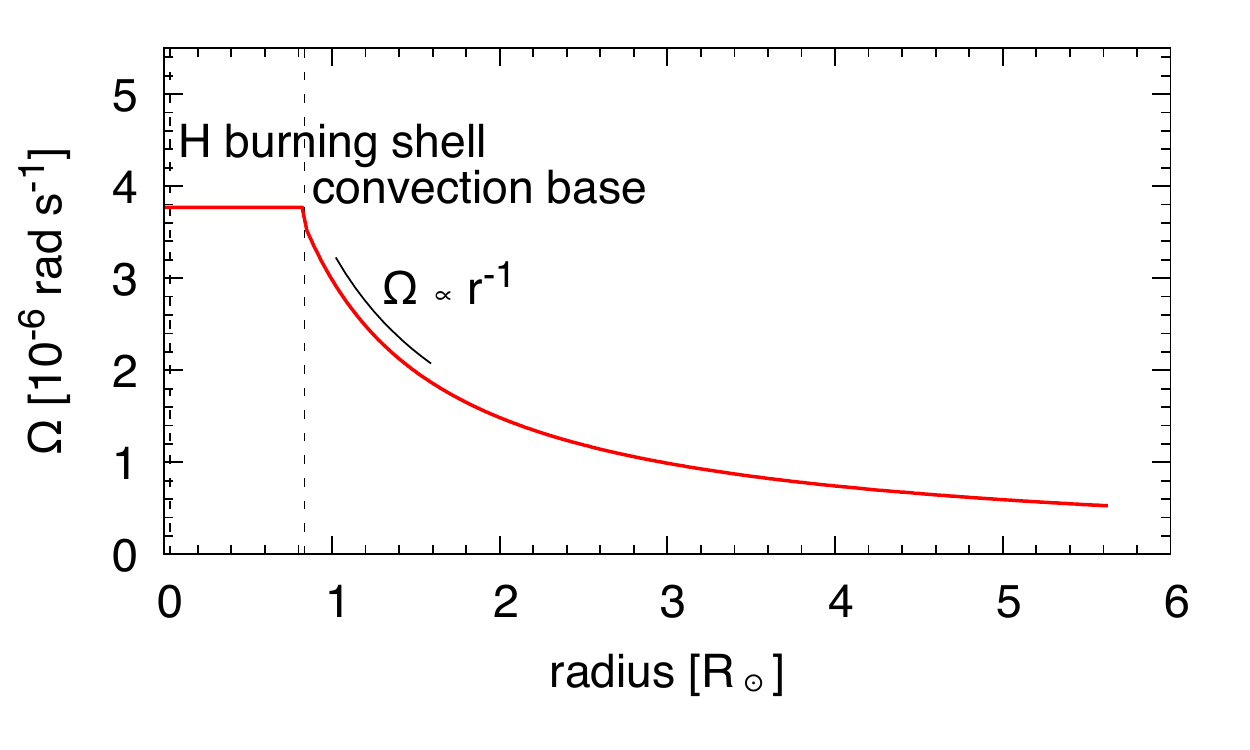}
	\caption{Angular velocity as function of radius for our magnetic red-giant model computed with $n_{\rm cv}=2$, at $\log R/{\rm R}_{\odot} = 0.75$. The left dashed line at a radius of $\sim 10^{-2}$ R$_\odot$ indicates the location of the hydrogen-burning shell, and the	right dashed line at $\sim 1$ R$_\odot$ shows the location of the base of the envelope convection.
	}
	\label{fig:Odist}
	\end{center}
\end{figure}
Interestingly, in the magnetic model with $n_{\rm cv}=2$, the core develops a shorter rotation period than the surface after the surface rotation period increases to $\sim20$ d and keeps its period roughly constant up to the point when $\log R/{\rm R}_{\odot} = 1.4$. Therefore, not only the core rotation period but also the surface period agrees with the observations. This is because this model develops an angular momentum distribution as $\Omega \propto r^{-1}$  in the convective envelope, while $\Omega$ is constant in the inner radiative region (Fig.\,\ref{fig:Odist}). The distribution $\Omega \propto r^{-1}$ results from the approximation $\nu_{\rm DS} = \nu_{\rm MLT}$ (Sect.\,\ref{sec:intro}). On the other hand, the rigid rotation in the inner radiative region is solely due to the magnetic effect.

Our $n_{\rm cv}=2$ result predicts that strong shear exists at the base of the convective layer, at $\sim$ 1 $R_\odot$, in a red-giant star. This is in contrast to the result of \citet{Fuller+19} (Fig. 5), in which the shear rotation mainly forms at the hydrogen-burning shell, at $\sim 3 \times 10^{-2}$ $R_\odot$, much deeper inside than ours. To reveal the differential rotation inside the convective envelope in a red-giant star by asteroseismology, the detection of quadrupole and octopole pulsation modes may be required \citep{Ahlborn+20}. We expect that future asteroseismic observations will discriminate these two cases.

We have also noticed an interesting correspondence between the very strong toroidal field of $\sim 10^6$ G obtained in our simulation at the core-envelope boundary ($B_\phi$ at $M_r \sim 0.4$ M$_\odot$, Fig.\ref{fig:RGevol}) and possible observational constraint for the internal magnetic field for red-giant stars of $\lesssim 10^{6.5}$ G \citep{Fuller+15, Stello+16, Cantiello+16}. More stringent simulation and discussion for the internal magnetic field of red-giant stars will be done in the future.

We note that more than two orders of magnitude acceleration happen in the cores of magnetic models at $\log R/{\rm R}_{\odot} = 1.4$, the mechanism of which has been discussed in section\ \ref{sec:result-RG}, whereas it does not take place in the model with the Tayler--Spruit dynamo. This is because the core has already a larger specific angular momentum than the accreted material, and besides, the timescale of the angular momentum transfer inside the core at this point is already too long to affect the rotation period at the center of the core.

\subsection{Rotation periods of white dwarfs} \label{sec:dis-WDrotation}

Finally, we discuss the rotation velocities of white dwarfs that are expected from our simulations. The $\sim2$ orders of magnitude acceleration of the core shown in Fig. \ref{fig:omega} is due to the matter accretion of the helium core, as explained in Section \ref{sec:result-RG}. Assuming that the angular momentum of a white dwarf chiefly originates from the accreting matter, the angular velocity of a white dwarf will be estimated as
\[
	\Omega_{\rm WD} = \frac{J_{\rm acc}}{I_{\rm WD}},
\]
where $J_{\rm acc}$ and $I_{\rm WD}$ are the angular momentum accreted onto the central core after rotational decoupling and the moment of inertia of the white dwarf, respectively. $I_{\rm WD} = 0.205 \times R_{\rm WD}^2 M_{\rm WD}$ and the radius of $R_{\rm WD} = 1.20 \times 10^{-2}$ R$_\odot$ and the mass of $M_{\rm WD} = 0.575$ M$_\odot$ are taken from a 1.5 M$_\odot$ model of \citet{Suijs+08}. Furthermore, we have $J_{\rm acc} \sim (2 R_{\rm base}^2 \Omega_{\rm base}/3) \Delta M_{\rm acc}$, where $R_{\rm base} \sim 1$ R$_\odot$ and $\Omega_{\rm base}$ are the radius and the rotation rate at the base of the convective envelope, and $\Delta M_{\rm acc}$ is the mass that is accreted onto the core after the decoupling. Therefore,
\[
	\Omega_{\rm WD} = 3.25 \times 10^{-4} \
		\left( \frac{\Omega_{\rm base}}{10^{-8} \ \mathrm{rad} \ \mathrm{s}^{-1}} \right)
		\left( \frac{0.01 \ R_{\rm base}}{R_{\rm WD}} \right)^2
		\left( \frac{\Delta M_{\rm acc}}{M_{\rm WD}} \right)
		\ \mathrm{rad} \ \mathrm{s}^{-1}
\]
is obtained. Our fiducial model has $\Omega_{\rm base} \sim 10^{-8}$ rad s$^{-1}$ and $\Delta M_{\rm acc} =$ 0.28 M$_\odot$, hence a white dwarf angular velocity of $\Omega_{\rm WD} \sim 1.3 \times 10^{-4}$ rad s$^{-1}$, and correspondingly a rotation velocity of the white dwarf of $v_{\rm WD} \sim 1.1$ km s$^{-1}$ is expected. The $n_{\rm cv}=2$ model has a rotation rate at the base of the envelope which is one order of magnitude faster, $\Omega_{\rm base} \sim 10^{-7}$ rad s$^{-1}$, and similar $\Delta M_{\rm acc} =$ 0.25 M$_\odot$, and thus $\Omega_{\rm WD} \sim 1.2 \times 10^{-3}$ rad s$^{-1}$ and $v_{\rm WD} \sim 9.9$ km s$^{-1}$ are expected. Both estimates are compatible with the spectroscopic upper limit of $\sim 10$ km s$^{-1}$ \citep{Berger+05}.

For the present models, the $\alpha$ effect has not yet been taken into account. A red-giant model with the $\alpha$ effect will show stronger magnetic fields than the present models, and the magnetic braking will reduce the angular momentum of accreted material and thus reduce the rotation rate of the white dwarf. This reduction will improve the comparison to observations, since rotation periods of both groups, non-magnetic white dwarfs with $P \sim 1$--$169$ h corresponding to $\Omega \sim 1.7 \times 10^{-3}$--$1.03 \times 10^{-5}$ rad s$^{-1}$, which are measured by asteroseismic observations, and magnetic white dwarfs of $P \sim 0.2$--$429$ h corresponding to $\Omega \sim 8.7 \times 10^{-3}$--$4.1 \times 10^{-6}$ rad s$^{-1}$, which are measured by spectroscopic modulation, include slow rotators \citep[][and references therein]{Kawaler15, Corsico+19}.

Another relevant question here is whether a large-scale poloidal field that links the whole radiative mantle region, which covers the central core and is covered by the convective envelope, can form in a red-giant star. Because the magnetic fields induced by the $\alpha$ effect will show a time variability and/or probably be dominated by small scale fields, the magnetic fields may soon cancel out each other after the material migrates from the convective region into the radiative region. If the magnetic link is lost, the rotational decoupling happens no matter how strong the field is formed in the convective envelope. The rotation velocity of the white dwarf depends on when the rotational decoupling takes place because the accreted mass on the helium core after the decoupling determines the angular momentum of the white dwarf.

\section{Conclusion} \label{sec:conclusion}

We have developed a new formalism for stellar evolution calculations, which includes the interaction of stellar rotation and stellar magnetic field as self-consistently as possible in 1D model. With this method, we have computed evolutionary models of 1.5 M$_\odot$ stars, adopting various initial magnetic field strengths and rotation rates. The theoretical models have been compared with relevant observations of 
1) ages, rotation rates, and magnetic field strengths of Ap stars,
2) surface rotation variations observed in Bp stars,
3) core and surface rotation periods of red-giant stars,
and 4) rotation periods of white dwarfs.
Even though we have not manipulated the model to explain any of these observations, there are in general good agreements between our modeling and observations.

This work demonstrates the first results in a series in which we intend to develop and apply a new scheme to simulate the evolution of magneto-rotating stars. In forthcoming papers, additional magnetic effects on the stellar evolution, such as the $\alpha$ effect, magnetic pressure, and magnetic modification of convective criterion will be included and analyzed. There are plenty of possible applications of the new scheme: evolution of solar-type stars, for which abundant works have been done; the evolution of massive stars is an interesting target as well because strong surface fields will significantly affect the evolution as they have a relatively fast rotation and strong wind mass loss and, of course, because they are the progenitors of (non-)magnetized compact remnants; the evolution of binary systems, such as a merger remnants \citep{Beloborodov14, Schneider+19}, and tidally interacting binaries \citep{Vidal+18, Vidal+19}, will also be an interesting target.

\begin{acknowledgements}
K.T. appreciate invaluable discussions with
John D. Landstreet,
Matthew E. Shultz,
Alexandre David-Uraz,
Jiri Krti\v{c}ka,
Jim Fuller,
G\"otz Gr\"afener,
Youhei Masada,
Munehito Shoda,
and Tomoya Takiwaki.
K.T. is grateful to
Zahra Mirzaiyan 
and Antonio Ferriz-Mas
for discussions of mathematical expression.

\end{acknowledgements}

\bibliographystyle{aa} 
\bibliography{biblio_200817}

\begin{appendix}

\section{Alfv\'{e}n's theorem} \label{sec:app-theorem}

{\bf Theorem:}
Let $\mathcal{M}$ be a 2d compact manifold with a boundary $\mathcal{C}$.
$\mathcal{M}$ is embedded in a 3d manifold and moves with time with a velocity field $\vec{v}(\vec{x}, t)$.
Consider a flux $\Phi$ of a time-dependent vector field $\vec{P}(\vec{x}, t)$ on $\mathcal{M}$.
If and only if $\vec{P}$ is divergence-free then
\begin{equation}
	\frac{ d \Phi }{d t} = \int_\mathcal{M} \left(
		\frac{\partial \vec{P}}{\partial t}
		- \nabla \times (\vec{v} \times \vec{P})
	\right) \cdot d \vec{S}.
\end{equation}

\begin{figure}[t]
	\begin{center}
	\includegraphics[width=0.5\textwidth]{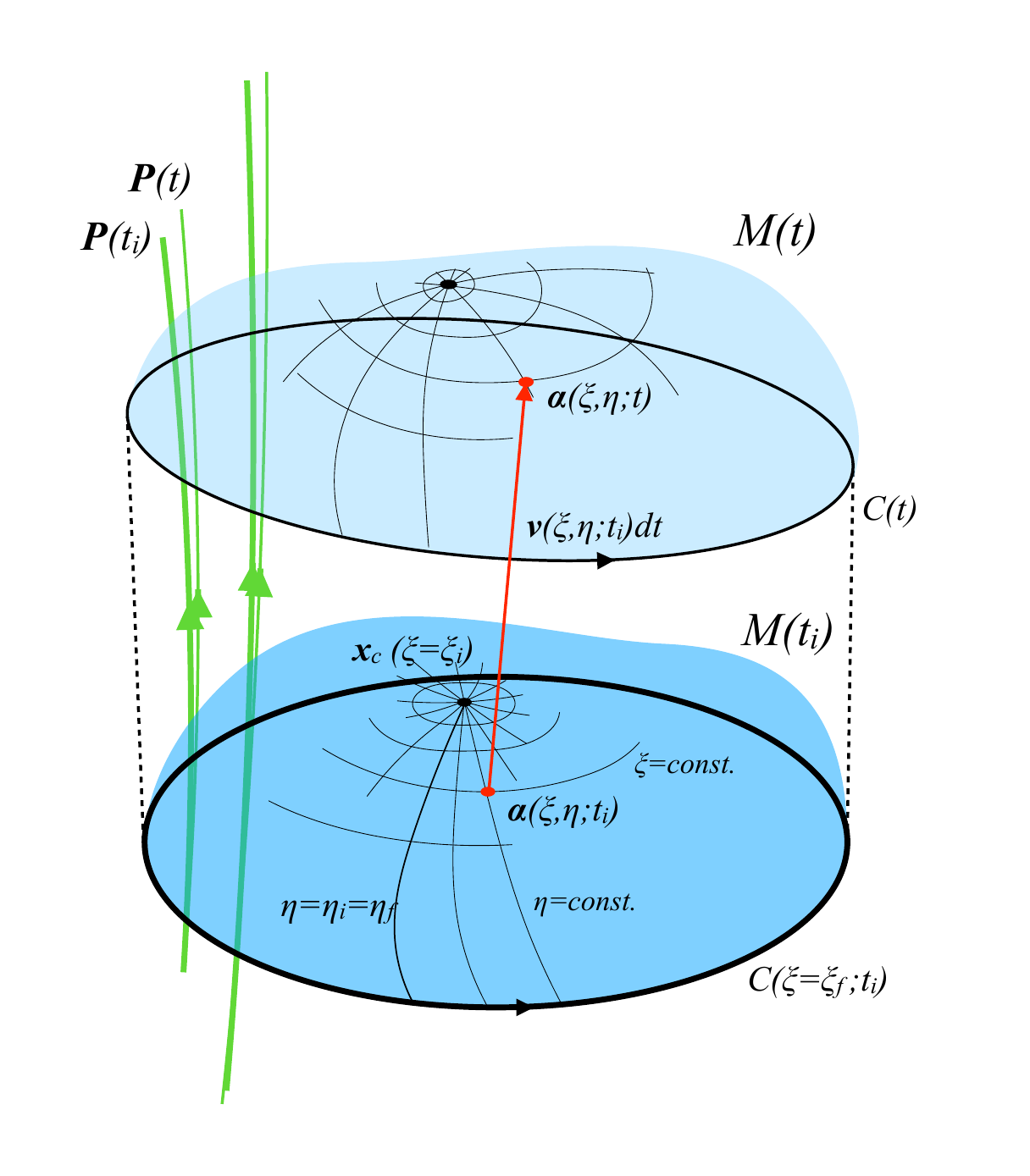}
	\caption{An illustration of the geometry.
	The 2d manifold $\mathcal{M}$ is shown by blue shaded areas. The below shows $\mathcal{M}$ at $t_i$,	and the above is $\mathcal{M}$ at $t$. An element of $\mathcal{M}$, which is parametrically specified as $\vec{\alpha}(\xi, \eta; t)$, is shown by a red point. It moves with velocity $\vec{v}(\xi, \eta; t)$. The divergence-free vector field $\vec{P}$ is shown as green lines. The thicker ones are $\vec{P}$ at $t_i$ and the thinner ones are $\vec{P}$ at $t$.
	}
	\label{fig:geometry2}
	\end{center}
\end{figure}

{\bf Proof:}
We express the position of the element of $\mathcal{M}$ at a given time $t$ in a parametric form by $\vec{x} = \vec{\alpha}(\xi, \eta; t)$, where $\xi \in [\xi_i, \xi_f]$ and $\eta \in [\eta_i, \eta_f]$. Then the velocity field is equated as $\vec{v} = \partial \vec{\alpha}/\partial t$. The parameters ($\xi$, $\eta$) are concentrically defined: the center of $\mathcal{M}$ at time $t$ is specified as $\vec{x} = \vec{\alpha}(\xi_i, \eta; t) = \vec{x}_c(t) $, likewise the boundary $\mathcal{C}$ as $\vec{x} = \vec{\alpha}(\xi_f, \eta; t)$. The geometry defined here is illustrated in Fig. \ref{fig:geometry2}.

In the parametric form, a flux $\Phi$ of $\vec{P}$ on $\mathcal{M}$ is defined as a function of time as
\begin{eqnarray}
	\Phi(t)	&\equiv&	\int_\mathcal{M} \vec{P} \cdot d \vec{S}\\
		&=& 
			\int_\mathcal{M} \vec{P}(\vec{x}=\vec{\alpha}(\xi, \eta; t), t) \cdot
				\left(
						\frac{\partial \vec{\alpha}}{\partial \xi}
				\times	\frac{\partial \vec{\alpha}}{\partial \eta}
				\right)
			d\xi d\eta.
\end{eqnarray}
So now we equate the total time derivative of $\Phi$,
\begin{equation}
	\frac{d \Phi}{d t}
		=	\frac{d}{d t} \left[
			\int_\mathcal{M} \vec{P} \cdot
				\left(
						\frac{\partial \vec{\alpha}}{\partial \xi}
				\times	\frac{\partial \vec{\alpha}}{\partial \eta}
				\right)
			d\xi d\eta
			\right].
\end{equation}

Let $V$ be the volume swept by $\mathcal{M}$ for $t \in [t_i, t]$. The surface of $V$ consists of three manifolds;
\begin{itemize}
	\item $S_1=\{ \vec{x} \in \vec{\alpha}(\xi,\eta,t) | \xi \in [\xi_i, \xi_f], \eta \in [\eta_i, \eta_f], t = t_i \}$
	\item $S_2=\{ \vec{x} \in \vec{\alpha}(\xi,\eta,t) | \xi \in [\xi_i, \xi_f], \eta \in [\eta_i, \eta_f], t = t \}$
	\item $S_3=\{ \vec{x} \in \vec{\alpha}(\xi,\eta,t) | \xi = \xi_f, \eta \in [\eta_i, \eta_f], t \in [t_i, t]  \}$.
\end{itemize}
$S_1$ and $S_2$ are $\mathcal{M}$ at $t_i$ and $t$, and $S_3$ composes the side of $V$.

In the parametric form, volume integral of div $\vec{P}$ on $V$ is done as
\begin{eqnarray}
	\int_V (\nabla \cdot \vec{P}) d V 
	&=&	\int_V (\nabla \cdot \vec{P}) \left[
				\left(
						\frac{\partial \vec{\alpha}}{\partial \xi}
				\times	\frac{\partial \vec{\alpha}}{\partial \eta}
				\right)
			\cdot
				\frac{\partial \vec{\alpha}}{\partial t}
		\right] d\xi d\eta dt \nonumber \\ 
	&=&		\int_{S_2} \vec{P} \cdot
				\left(
						\frac{\partial \vec{\alpha}}{\partial \xi}
				\times	\frac{\partial \vec{\alpha}}{\partial \eta}
				\right)
			d\xi d\eta
		-	\int_{S_1} \vec{P} \cdot
				\left(
						\frac{\partial \vec{\alpha}}{\partial \xi}
				\times	\frac{\partial \vec{\alpha}}{\partial \eta}
				\right)
			d\xi d\eta  \nonumber \\ 
	&& +		\int_{S_3} \vec{P} \cdot
				\left(
						\frac{\partial \vec{\alpha}}{\partial \eta}
				\times	\frac{\partial \vec{\alpha}}{\partial t}
				\right)
			d\eta dt. \label{eq:gauss}
\end{eqnarray}
Here Gauss's theorem is used. By taking the total time derivative,
\begin{eqnarray}
	\frac{d}{d t} \left( \int_V (\nabla \cdot \vec{P}) d V \right)
	&=&		\frac{d}{d t} \left[
			\int_{S_2} \vec{P} \cdot
				\left(
						\frac{\partial \vec{\alpha}}{\partial \xi}
				\times	\frac{\partial \vec{\alpha}}{\partial \eta}
				\right)
			d\xi d\eta
			\right] \nonumber \\
	&&	-	\frac{d}{d t} \left[
			\int_{S_1} \vec{P} \cdot
				\left(
						\frac{\partial \vec{\alpha}}{\partial \xi}
				\times	\frac{\partial \vec{\alpha}}{\partial \eta}
				\right)
			d\xi d\eta
			\right] \nonumber \\
	&&	+	\frac{d}{d t} \left[
			\int_{S_3} \vec{P} \cdot
				\left(
						\frac{\partial \vec{\alpha}}{\partial \eta}
				\times	\frac{\partial \vec{\alpha}}{\partial t}
				\right)
			d\eta dt
			\right] \label{eq:tderiv}
\end{eqnarray}
is obtained. The left hand side becomes zero because $\vec{P}$ is divergence-free. The first term of the right hand side is $d\Phi/dt$. The second and third terms are equated as
\begin{eqnarray}
		-	\frac{d}{d t} \Biggl[
			\int_{S_1} \vec{P} &\cdot&
				\left(
						\frac{\partial \vec{\alpha}}{\partial \xi}
				\times	\frac{\partial \vec{\alpha}}{\partial \eta}
				\right)
			d\xi d\eta
			\Biggr]  \nonumber \\
	&=&	- \int_{S_1} 
				\frac{\partial \vec{P}}{\partial t} \cdot
				\left(
						\frac{\partial \vec{\alpha}}{\partial \xi}
				\times	\frac{\partial \vec{\alpha}}{\partial \eta}
				\right)
			d\xi d\eta 
\end{eqnarray}
and
\begin{eqnarray}
			\frac{d}{d t} \Biggl[
			\int_{S_3} \vec{P} &\cdot&
				\left(
						\frac{\partial \vec{\alpha}}{\partial \eta}
				\times	\frac{\partial \vec{\alpha}}{\partial t}
				\right)
			d\eta dt
			\Biggr] \nonumber \\
	&=&		\int_{S_3} 
				\frac{\partial \vec{P}}{\partial t} \cdot
				\left(
						\frac{\partial \vec{\alpha}}{\partial \eta}
				\times	\frac{\partial \vec{\alpha}}{\partial t}
				\right)
			d\eta dt 
		+	\oint_{\xi=\xi_f, t=t} \vec{P} \cdot
				\left(
						\frac{\partial \vec{\alpha}}{\partial \eta}
				\times	\frac{\partial \vec{\alpha}}{\partial t}
				\right)
			d\eta \nonumber \\
	&=&		\int_{S_3} 
				\frac{\partial \vec{P}}{\partial t} \cdot
				\left(
						\frac{\partial \vec{\alpha}}{\partial \eta}
				\times	\frac{\partial \vec{\alpha}}{\partial t}
				\right)
			d\eta dt \nonumber \\
	&& +	\int_{S_2} 
			\left[
				\nabla \times 
				\left(	\frac{\partial \vec{\alpha}}{\partial t}
						\times	\vec{P}
				\right)
			\right] \cdot
				\left(
						\frac{\partial \vec{\alpha}}{\partial \xi}
				\times	\frac{\partial \vec{\alpha}}{\partial \eta}
				\right)
			d \xi d \eta,
\end{eqnarray}
where, Stokes's theorem is used. Notice that
$
		\vec{P} \cdot
				\left(
						\frac{\partial \vec{\alpha}}{\partial \eta}
				\times	\frac{\partial \vec{\alpha}}{\partial t}
				\right)
	=			\left(
						\frac{\partial \vec{\alpha}}{\partial t}
				\times	\vec{P}
				\right) \cdot
				\frac{\partial \vec{\alpha}}{\partial \eta}.
$
Therefore, eq. (\ref{eq:tderiv}) equates with
\begin{eqnarray}
	0 &=& \frac{d \Phi}{d t}
			- \int_{S_1} 
				\frac{\partial \vec{P}}{\partial t} \cdot
				\left(
						\frac{\partial \vec{\alpha}}{\partial \xi}
				\times	\frac{\partial \vec{\alpha}}{\partial \eta}
				\right)
			d\xi d\eta 
	 +		\int_{S_3} 
				\frac{\partial \vec{P}}{\partial t} \cdot
				\left(
						\frac{\partial \vec{\alpha}}{\partial \eta}
				\times	\frac{\partial \vec{\alpha}}{\partial t}
				\right)
			d\eta dt \nonumber \\
	&& +		\int_{S_2} 
			\left[
				\nabla \times 
				\left(	\frac{\partial \vec{\alpha}}{\partial t}
						\times	\vec{P}
				\right)
			\right] \cdot
				\left(
						\frac{\partial \vec{\alpha}}{\partial \xi}
				\times	\frac{\partial \vec{\alpha}}{\partial \eta}
				\right)
			d \xi d \eta
			 \nonumber \\
	&=&		\frac{d \Phi}{d t}
	-		\int_{S_2} 
				\left[
					\frac{\partial \vec{P}}{\partial t}
				-	\nabla \times 
					\left(	\frac{\partial \vec{\alpha}}{\partial t}
							\times	\vec{P}
					\right)
				\right] \cdot
				\left(
						\frac{\partial \vec{\alpha}}{\partial \xi}
				\times	\frac{\partial \vec{\alpha}}{\partial \eta}
				\right)
			d \xi d \eta.
\end{eqnarray}
Here Gauss's theorem is again applied to the divergence-free vector field $\partial {\vec{P}}/\partial t$.

So we obtain
\begin{equation}
	\frac{d \Phi}{d t}
	= \int_{\mathcal{M}} 
			\left(
				\frac{\partial \vec{P}}{\partial t}
			-	\nabla \times 
				\left(	\vec{v} \times \vec{P}
				\right)
			\right) \cdot d\vec{S}.
\end{equation}

\section{Viscosity of the Pitts--Tayler instability} \label{sec:app-PTvisc}

The vertical and horizontal length scales of the fluctuation driven by the Pitts--Tayler instability are respectively defined as $l_v$ and $l_h$. The horizontal scale will be replaced with $r$. The vertical sale is limited to be
\begin{equation}
	\frac{l_v}{l_h} = \frac{l_v}{r} < \frac{\omega_A}{N},
\end{equation}
where $\omega_A \equiv B_{\phi}/\sqrt{ 4 \pi \rho } r$ is the \Alfven frequency of the toroidal magnetic field and $N$ is the Brunt-V\"{a}is\"{a}l\"{a} frequency, since the vertical replacement should work against the restoring force of buoyancy. This gives the maximum scale for the vertical fluctuation. On the other hand, the fluctuation will be damped by the dissipation, if the vertical fluctuation is too small. Therefore, the dissipation time should be longer than the growth time of the Pitts--Tayler instability,  $\tau_{\rm PT}$. This gives
\begin{equation}
	\frac{l_v^2}{\nu} > \tau_{\rm PT} = \frac{\tilde{\Omega}}{\omega_A^2}.
\end{equation}
The two inequalities yield a relation
\begin{equation}
	\left( \frac{\omega_A}{N} \right)^2 > \left( \frac{\tilde{\Omega}}{\omega_A} \right)^2 \left( \frac{\nu}{\tilde{\Omega} r^2} \right).\label{eq-PT1}
\end{equation}
$\tilde{\Omega} = \Omega$ has been used in previous works, but in such a case $\tau_{\rm PT}$ becomes zero when $\Omega$ is zero. In order to avoid too rapid growth, we apply $\tilde{\Omega} = \Omega (1+\omega_A/\Omega)$ instead so that the growth time approaches $1/\omega_A$ in the limit of $\Omega \rightarrow 0$.

Following the original discussion by \citet{Spruit02}, we assume that the inequality (\ref{eq-PT1}) reaches the equipoise situation when the instability saturates. This is because a turbulent viscosity induced by the Pitts--Tayler instability starts to act as the effective viscosity in the right-hand side of the inequality. This results in
\begin{equation}
	\left( \frac{\omega_A}{N} \right)^2 = \left( \frac{\tilde{\Omega}}{\omega_A} \right)^2 \left( \frac{\nu_{PT}}{\tilde{\Omega} r^2} \right).
\end{equation}

Besides, according to \citet{Maeder&Meynet04}, we assume the relation between the Brunt-V\"{a}is\"{a}l\"{a} frequency and the effective viscosity as
\begin{equation}
	N^2 = \frac{\nu_{\rm PT}/K}{\nu_{\rm PT}/K + 2} N_T^2 + N_\mu^2,
\end{equation}
where $K$ is the thermal diffusivity and $N_T$ and $N_\mu$ are oscillation frequencies associated with thermal and chemical gradients, respectively.

Finally, the two equations yield a quadratic equation for $\nu_{\rm PT}/K$,
\begin{equation}
\begin{split}
	0 &=	\left( \frac{\nu_{PT}}{K} \right)^2
		\left\{ 	\left( \frac{N_T}{K/r^2} \right)^2
			+	\left( \frac{N_\mu}{K/r^2} \right)^2 \right\}
		\left( \frac{\tilde{\Omega}}{K/r^2} \right) \nonumber \\
	&\quad +	\left( \frac{\nu_{PT}}{K} \right)
		\left\{ 	2 \left( \frac{N_\mu}{K/r^2} \right)^2 \left( \frac{\tilde{\Omega}}{K/r^2} \right) 
			-	\left( \frac{\omega_A}{K/r^2} \right)^4
		\right\}
	-	2 \left( \frac{\omega_A}{K/r^2} \right)^4.
\end{split}
\end{equation}
Since it has a negative zeroth-degree coefficient, it has one positive solution when $N_T^2 + N_\mu^2$ is positive. We solve this equation for the estimate of the turbulent viscosity of the Pitts--Tayler instability.

\section{Wind-magnetic field interaction} \label{sec:app-windmag}

According to 2D axisymmetric MHD simulations with an aligned dipole magnetic field by \citet{udDoula&Owocki02, udDoula+08, udDoula+09}, we estimate the effects of the magnetic wind confinement and the magnetic braking as follows.

First, the magnetic confinement parameter $\eta_*$ is calculated as
\begin{equation}
	\eta_* \equiv \frac{B_{\rm eq}^2 R^2}{\dot{M}_{B=0} v_{\infty, B=0}},
\end{equation}
where $B_{\rm eq} = B_{\theta}(r=R, \theta = \pi/2)$ is the surface magnetic field strength at the equator and $\dot{M}_{B=0}$ and $v_{\infty, B=0}$ are the wind mass-loss rate and the terminal wind velocity for a nonmagnetic model. In the present work we approximately estimate the terminal wind velocity as $v_{\infty, B=0} = 2 \sqrt{GM/R}$. Next, the \Alfven radius, $R_A$, where the radial components of the field and the matter flow have an equal energy density, is estimated as
\begin{equation}
	\frac{R_A}{R} = 1 + ( \eta_* + 1/4 )^{1/4} - (1/4)^{1/4}.
\end{equation}

Efficiencies of the magnetic confinement and the magnetic braking are estimated as 
\begin{equation}
	f_{\rm conf} 
	= \left( 1 - \sqrt{ 1 - \frac{R}{R_c} } \right)
\end{equation}
and
\begin{equation}
	f_{\rm break} 
	= \left( \frac{R_A}{R} \right)^2,
\end{equation}
where $R_c = R + 0.7 ( R_A - R )$ is a maximum closure radius of magnetic loops. We note that an additional term of $\left( 1 - \sqrt{ 1 - 0.5 R/R_K } \right)$, in which $R_K = R (v_{\rm K}/v_{\rm rot})^{2/3}$ is the Kepler corotation radius, is included in the original formula of the magnetic confinement in \citet{udDoula+08}. It is discussed that this term accounts for the breakout of the gas from the closed loops due to the fast rotation of the star. However, since the enhancement happens even for a non-magnetic ($\eta_*=0$) model, this additional term likely partly accounts for the $\Omega$ effect, which is already taken into account in our simulation by eq.(\ref{eq:omegagamma}). To avoid double-counting of the $\Omega$ effect, this term is omitted from our simulation.

\section{Code test}  \label{sec:app-codetest}

\subsection{Magnetic flux conservation}
\begin{figure}[t]
	\begin{center}
	\includegraphics[width=0.5\textwidth]{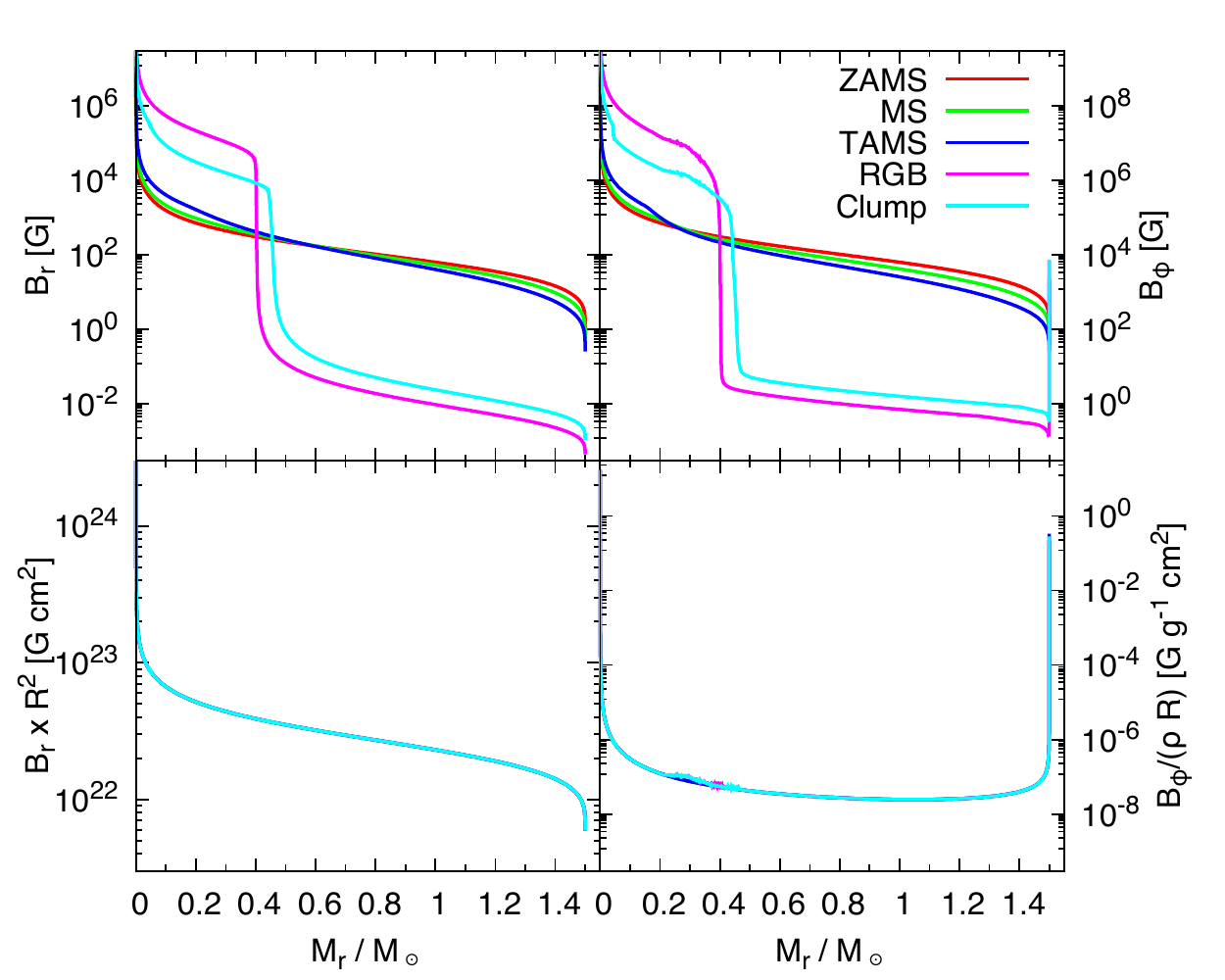}
	\caption{Internal magnetic field distributions of 1.5 $M_\odot$ magneto-rotational star with 5 evolutionary stages, at ZAMS (red), at the middle of the main-sequence phase (MS; green), at TAMS (blue), at the middle of the red-giant phase (RG; magenta), and at the red clump phase (Clump; cyan) are shown. In this test case, $\eta$- and $\Omega$-effects are switched off to confirm magnetic flux conservation, and the Maxwell stress is also neglected. The top two panels show the evolution of the radial component ($B_r$; left) and the toroidal component ($B_\phi$; right). The bottom two panels show the evolution of corresponding conserved quantities, $B_r r^2$ by the left, and $B_\phi / \rho r$ by the right.
	}
	\label{fig:Bcons}
	\end{center}
\end{figure}

As the simplest test case, we have calculated a 1.5 M$_\odot$ stellar evolution with the magnetic field but switching off the magnetic dissipation and the $\Omega$ effect to test whether the magnetic field satisfies the flux conservation. The initial magnetic field is arbitrarily set to have a $r^{-3}$ radial dependence for both the poloidal and toroidal components. The evolution is followed from the ZAMS phase through the TAMS and the red-giant phase until the star experiences helium flash and starts core helium burning, entering into the red clump in the Hertzsprung-Russell Diagram. During the evolution, the star experiences significant contraction in the central core and expansion in the outer envelope.

Figure \ref{fig:Bcons} shows the resulting evolution of the internal magnetic field. Note that $B_r$ in the figure shows the polar value of the radial magnetic component, thus $B_r = 2A(r)/r$, and, $B_\phi$ in the figure is the toroidal component at $\theta=\pi/4$, thus $B_\phi = B(r)$. The top two panels showing radial and toroidal magnetic field components exhibit the effect of core contraction and envelope expansion: the magnetic field in the central core of $\la 0.4$ M$_\odot$ is amplified by about two orders of magnitude, while that in the outer envelope is reduced by about four orders of magnitude. Nevertheless, the two conserved quantities, which are shown in the bottom panels, are entirely conserved during the whole evolutionary phases. Only a small fluctuation is seen for $B_\phi / \rho r$ at $\sim$0.3--0.4 M$_\odot$. This results from automated mesh refinement, which is done to capture a thin structure of the hydrogen-burning shell that surrounds the helium core.

\subsection{Magnetic dissipation}
\begin{figure}[t]
	\begin{center}
	\includegraphics[width=0.5\textwidth]{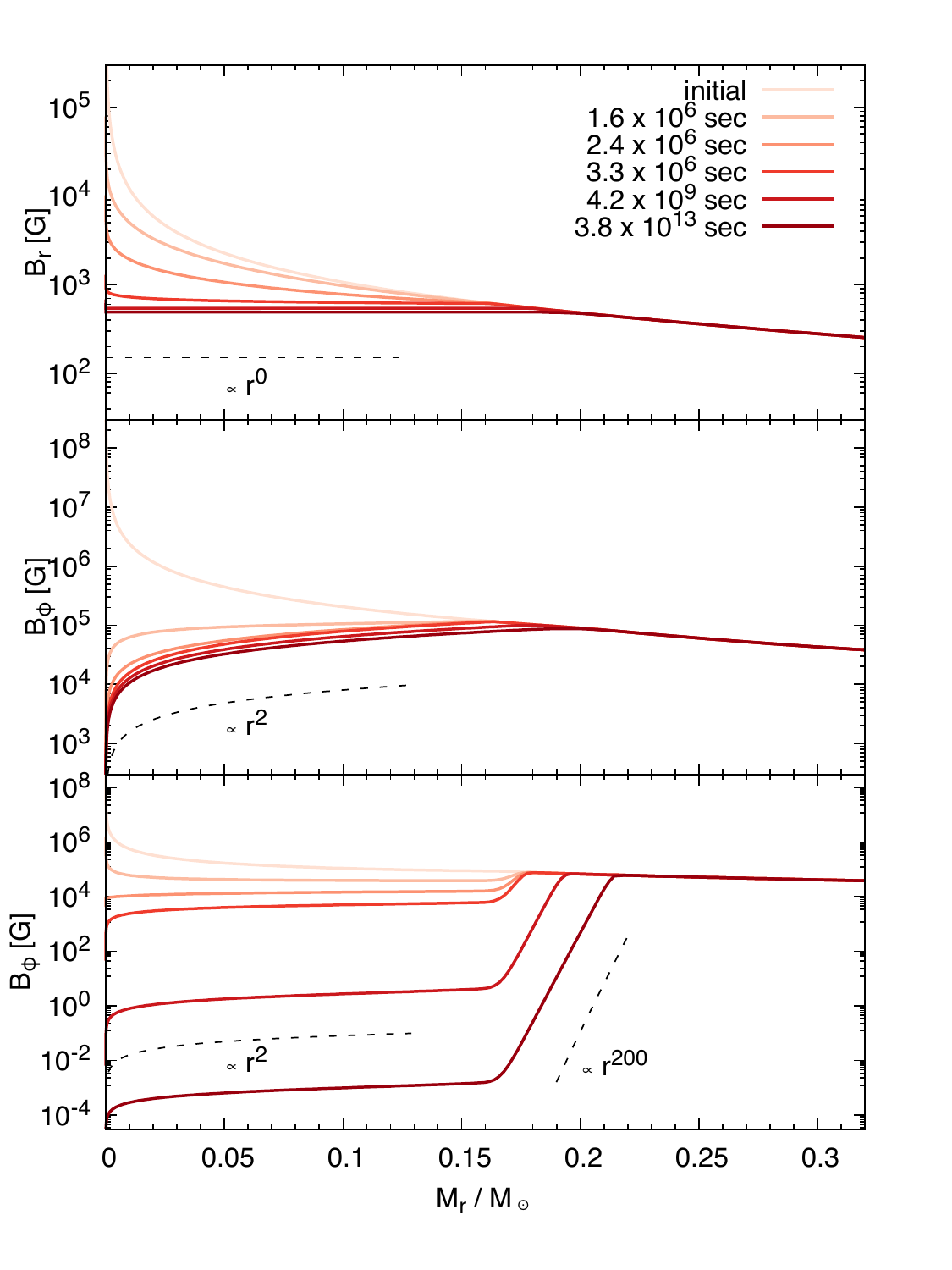}
	\caption{For five different epochs, which are indicated by the legends, the evolution of magnetic field distributions during the core hydrogen-burning phase is shown for the inner 0.34 $M_\odot$ region of a 1.5 $M_\odot$ magneto-rotational star. In this test case, magnetic dissipation by the $\eta$ effect is included and accounts for the short timescale evolution. The top panel shows the radial component of the field, $B_r$. In the middle panel, the toroidal component, $B_\phi$, is shown for a case in which only the diffusion term of the magnetic dissipation is included. Meanwhile, $B_\phi$ evolution for a case with both diffusion and advection terms of the magnetic dissipation is shown in the bottom panel.
	}
	\label{fig:Bdiff}
	\end{center}
\end{figure}

Here the effect of magnetic dissipation is tested. Firstly, magnetic dissipation due to core convection is calculated taking the initial condition from the above test calculation. It is in the main-sequence phase with the central hydrogen mass fraction of 0.3. Both field components initially have a radial dependence of $\sim r^{-3}$. The star at this phase has a convective hydrogen-burning core of $\sim 0.2$ M$_\odot$. The convective turbulence explains the magnetic diffusivity of $\sim 10^{12}$ cm$^2$ s$^{-1}$, which is large enough to establish a steady state for the magnetic field within a much shorter timescale than the evolutionary time.

Figure \ref{fig:Bdiff} shows the result of the evolution of the radial (top) and toroidal (middle and bottom) magnetic field components. As a result of the magnetic diffusion, the radial component reaches a steady state inside the convective region within a timescale of $\sim 10^{9}$ s, where the time derivative of the poloidal field becomes uniformly nearly zero. This steady state is achieved as the magnetic diffusive flux becomes uniform in the convective region: 
\begin{equation}
	\frac{1}{r^2} \frac{ \partial }{ \partial r} (Ar^2) = \rm{(const.)}, \nonumber
\end{equation}
with $B_r \propto r^0$ and $A \propto r$. The convective region is surrounded by an overshoot region, in which magnetic diffusivity exponentially decreases with radius. The small magnetic diffusivity limits the magnetic diffusive flux, explaining the longer timescale of $\sim 10^{13}$ s for the further extension of the steady region. Similarly to the poloidal component, the toroidal component also reaches a steady state at $\sim 10^{9}$ s, and the steady region extends further with a longer timescale of $\sim 10^{13}$ s. The middle panel shows the calculation result in which the magnetic diffusion term in eq.(\ref{eq-jflux3}) is considered but the magnetic advection term is not. Because of the $B_\phi \propto r^2$ distribution, the diffusive flux of the toroidal field becomes uniform as well.

In the bottom panel, the toroidal field evolution of a calculation in which both the magnetic diffusion and the magnetic advection are taken into account is shown. One may suspect that the toroidal component in this case does not reach the steady state because $B_\phi$ at late times does not converge. However, the rate of the change of $B_\phi$ is significantly smaller than the fluxes of the diffusion and the advection. In fact, both fluxes cancel each other to achieve $\partial B/\partial t \sim 0$. In this meaning, again, we consider that the toroidal field evolves keeping a steady state. In this steady state, the radial gradient of the toroidal magnetic field can be determined by solving the equation
\begin{equation}
	0	=\eta r^2 \frac{\partial}{\partial r}\left(
				\frac{1}{r^4} \frac{\partial}{\partial r} (B r^3)
			\right)
		+ \frac{\partial \eta}{\partial r} \frac{\partial Br}{\partial r}, \nonumber
\end{equation}
which yields $0 = m^2 - (n-1) m - (n+6)$, where $m \equiv \partial \ln B/\partial \ln r$ and $n \equiv -\partial \ln \eta/\partial \ln r$. The dissipative region can be divided into the fully convective region of $\lesssim$ 0.17 M$_\odot$ and the surrounding overshooting region at $\sim$ 0.17--0.21 M$_\odot$. In the former region, $n$ is so small that $m \sim 2$ is achieved. Meanwhile, $n$ is as large as $\sim$ 200 in the latter region, resulting in $m \sim 200$ to compensate for the advection flux by the diffusion flux.

\begin{figure}[t]
	\begin{center}
	\includegraphics[width=0.5\textwidth]{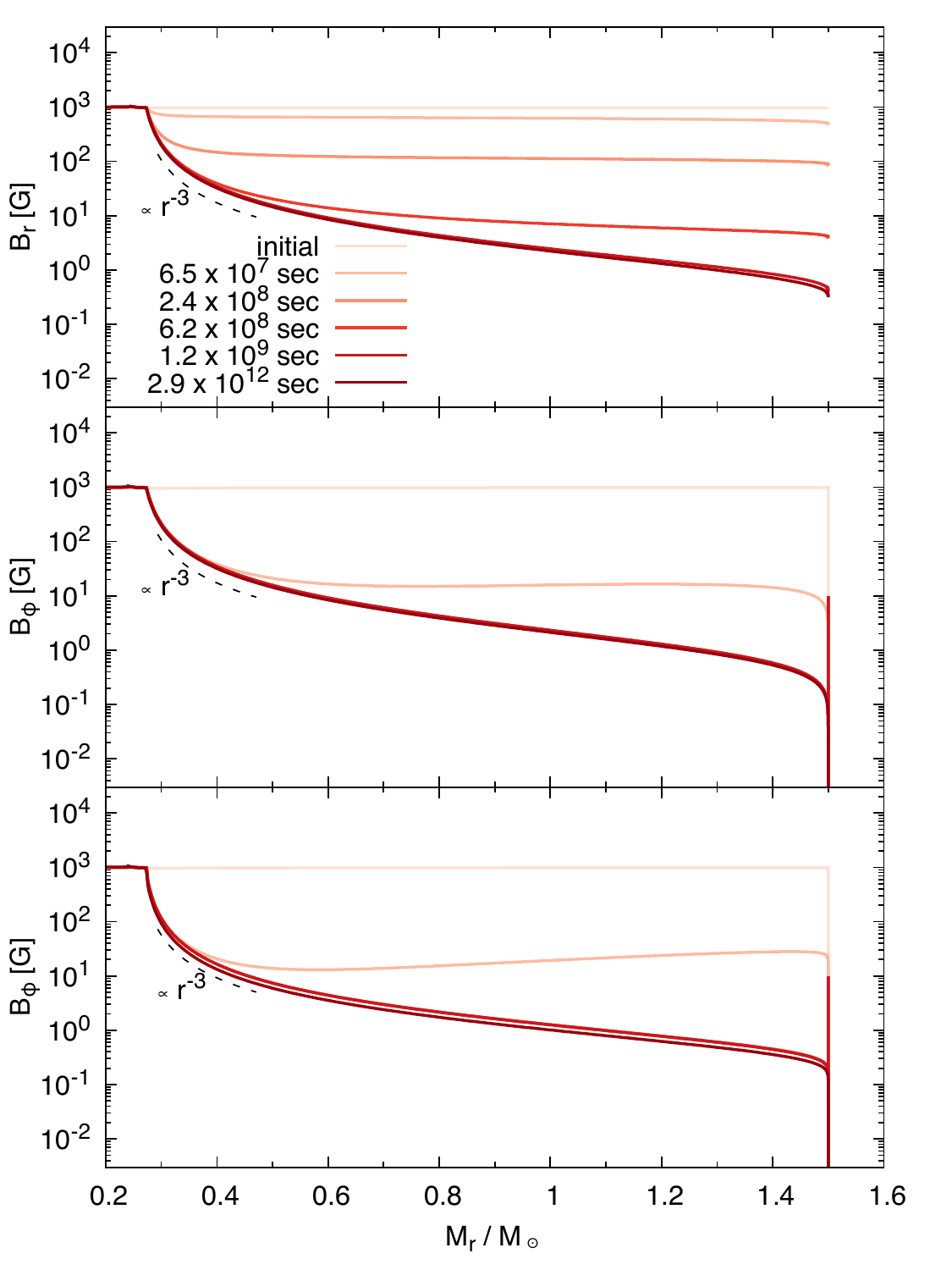}
	\caption{Same as Fig.\ref{fig:Bdiff} but showing magnetic dissipation in a convective envelope during the red-giant phase of a 1.5 $M_\odot$ magneto-rotational star.
	}
	\label{fig:Bdiff-env}
	\end{center}
\end{figure}
Secondly, magnetic dissipation due to the envelope convection is tested in a similar way taking an initial condition with a red-giant envelope of a radius of $R = 10$ R$_\odot$. The uniform initial distributions of $B_r = B_\phi = 1$ kG are applied for the magnetic components in this case. The results are shown in Fig.\ref{fig:Bdiff-env}. Within a diffusion timescale of $\sim 10^{9}$ s, the magnetic field finds a steady state of $B_r \propto B_\phi \propto r^{-3}$, which corresponds to having a zero magnetic diffusive flux. The advection term in this case has only a minor effect, probably due to the much narrower convective overshooting region located at the base of the convective envelope.

In conclusion, the magnetic field is destined to reach a steady state under the efficient magnetic dissipation effects of convection. In a convective core, the magnetic field distributes such that the magnetic flux uniformly distributes. In contrast, the magnetic field distributes such that the magnetic flux becomes zero in a convective envelope.

\end{appendix}

\end{document}